\documentclass[a4paper,11pt]{article}
\pdfoutput=1 

\usepackage{jcappub} 
\usepackage[utf8]{inputenc}
\usepackage[T1]{fontenc}

\usepackage[mathscr]{eucal}
\usepackage[Symbolsmallscale]{upgreek}
\usepackage{xcolor}
\usepackage{slashed}
\usepackage{hyperref}
\usepackage{placeins}
\usepackage{listings}
\usepackage{float}

\usepackage{fontawesome}
\usepackage{xcolor}  

\definecolor{codegreen}{rgb}{0,0.6,0}
\definecolor{codegray}{rgb}{0.5,0.5,0.5}
\definecolor{codepurple}{rgb}{0.58,0,0.82}
\definecolor{backcolour}{rgb}{0.95,0.95,0.92}
\definecolor{linkblue}{rgb}{0.0, 0.0, 0.5}

\hypersetup{
    colorlinks=true,
    linkcolor=linkblue,    
    urlcolor=linkblue,
    citecolor=linkblue
    }

\lstdefinestyle{mystyle}{
    backgroundcolor=\color{backcolour},   
    commentstyle=\color{codegreen},
    keywordstyle=\color{magenta},
    numberstyle=\tiny\color{codegray},
    stringstyle=\color{codepurple},
    basicstyle=\ttfamily\footnotesize,
    breakatwhitespace=false,         
    breaklines=true,                 
    captionpos=b,                    
    keepspaces=true,                 
    numbers=left,                    
    numbersep=5pt,                  
    showspaces=false,                
    showstringspaces=false,
    showtabs=false,                  
    tabsize=2
}

\definecolor{darkgreen}{rgb}{0.0,0.5,0.0}

\title{A Precise Fitting Formula for Gravitational Wave Spectra from Phase Transitions}

\author[a]{Huai-ke Guo,}
\author[b]{Fazlollah Hajkarim,}
\author[b]{Kuver Sinha,}
\author[c]{\\Graham White, }
\author[d,e]{Yang Xiao }

\affiliation[a]{International Centre for Theoretical Physics Asia-Pacific, University of Chinese Academy of Sciences, 100190 Beijing, China}
\affiliation[b]{Department of Physics and Astronomy, University of Oklahoma, Norman, OK 73019, USA}
\affiliation[c]{School of Physics and Astronomy, University of Southampton, Southampton SO17 1BJ, United Kingdom}
\affiliation[d]{CAS Key Laboratory of Theoretical Physics, Institute of Theoretical Physics, Chinese Academy of Sciences, Beijing 100190, P. R. China}
\affiliation[e]{School of Physical Sciences, University of Chinese Academy of Sciences, Beijing 100049, P. R. China}

\emailAdd{guohuaike@ucas.ac.cn}
\emailAdd{fazlollah.hajkarim@ou.edu}
\emailAdd{kuver.sinha@ou.edu}
\emailAdd{g.a.white@soton.ac.uk}
\emailAdd{xiaoyang@itp.ac.cn}

\abstract{
Obtaining a precise form for the predicted gravitational wave (GW) spectrum from a phase transition is a topic of great relevance for beyond Standard Model (BSM)  physicists.  Currently, the most sophisticated semi-analytic framework for estimating the dominant contribution to the spectrum is the sound shell model; however, full calculations within this framework can be computationally expensive, especially for large-scale scans. The  community therefore generally manages with fit functions to the GW spectrum, the most widely used of which is a single broken power law. We provide a more precise fit function based on the  sound shell model: our fit function features a double broken power law with two frequency breaks corresponding to the two characteristic length scales of the problem -- inter-bubble spacing and thickness of sound shells, the second of which is neglected in the single broken power law fit. Compared to previously proposed fits, we demonstrate that our fit function more faithfully captures the GW spectrum coming from a full calculation of the sound shell model, over most of the  space of the thermodynamic parameters governing the phase transition. The physical origins of the fit parameters and their dependence on the thermodynamic parameters are studied in the underlying sound shell model: in particular, we perform a series of detailed scans for these quantities over the plane of the 
strength of the phase transition ($\alpha$) and the bubble wall velocity ($v_w$). Wherever possible, we comment on the physical interpretations of these scans.  
The result of our study can be used to generate accurate GW spectra with our fit function, given initial inputs of $\alpha$, $v_w$, $\beta/H$ (nucleation rate parameter) and $T_n$ (nucleation temperature) for the relevant BSM scenario.
\href{https://github.com/SFH2024/precise-fit-fopt-gw}{\faGithub}
}

\begin{document}
\maketitle
\flushbottom

\section{Introduction}
\label{sec:intro}

One of the most plausible cosmological mechanisms for producing a stochastic gravitational wave (GW) background is a first order phase transition \cite{Caprini:2019egz}
. The most common motivation for such an epoch is electroweak symmetry breaking, which can become strongly first order in extensions to the Standard Model \cite{Caldwell:2022qsj, Alves:2019igs, Dorsch:2013wja,Basler:2016obg,Dorsch:2016tab,Dorsch:2016nrg,Bernon:2017jgv,Dorsch:2017nza,Andersen:2017ika,Kainulainen:2019kyp,Wang:2019pet,Su:2020pjw,Davoudiasl:2021syn,Biekotter:2021ysx,Zhang:2021alu,Aoki:2021oez,Goncalves:2021egx,Phong:2022xpo,Biekotter:2022kgf,Anisha:2022hgv,Atkinson:2022pcn,Biekotter:2023eil,Goncalves:2023svb,Graf:2021xku,Arcadi:2022lpp,Blinov:2015vma,Huang:2017rzf,Paul:2019pgt,Fabian:2020hny,Benincasa:2022elt,Paul:2022nzx,Jiang:2022btc,Astros:2023gda,Benincasa:2023vyp,Carena:1996wj,Espinosa:1996qw,Delepine:1996vn,Cline:1996cr,Losada:1996ju,Laine:1996ms,Bodeker:1996pc,deCarlos:1997tma,Cline:1998hy,Losada:1998at,Laine:2000rm,Carena:2008vj,Delgado:2012rk,Carena:2012np,Chung:2012vg,Huang:2012wn,Laine:2012jy,Liebler:2015ddv,Patel:2012pi,Chala:2018opy,Baum:2020vfl,Kazemi:2021bzj,Baldes:2021vyz,Azatov:2021irb,Jiang:2015cwa,Chiang:2017nmu,Ahriche:2018rao,Chen:2019ebq,Cho:2021itv,Schicho:2022wty,DiBari:2023upq,Espinosa:1993bs,Profumo:2007wc,Cline:2009sn,Cline:2012hg,Cline:2013gha,Katz:2014bha,Profumo:2014opa,Vaskonen:2016yiu,Hashino:2016xoj,Chao:2017vrq,Matsui:2017ggm,Kang:2017mkl,Alves:2018jsw,Shajiee:2018jdq,Beniwal:2018hyi,Matsui:2018tpp,Kannike:2019mzk,Demidov:2021lyo,Cline:2021iff,Cao:2022ocg,Athron:2023xlk,Alanne:2020jwx,Bian:2019kmg,Xie:2020bkl,Grojean:2004xa,Bodeker:2004ws,Delaunay:2007wb,Balazs:2016yvi,deVries:2017ncy,DeVries:2018aul,Chala:2018ari,Ellis:2019flb,Postma:2020toi,Batell:2023wdb,Ghosh:2023aum}. Other motivations for a cosmic first order phase transition include dark matter \cite{Schwaller:2015tja,Baldes:2018emh,Breitbach:2018ddu,Croon:2018erz,Hall:2019ank,Baldes:2017rcu,Croon:2019rqu,Hall:2019rld,Chao:2020adk,Dent:2022bcd,Helmboldt:2019pan,Aoki:2019mlt,Helmboldt:2019pan,Croon:2019ugf,Croon:2019iuh,Alanne:2020jwx,Garcia-Bellido:2021zgu,Huang:2020crf,Halverson:2020xpg,Kang:2021epo,Fornal:2022qim,Costa:2022oaa,Benincasa:2023vyp}, symmetry breaking chains from a grand unified group \cite{Hashino:2018zsi,Huang:2017laj,Croon:2018kqn,Brdar:2019fur,Huang:2020bbe,Graf:2021xku,Fornal:2023hri}, or the QCD transition \cite{Costa:2008gr,Marquez:2017uys,Vovchenko:2018eod,CamaraPereira:2020rtu,Gao:2020qsj,Gao:2020fbl,Gao:2021nwz,Gao:2023djs,Halverson:2020xpg,Bigazzi:2020phm,Bigazzi:2020avc,Huang:2020crf,Wang:2020zlf,Kang:2021epo,Garcia-Bellido:2021zgu,Morgante:2022zvc}. The GW spectrum resulting from a cosmological phase transition has three contributions - a term arising from the collision of scalar shells, a term arising from crashing sound waves in the plasma and the after party of turbulence. For most of the parameter space, it is widely accepted that the sound shell source produces the dominant contribution to the stochastic GW background \cite{Caprini:2019egz}. 

 At the present time, the most widely used  semi-analytical framework for calculating the velocity spectrum sourced by sound waves  is provided by the sound shell model \cite{Hindmarsh:2016lnk}. The corresponding GW spectrum was  calculated for a non-expanding Universe in \cite{Hindmarsh:2019phv}, and subsequently extended to an expanding Universe in \cite{Guo:2020grp}. Alongside the semi-analytical framework,  simulations have recently also made great progress in this area \cite{Cutting:2018tjt, Cutting:2020nla, Gould:2021dpm} and motivate the use of a broken power law for the GW spectrum, which has been adopted by the LISA cosmology working group \cite{Caprini:2019egz}.  
 The single broken power law comes from the simplifying assumption of ignoring the the details of the velocity profile of the sound waves, and instead taking the RMS \footnote{Root mean square.} 
 of the fluid velocity, $U_f$; the peak frequency is then set by the mean bubble separation \cite{Weir:2017wfa}. On the other hand, recent work with the sound shell model \cite{Gowling:2021gcy, Gowling:2022pzb} motivates using a double broken power law, which comes from including the full velocity profile. The use of the full velocity profile results in qualitatively different physics: the spectrum is  better described by a double broken power law since the velocity profile now contains information about the thickness of the profile, which implies that there is more than one physically significant scale.

 It should be noted that the dependence of the spectral features of a fit on the underlying thermodynamic parameters of the phase transition (the nucleation temperature $T_n$, nucleation rate parameter $\beta$, phase transition strength $\alpha$, and wall speed $v_w$) is a difficult question and progress can be made by a combination of analytical calculations and numerical fits. In principle, one could imagine the possibility of reconstructing the values of four thermal parameters from a hypothetical spectrum \cite{Gowling:2021gcy}, (though a more conservative study argued for three \cite{Giese:2021dnw}). Such a wealth of information, if it were to be available, would render GW detectors almost competitive with colliders in the extent of details of the underlying model they can reveal \cite{Alves:2018jsw}. In a sense, the community is actually already in the era where the degree of precision we use in calculating GW spectra becomes important, since LIGO currently puts constraints on the Pati-Salam unification scenario \cite{Athron:2023aqe}. This urgency will grow as next generation detectors  cover many more decades in frequency at a much high strain sensitivity (for a review see \cite{Roshan:2024qnv}). 
From a beyond Standard Model (BSM) physicist's point of view, the ideal workflow would be the following: for every point on the parameter space of the model, obtain the thermodynamic parameters $(T_n,\, \beta,\, \alpha,\, v_w)$, and compute the GW spectrum using  the sound shell model (or better still, use actual simulations); then check how the parameter space is constrained by GW detectors. Needless to say,  full calculations of the sound shell model, not to speak of actual simulations, are difficult and expensive; to perform large-scale scans over the parameter space of   particle physics models and obtain the predicted GW spectrum, portable and simple fits  to the spectrum, and the interpretation of the fit parameters, become necessary. 

The main purpose of this paper is to provide such a precise fit function. Elaborating further, our goal in the current paper will be the following: 
given a particle physics model with input values of $(T_n,\, \beta,\, \alpha,\, v_w)$, an appropriate fit function  for the GW spectrum is provided, along with a public code.  A historical progression of fit functions can help contextualize our results: the cornerstone of GW phenomenology has been a broken power law  with a peak set by the mean bubble separation; this was recently improved upon by the double broken power law fit function  introduced in~\cite{Gowling:2021gcy,Gowling:2022pzb} with fit  parameters corresponding to the peak amplitude, two frequency breaks and the spectral slope between them; our fit function is a further improvement, by introducing an extra parameter that governs the IR behavior of the spectrum. Scanning over $(\alpha, v_w)$, we show that our fit function reproduces, much more faithfully, the GW spectrum coming from the sound shell model (Fig.~\ref{fig:hist}). We provide data files and scripts in \texttt{Python} and \texttt{Mathematica} that can be directly used by a front-end user to generate accurate GW spectra, given inputs of $(T_n,\, \beta,\, \alpha,\, v_w)$ for a given beyond-SM scenario. This tool is expected to allow the community to fully investigate the extent to which each individual BSM model can be constrained by future detectors, without actually performing calculations with the sound shell model.

The second purpose of our paper is to  attempt to study the physical interpretation of the fit parameters by way of detailed scans on the plane of $(\alpha, v_w)$. We provide scans over $(\alpha, v_w)$ in the deflagration, hybrid, and detonation regimes and study the velocity and enthalpy profiles, as well as a host of quantities: the maximum velocity $v_{max}$, the thermal efficiency factor $\kappa$, the peak frequency $f_p$, the maximum relic density $\Omega_p$, the width of the velocity  profile $\delta_{\xi}$, and the fit parameters $\tilde{b}$ and $\tilde{a}$. The dependence of these quantities on $(\alpha, v_w)$ can be highly non-linear. Wherever possible, we make qualitative statements about this dependence, as well as  correlations with the relative behaviour of other parameters.

This paper is organized as follows. In Section~\ref{generalform},  we discuss the general framework for the calculation of GWs in a first order phase transition and show, in turn, the standard broken power law approximation (Section~\ref{brokenpower}) and the improved double broken power law fit (Section~\ref{doublebroken}). 
 In Section \ref{sec:newfit} we introduce our new fit function and discuss its faithfulness in capturing the GW spectrum compared to the older fits. The dependence of various physical and fit parameters on $(\alpha, v_w)$ is discussed in  Section \ref{sec:fitfopt}, where we also display the results of our scans. Discussions and code for using the new fit function \footnote{It can be downloaded from this  \href{https://github.com/SFH2024/precise-fit-fopt-gw}{{\tt GitHub} link}.} are relegated to the Appendix \ref{app-a} and \ref{app-b}. Finally we summarize and conclude in Section~\ref{conclu}.

\section{General Framework for First Order Phase Transition}\label{generalform}
In this section we briefly review the framework for the calculation of GWs from sound waves during a first order phase transition (FOPT) \cite{Mazumdar:2018dfl,Hindmarsh:2013xza,Hindmarsh:2015qta,Espinosa:2010hh,Caprini:2009yp,Hindmarsh:2017gnf,Hindmarsh:2019phv,Guo:2021qcq,Guo:2020grp,Gowling:2021gcy,Hindmarsh:2016lnk}.
The details of a gravitational wave spectrum produced by sound waves depend upon a number of macroscopic features of the phase transition. For instance, it matters how far apart the bubbles are during the phase transition, how fast the bubbles expand, the fraction of energy released during the phase transition that is dumped into kinetic energy modes, the details of the fluid veloicty and of course the temperature at which all this occurs. 

Since the background is expanding, the key parameter for describing the nucleation of bubbles during the phase transition is not the nucleation rate - as the phase transition always begins when the nucleation rate is fast enough so that the nucleation of bubbles can keep pace with the expanding Universe. Instead, it is customary to consider how fast the nucleation rate increases from this bare minimum rate
\begin{equation}
    \Gamma \sim e^{-S(t)}\sim e^{-S_0 + \beta (t-t_n)} 
\end{equation}
where $t_n$ is the time of nulceation at $S_0$ is the value needed to have at least one bubble per hubble volume
. Unless there is a significant amount of supercooling \cite{Athron:2023aqe}, it is $\beta$ that actually controls the time scale of the transition. It is straightforward to relate this parameter to the mean bubble separation \cite{Hindmarsh:2019phv,Gowling:2021gcy}
, $R_\ast$  
\begin{eqnarray}
\frac{\beta}{H_n}= \left(8 \pi\right)^{\frac13} \frac{v_w}{r_{*}}\,. 
\end{eqnarray}
Here, the wall velocity is denoted by $v_w$ and 
\begin{eqnarray}
\label{eq:rstar}
r_* = H_n R_*\,,
\end{eqnarray}
where $H_n$ is the Hubble rate at the nucleation time.
The fraction of the kinetic energy dumped into the plasma, as well as the details of the velocity profile tend to depend upon how strong the transition is. This is governed by the change in the trace anomaly, $\Delta \theta$ in the energy momentum tensor between the phases, normalized to the enthalpy of the symmetric phase, $w_s$, %
\begin{equation}\label{Eq:alpha}
 \alpha=\left.\frac{4}{3} \frac{\Delta \theta}{w_{\mathrm{s}}}\right|_{T=T_{\mathrm{n}}}\,.
 \end{equation}
Making predictions in the sound shell model require solving the hydrodynamic equations and using the macroscopic thermal parameters of a phase transition as boundary conditions. We use a bag model to match solutions of thermodynamic variables at the boundary between the two phases. We consider a non-vanishing bag constant, which we denote as $\epsilon$, \cite{Giese:2020znk} which has the value %
\begin{eqnarray}
    \label{eq:bag-eps}
    \alpha = \frac{4}{3}\frac{\epsilon}{ w_{n}}\,.
\end{eqnarray}
There are three qualitatively distinct types of phase transitions which are delineated by the value of the bubble wall velocity, compared to the Jouget velocity, $v_j$ and the speed of sound, $c_s$ \cite{Weir:2017wfa,Krajewski:2024gma,Hindmarsh:2019phv}. Specifically, the Jouget velocity is defined as
\begin{equation}\label{Eq:Jouguet}
c_J = c_\text{s} \frac{\left(1 + \sqrt{\alpha(2 + 3\alpha) }\right)}{\left(1 + \alpha\right)}\,, 
\end{equation}
where the detonation regime occurs when $v_w>c_J>c_s$
, a deflagration of subsonic regime is when $v_w<c_J<c_s$. 
A hybrid transition involves both deflagration and detonation phases. Initially, $v_w<c_s$ and $v_w<c_J$ indicating a deflagration. However, as the phase transition evolves, energy accumulates in the bubble wall or in a shock front preceding the wall, potentially allowing the bubble to accelerate to a velocity $v_w$ such that $c_s<v_w<c_J$ or even $v_w>c_J$, at which point a detonation may occur \cite{Weir:2017wfa,Krajewski:2024gma,Hindmarsh:2019phv}. 

The relative density of GW for each value of $z=k R_*$ corresponding to a specific frequency $k=2 \pi f$ is controlled by the kinetic energy fraction $K$, as well as \cite{Hindmarsh:2019phv,Gowling:2021gcy} the lifetime of the soundshell source, $H_n \tau _v$ and its characteristic scale, $H_n R_\ast$, for which we use the shorthand
\begin{equation}\label{Eq:scaling_factors_GW}
J = H_n R_* H_n\tau_v  = r_* \left(1 -  \frac{1}{\sqrt{1 + 2x}} \right) \equiv r_* \Upsilon \, ,
\end{equation}
where $x = H_n R_* / \sqrt{K} $ and $r_*$ defined in Eq.~(\ref{eq:rstar}). 
We can then write the gravitaitonal wave power spectrum at the time it is produced as,
\begin{equation}\label{Eq:Omgw_ssm}
 \Omega_{\mathrm{GW}}(z) = 3K^{2}(v_{w},\alpha)J \frac{z^{3}}{2 \pi^{2}} \tilde{P}_{\mathrm{GW}}\left(z\right) . 
 \end{equation}
If one takes the RMS fluid velocity, rather than calculating the gravitational wave spectrum from the full velocity profile, one can write the frequency dependent analytical form of spectral density $z^3 \tilde{P}_{\mathrm{GW}}$ for GW from Refs.~\cite{Huber:2008hg,Weir:2017wfa} as a broken power law %
\begin{eqnarray}
	S_{\rm SW} (f) = \left( \frac{f}{f_{\rm SW}}\right)^3 \left[ \frac{7}{4 + 3(f/f_{\rm SW})^2}\right]^{7/2} .
\end{eqnarray}
We will be comparing the predictions of the soundshell model to this single broken power law throughout, as this simplified case is ubiquitous in the literature. 

 In such a case we take the kinetic energy fraction as $K=\Gamma \bar{U_f}^2$ where $\Gamma =\bar{w}/\bar{v}$ is the ratio of the average enthalpy to the average fluid velocity. Further, the mean squared velocity defines \cite{Ellis:2020awk,Hindmarsh:2019phv}
 \begin{equation}
\overline{U}_f^2 = \frac{3}{\bar{w} v_w^3} \int w \xi^2 \frac{v^2}{1 - v^2} d\xi
\end{equation}
that can be approximated $\overline{U}_f \approx \sqrt{\frac{3}{4} \frac{\kappa \alpha}{1 + \alpha}}$ and the quantity $\kappa$ is given by \cite{Hindmarsh:2019phv}
\begin{eqnarray}
    \kappa = \frac{3}{\epsilon w^3} 
    \int
    d\xi \, \xi^2 w^2 \gamma^2 \nu^2 \ .
    \label{eq:kappa}
\end{eqnarray}
where we have defined
\begin{equation}
\xi \, \equiv \, r/t \,\,,
\end{equation}
with $r$ being the distance from the center of the bubble and $t$ being the time from the onset of the phase transition and nucleation.

The sound shell model differs from the results of numerical simulations as some sound shells collide before they reach a self similar solution, that is a solution that depends upon the ratio $\xi=r/t$ only. The effect of this was numerically approximated in Ref.~\cite{Gowling:2021gcy} by introducing an error factor
\begin{equation}\label{Eq:SSM_suppressed}
\Omega_{ \rm GW}(z) = \Omega_{\mathrm{GW}}^{\mathrm{SSM}}(z)\Sigma(v_{w},\alpha),
\end{equation}
where $\Omega_{\mathrm{GW}}^{\mathrm{SSM}}(z)$ is the relic from sound wave model at redshift $z$ and $\Sigma(v_{w},\alpha)$ compensates from the sound shell model's overestimation of the gravitational waves due to energy lost into vorticity modes as demonstrated in Ref.~\cite{Cutting:2019zws}. As the precise nature of this factor for all possible thermal parameters is not known, we assume it as $\Sigma(v_{w},\alpha)=1$ in order to focus on other aspects of sound shell model that affect the GW signal. The reader should note that it is a simple fix to include the results of Ref. \cite{Cutting:2019zws} if one is working in a regime where interpolations of their result can be used.\par
After formation the power spectra will redshift like radiation leading to its dilution and a shift in the peak frequency such that,
\begin{equation}
\label{eq:gwrelic}
    \Omega^0_{\mathrm{GW}}(f) = F_{\mathrm{GW}}^0 \Omega_{\text{GW}}(z(f)),
\end{equation}
where the $F_{\mathrm{GW}}^0 $ factor that redshifts the GW relic from its  moment of production in the early Universe to today is 
\begin{equation}\label{Eq:Fgw0_def}
F_{\mathrm{GW}}^0=\Omega_{\gamma, 0}\left(\frac{g_{*s,0}}{g_{*s }}\right)^{\frac{4}{3}} \frac{g_{*}}{g_{*,0}} \simeq (3.57 \pm 0.05) \times 10^{-5} {\bigg( \frac{100}{g_*}\bigg)}^{\frac{1}{3}} \,\,,
\end{equation}
assuming $g_* \simeq g_{*s}$ at $T\gg T_{\nu}$.The peak frequency is set by the mean bubble separation and a redshift factor
\begin{equation}
\label{eq:fz}
f =\frac{z }{r_*} f_{*,0},
\end{equation}
where $z=k R_*$ and the redshift factor is given by
\begin{equation} 
\label{eq:f0}
f_{*,0}=  2.6 \times 10^{-6} \,\textrm{Hz} \left(\frac{T_{n}}{100\,\textrm{GeV}}\right)\left(\frac{g_*}{100}\right)^{\frac{1}{6}}\,.
\end{equation}
Here we use the degrees of freedom of the thermal background of SM particles for energy density $g_*$ and entropy density  $g_{*s}$ from Ref.~\cite{Drees:2015exa}.

Putting everything together, the relic density of GW from sound waves is given by \cite{Guo:2020grp,Weir:2017wfa}
\begin{equation}
	 \Omega_{\rm GW} h^2 \simeq  1.2 \times 10^{-6} \left( \frac{100}{g_*}\right)^{1/3} K^2 \Upsilon  \left( \frac{H_n}{\beta}\right) v_{w}  S_{\rm SW}(f)\,.
	\label{eq:omega_suppressed} 
\end{equation}
The low frequency tail of the GW spectrum from sound waves can either have $k^3$ or $k^9$ dependence based on recent studies \cite{RoperPol:2023dzg,RoperPol:2023bqa,Sharma:2023mao}. We will address this issue  in the future.

\subsection{Broken Power Law Approximation}\label{brokenpower}
If the velocity profile is replaced with the RMS fluid velocity, there is a single scale left in the mean bubble separation. Thus under this approximation, the gravitational wave power spectrum is a broken power law (BPL) with a peak set by the mean bubble separation. This approximation has been the cornerstone of gravitational wave phenomenology and we will briefly review it here, following Ref.~\cite{Espinosa:2010hh}. The peak amplitude is set by a thermal efficiency factor, which ref \cite{Espinosa:2010hh} develop an analytic approximation for,
\begin{eqnarray}
\kappa(\alpha, v_w) \equiv \begin{cases} 

v_w^{6/5} \frac{6.9 \alpha}{1.36 - 0.037 \sqrt{\alpha} + \alpha}~, & ~~0\lesssim v_w \lesssim 0.2, \\
\frac{\alpha^{2/5}}{0.017 + (0.997 + \alpha)^{2/5}}~, & 0.2 \lesssim v_w \lesssim 0.8, \\ \frac{\alpha}{0.73 + 0.083 \sqrt{\alpha} + \alpha}~, &  0.8\lesssim v_w \lesssim 1.
\end{cases}
\end{eqnarray}
In the above we have given (very) rough limits of validity of each expression with approximate values of $v_w$.
The peak frequency of the  spectrum is given by 
\begin{eqnarray}
    f_{p}^{\rm BPL}= 1.19 \times 10^{-6} \frac{1}{v_w} \left(\frac{g_{*}}{100}\right)^{1/6} \left(\frac{T}{100}\right) \times 0.7 \times \frac{\beta}{H_n}\,,
\end{eqnarray}
with the mean bubble separation being the only scale in the problem
\begin{eqnarray}
    R_* = (8 \pi)^{1/3}  \frac{\beta}{v_w}\,.
\end{eqnarray}
The peak assuming adiabatic index $\Gamma \approx 4/3$ is then set in our approximation mostly by the RMS fluid velocity \cite{Weir:2017wfa}
\begin{eqnarray}
\label{eq:om_bpl}
    \Omega_{p}^{\rm BPL} \approx 1.2 \times 10^{-6} \left(\frac{100}{g_{*}}\right)^{1/3} \left(\frac{4}{3}\right)^2 \bar{U}_f(\alpha, v_w)^4 \frac{H_n}{\beta} v_w 
    \Upsilon \,,
\end{eqnarray}
which can be related to the thermal efficiency factor and trace anomaly,
\begin{eqnarray}
    \bar{U}_f \approx \sqrt{\frac{3}{4} \kappa(\alpha, v_w) \alpha}\,.
\end{eqnarray}

\begin{figure}[H]
  \centering
  \includegraphics[width=0.45\linewidth]{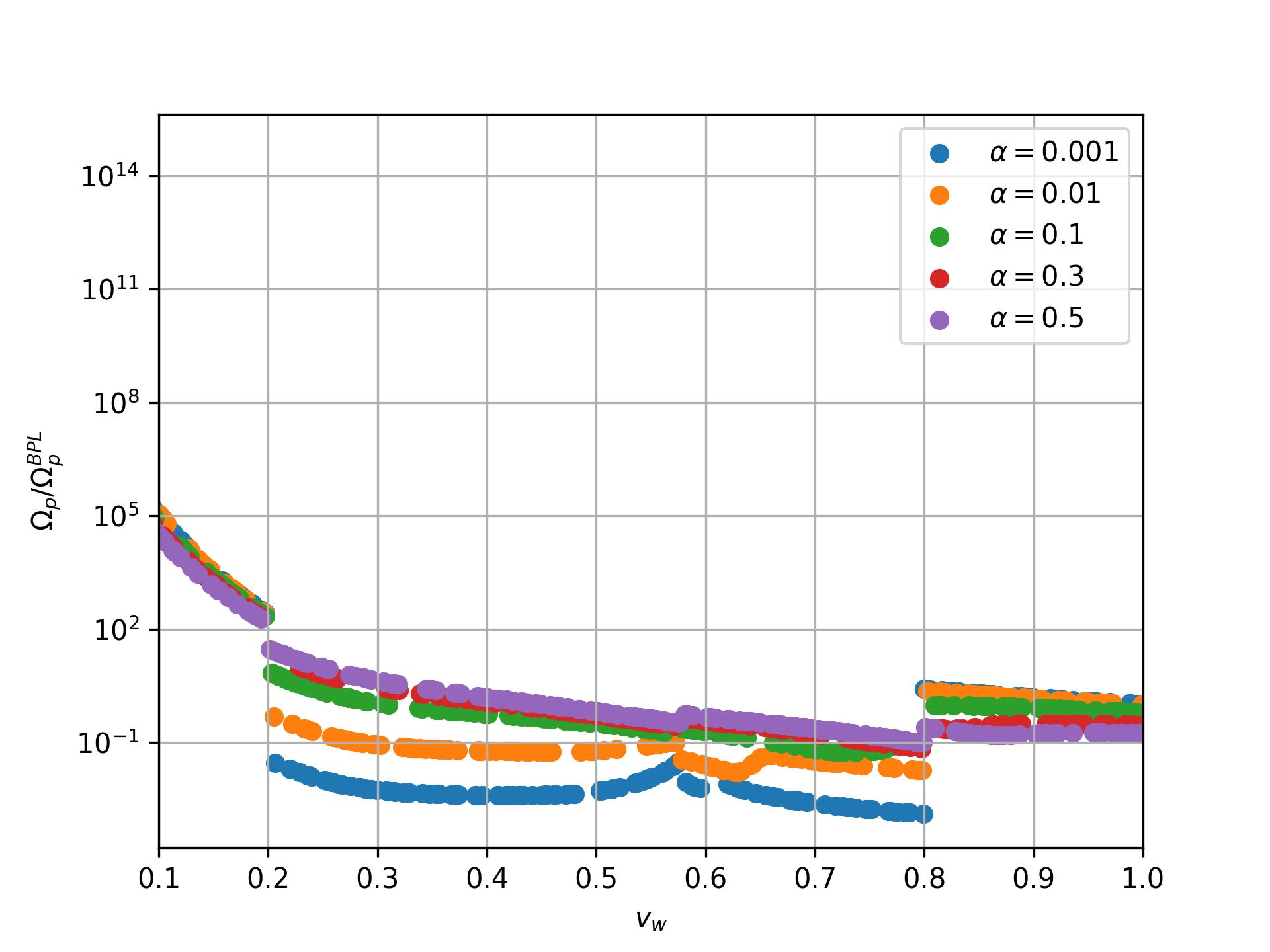}
  \includegraphics[width=0.45\linewidth]{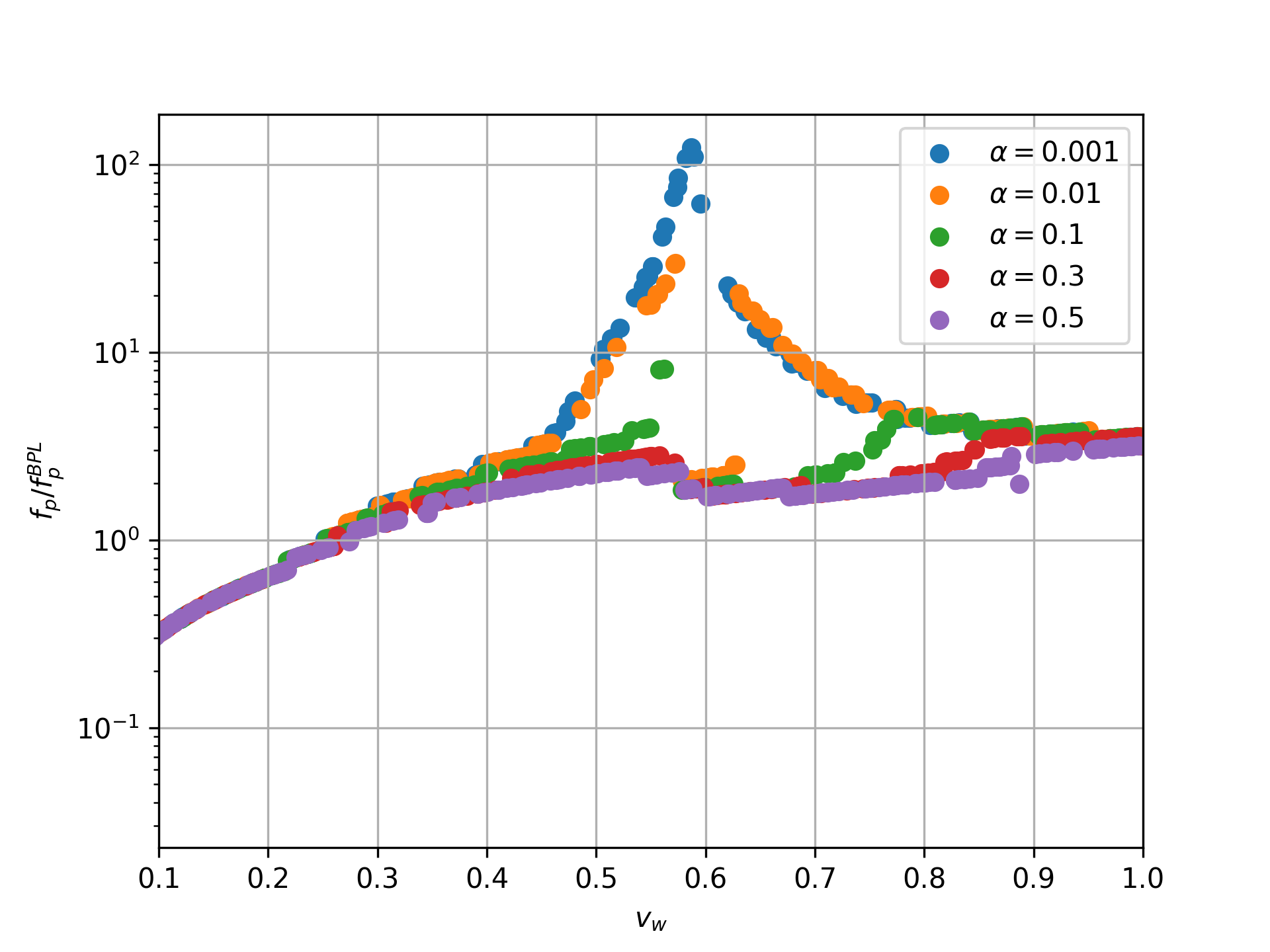}
    \includegraphics[width=0.45\linewidth]{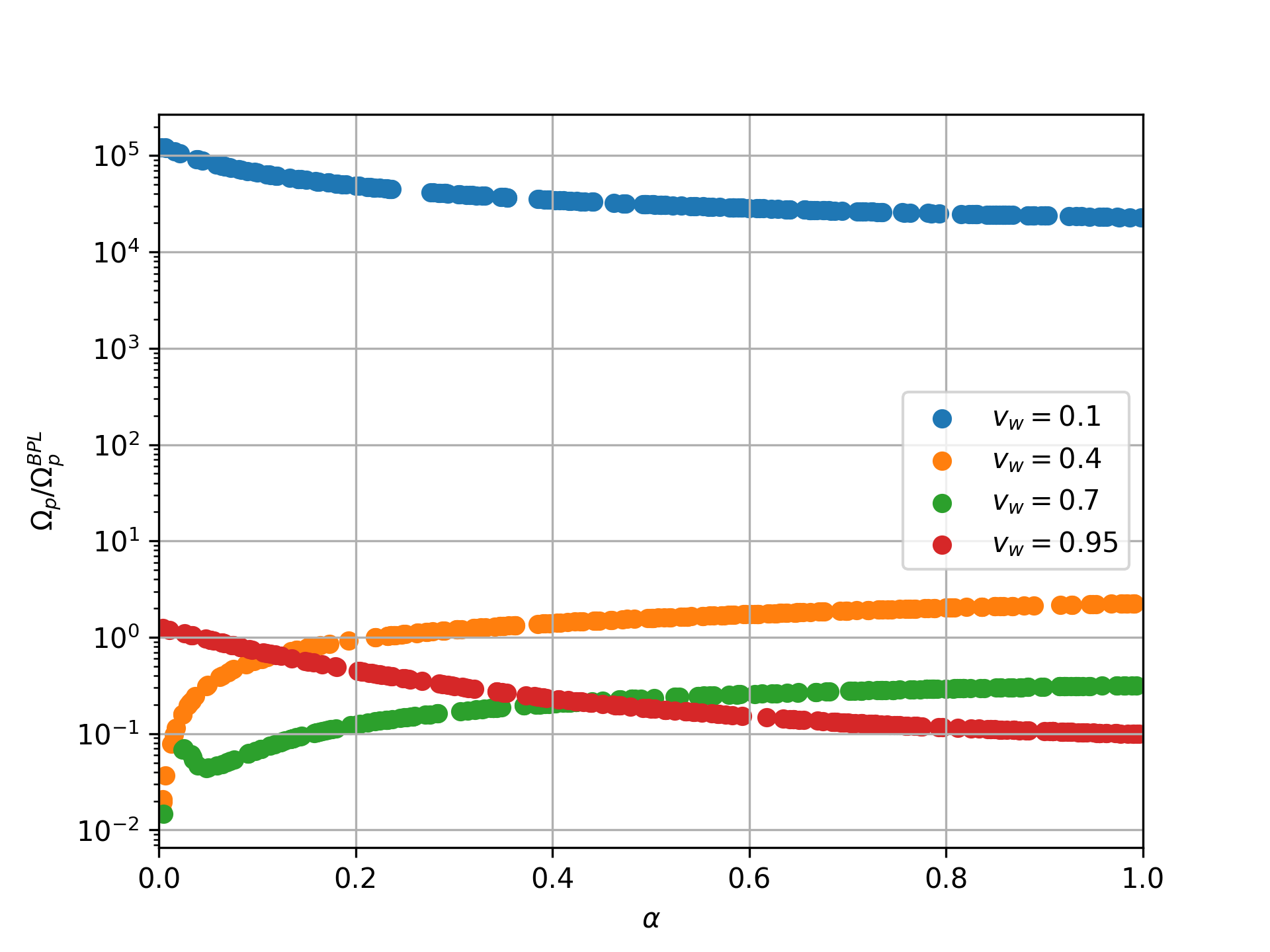}
  \includegraphics[width=0.45\linewidth]{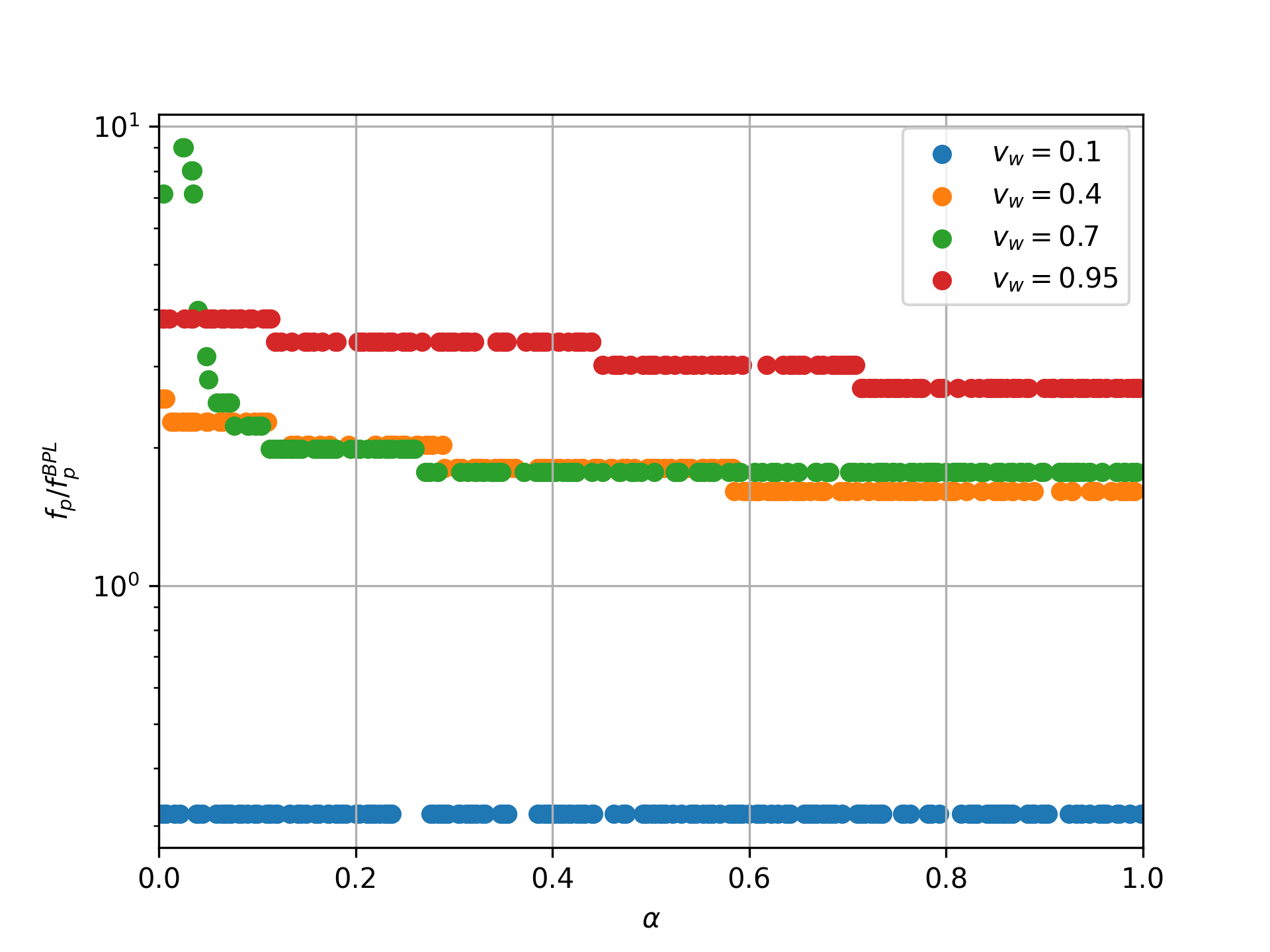}

    \caption{Ratio of $\Omega_p/\Omega_p^{\rm BPL}$ and $f_p/f_p^{\rm BPL}$ in different cases of either fixed $\alpha$'s or $v_w$'s. For the definition of $\Omega_p^{\rm BPL}$ see Eq.~(\ref{eq:om_bpl}).}
  \label{fig:bpl-comp}
\end{figure}

The approximation can be rather poor. In Fig.~\ref{fig:bpl-comp}, we show a comparison of the GW relic density coming from the full sound shell model and that calculated using  the broken power law given by Eq.~(\ref{eq:om_bpl}). The left (right) panel of the first row shows the ratio of the peak relic densities (frequencies) as a function of $v_w$ for several fixed values of $\alpha$. The second row depicts the same quantities as a function of $\alpha$, for several fixed values of $v_w$. There are two metrics of comparison to keep track of. Firstly, it is clear that $\Omega_p/\Omega_{p}^{\rm BPL} \sim 1$ and $f_p/f_{p}^{\rm BPL} \sim 1$ is only achieved for a narrow range of values of $\{v_w, \alpha\}$. For example, from the top and bottom left panels, it is clear that $\Omega_p/\Omega_{p}^{\rm BPL} \sim 1$ is achieved for $\alpha \sim 0.2 - 0.4$ with $v_w \sim 0.4$. Similarly, from the top and bottom right panels, it is clear that $0.1 \lesssim  f_p/f_{p}^{\rm BPL} \lesssim 1$ is  achieved for $v_w \lesssim 0.3$ and for $f_p/f_{p}^{\rm BPL} \sim 0.3$ for  $v_w \gtrsim 0.9$  while being relatively stable across values of $\alpha$.

\subsection{Double Broken Power Law Fit}\label{doublebroken}
Using the full velocity profile in calculating the gravitational wave spectrum (re)introduces a second scale, the thickness of the sound shell. A double broken power law fit function to match the calculated GW   spectrum was  therefore introduced in Refs.~\cite{Gowling:2021gcy,Gowling:2022pzb}
\begin{eqnarray}
\label{eq:oldfit}
       \Omega_{\rm GW, fit} =F^0_{\rm GW}\Omega_{\rm p} M(s,r_b, b) \,,
\end{eqnarray}
Here, $\Omega_p$ is the  peak value of the GW relic density, and the corresponding frequency is $f_p$. Also, $F^0_{\rm GW}$ is determined in Eq.~(\ref{Eq:Fgw0_def}). 
The parameter
   $r_b =  f_{\text{b}} /f_p$ gives the ratio between the two breaks in the spectrum. The parameter $b$ gives the spectral slope between the two breaks. The variable $s = f/f_p$ is introduced to normalize the frequency with respect to the peak frequency. The function $M$ is given by~\cite{Hindmarsh:2019phv,Gowling:2021gcy,Gowling:2022pzb}
\begin{eqnarray}
\label{eq:fitfun}
        M ( s, r_b , b ) = s^ { 9 } {\left( \frac { 1 + r_b^4 } { r_b^4 + s^4}\right)}^{(9 -b)/4}  \left( \frac { b +4 } { b + 4 - m + m s ^ { 2 } } \right) ^ { (b +4) / 2 } \,,
\end{eqnarray}
where the choice of $m$ ensures that for $r_b<1$ one gets a peak at $s=1$  and $M(1,r_b,b) = 1$. This leads to
\begin{eqnarray}
    m = \left( 9 {r_b}^4+ b\right) / \left( {r_b}^4 +1 \right)\,.
\end{eqnarray}
In our work, when using the fit formula from~\cite{Gowling:2021gcy} we use {\tt PySwarm} optimization package to determine $b$ and $r_{b}$. We consider a range of values for $b$ and $r_{b}$: $b \in [-200,200]$ and $r_b \in [0,200]$.

\section{New Fit Formula }
\label{sec:newfit}
We will find that the fit function Eq.~\ref{eq:fitfun} described in the previous Section gives a poor approximation when calculating the accoustic contribution to the gravitational wave spectrum where the hydrodynamic equations are solved using a non-vanishing bag constant $\epsilon$. However, physically we still expect a doubly broken power law to work. We therefore, consider the following new fit function that is different from~\cite{Gowling:2021gcy}
\begin{equation}
\label{eq:newfit}
\Omega_{\rm GW} h^2 = \Omega_p \cdot \left(\frac{f}{\tilde{s}_0}\right)^9 \cdot \frac{\left(2 + \tilde{r}_b^{-12 + \tilde{b}}\right)}{\left[\left(\frac{f}{\tilde{s}_0}\right)^{\tilde{a}} + \left(\frac{f}{\tilde{s}_0}\right)^{\tilde{b}} + \tilde{r}_b^{-12 + \tilde{b}} \cdot \left(\frac{f}{\tilde{s}_0}\right)^{12}\right]}\,\,.
\end{equation}
Here  the parameters $\Omega_p$, ${\tilde s}_{0}$, $\tilde{a}$, $\tilde{b}$ and ${\tilde r}_b = f_b / f_p$ are calculated by fitting to the numerical solution to the GW spectrum. The parameter, $\tilde{r}_b$ controls the ratio between the peaks, $\tilde{b}$ adjusts the spectral slope between the two breaks in the frequency, while $\tilde{a}$ governs the IR behavior of the GW spectrum. 
It should be noted that the two spectral breaks correspond to the two characteristic length scales in the system: the peak frequency $f_p$ corresponds to the mean bubble separation $R_*$, while the frequency $f_{\rm b}$ corresponds to the thickness of the sound shell $\Delta R_*$. Compared to the fit formula in Eq.~\ref{eq:fitfun}, an extra parameter $\tilde{a}$ has been introduced. The values of all parameters in Eq.~\ref{eq:newfit} are   produced from our  calculations of the GW relic density based on the sound shell model~\cite{Hindmarsh:2019phv,Gowling:2021gcy,Gowling:2022pzb}. 
The dependence of the fit parameters on the underlying physical parameters such as $\alpha$ and $v_w$ governing the phase transition are explored in Section~\ref{sec:fitfopt}.

\begin{figure}[h]
  \centering
  \includegraphics[width=0.45\linewidth]{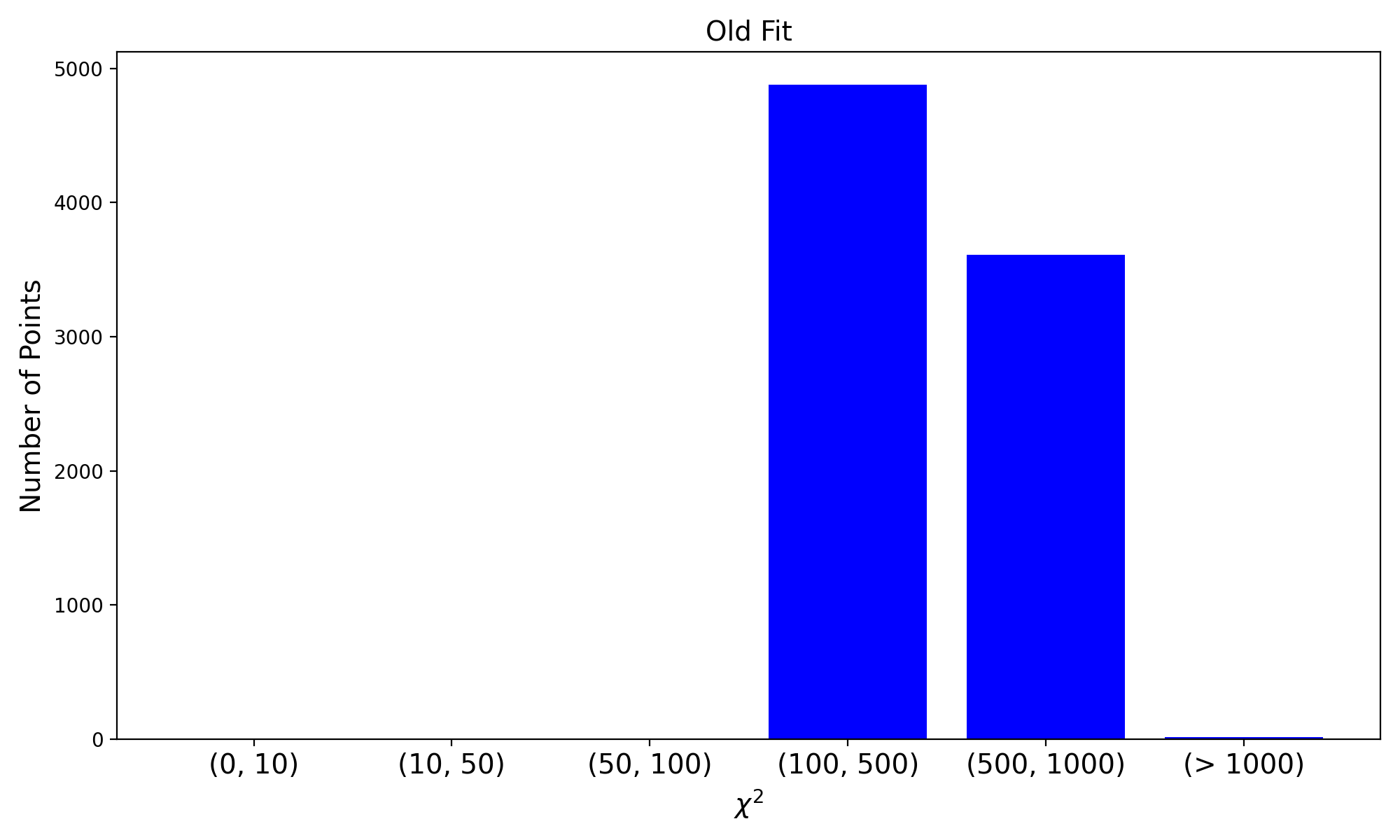}
  \includegraphics[width=0.45\linewidth]{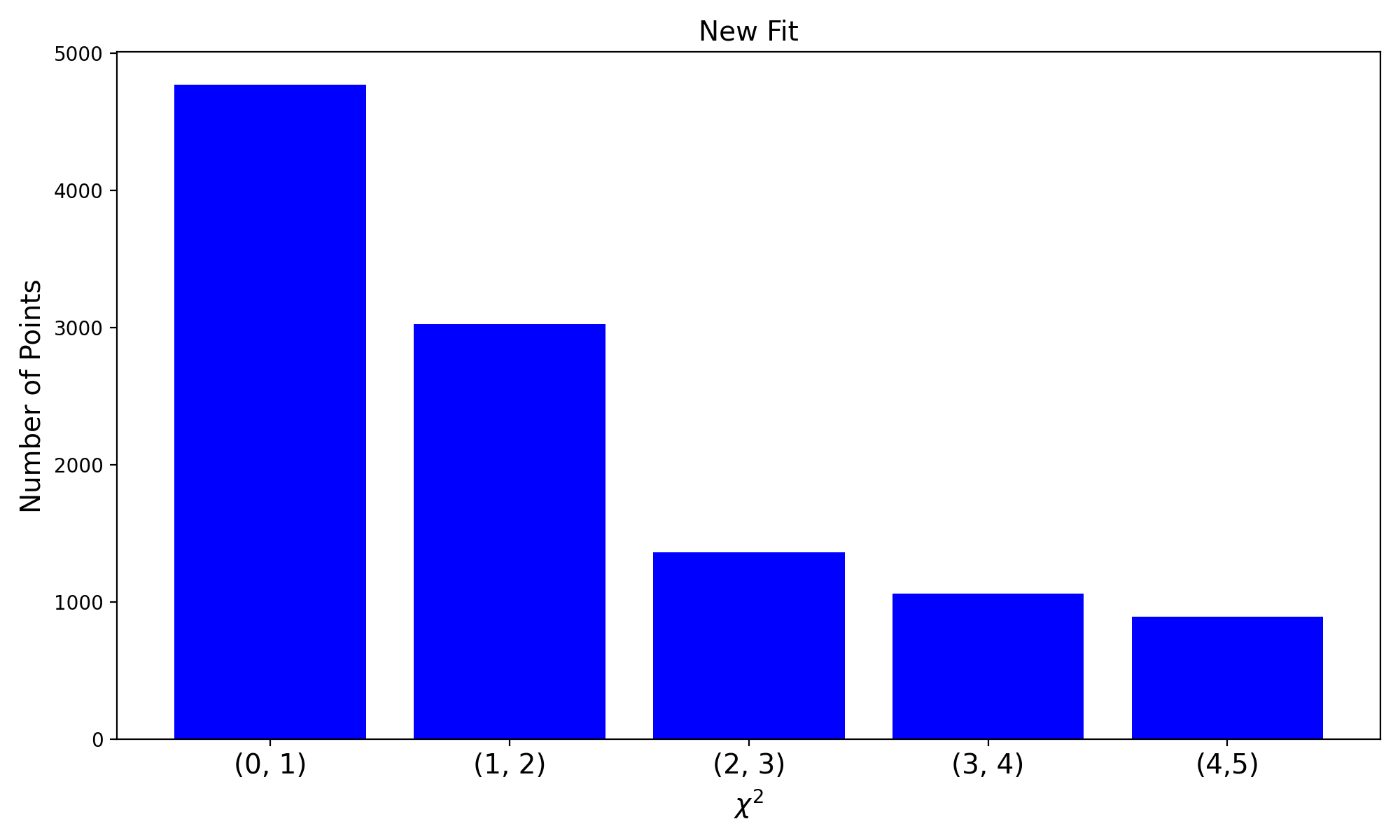}
    \caption{Histogram for comparison of fits using the fit function in Eq.~(\ref{eq:oldfit}) motivated by Ref.~\cite{Hindmarsh:2019phv,Gowling:2021gcy,Gowling:2022pzb} and our proposed fit formula in Eq.~(\ref{eq:newfit}). }
  \label{fig:hist}
\end{figure}

Before proceeding to compare the different fit formulas, we pause to give details about our computation procedure. 
A Monte Carlo approach to fitting Eq.~\ref{eq:newfit} to the GW spectrum  of the sound shell model is adopted. Relevant variables $\Omega_p$, ${\tilde s}_{0}$, $\tilde{a}$, $\tilde{b}$ and ${\tilde r}_b$ are first initialized.
Our algorithm then executes a loop 100,000 times, each iteration randomly generating the parameters introduced in Eq.~\ref{eq:newfit} to match the 
GW spectrum from the sound shell model data. The appropriateness of the fit is measured by calculating residuals - the squared differences between the logarithms of actual data for $\Omega_{\rm GW} h^2$ and the 
predictions from Eq.~\ref{eq:newfit}. The program continually updates the parameters that result in the lowest residuals, storing the best-fit parameters. After completing all iterations, it uses these parameters to generate the final predicted model values and calculates the total error between this  fit model and the original sound shell data. The objective is to robustly determine the parameters that best explain how the GW spectrum behaves across various frequencies, optimizing the fit to minimize error. 

A comparison of the fit to GW spectrum from \cite{Hindmarsh:2019phv} as given by Eq.~\ref{eq:oldfit} to our new fit in Eq.~\ref{eq:newfit} is illuminating. We do a $\chi^2$ analysis based on the fit equations and the GW spectrum produced by the sound shell model, using the full bubble velocity profile based on ~\cite{Hindmarsh:2019phv,Guo:2020grp}. The two histograms in Fig.~\ref{fig:hist} show the results. 

\begin{figure}[H]
  \centering
  \includegraphics[width=0.45\linewidth]{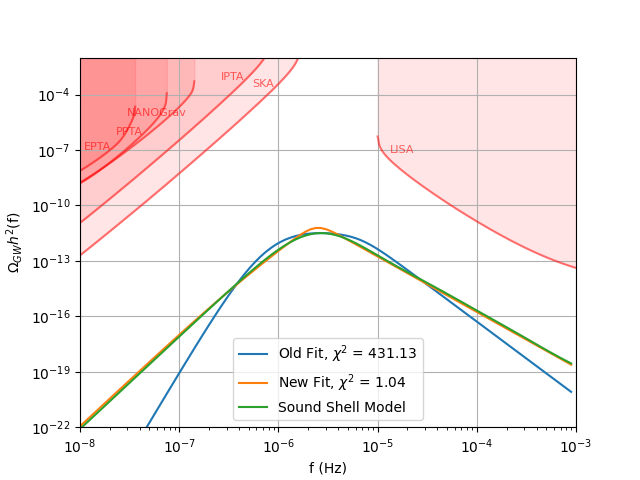}
  \includegraphics[width=0.45\linewidth]{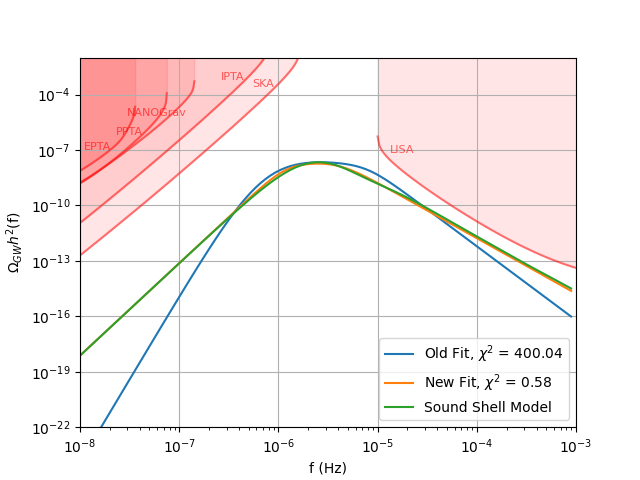}\\
  \includegraphics[width=0.45\linewidth]{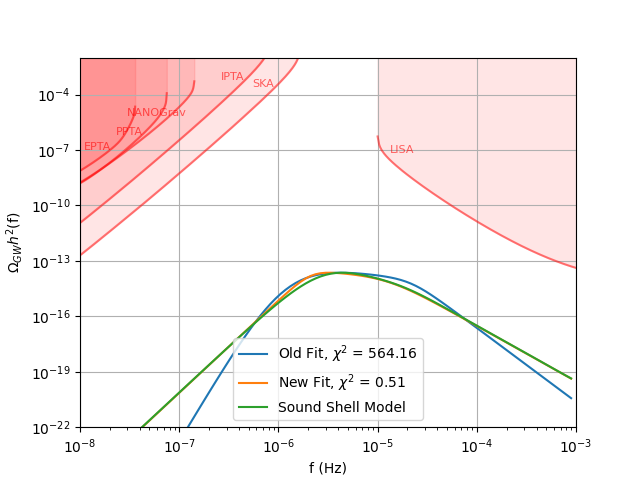}
  \includegraphics[width=0.45\linewidth]{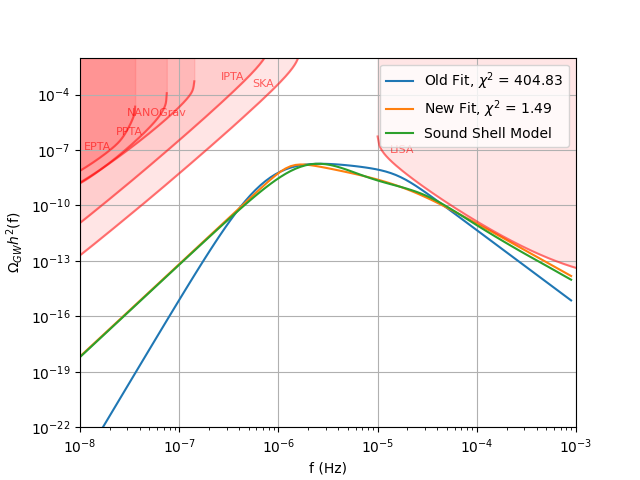}
    \caption{A benchmark point of GW from FOPT showing  the relic density of GW with respect to frequency for ``old fit" (\cite{Gowling:2021gcy}), ``new fit" (the current work), and sound shell model (\cite{Hindmarsh:2019phv,Gowling:2021gcy,Gowling:2022pzb}) is plotted.  Various current and future observational constraints are also shown and labeled using the integrated sensitivity curves from~\cite{Schmitz:2020syl} for different experiments. Here we assume $T_{n} = 100$~ GeV and $\beta/H = 1$. }
  \label{fig:bench}
\end{figure}

As can be seen from the left panel of Fig.~\ref{fig:hist}, the $\chi^2$ values for Eq.~\ref{eq:oldfit} are much larger than one and most values are above a hundred. On the other hand, the right panel of Fig.~\ref{fig:hist} shows that most data points have $\chi^2$ values  smaller than one for the new fit function in Eq.~\ref{eq:newfit} introduced in this paper. While some data points do have $\chi^2$  large than one, the number of such points decreases as $\chi^2$ increases. In Fig.~\ref{fig:bench}, we consider four benchmark points  and show the $\chi^2$ values for the old and new fits in each panel. These benchmarks have these set of FOPT parameters: $[6.13\times 10^{-2},3.08\times 10^{-3},1,100]$, $[1.28\times 10^{-1},8.66\times 10^{-2},1,100]$, $[3.32\times 10^{-1},6.36\times 10^{-2},1,100]$ and $[5.99\times 10^{-1},3.47\times 10^{-1},1,100]$ for this FOPT parameters $[v_w, \alpha, \beta/H, T_n ({\rm GeV})]$.  For these benchmarks it is apparent  that the new fit in Eq.~\ref{eq:newfit}  matches better with the  sound shell model than the fit formula in Eq.~\ref{eq:oldfit}. Overall, based on these figures, we have validated the quality of the fit function for most choices of $v_w$ and $\alpha$. This implies that our proposed fit in Eq.~\ref{eq:newfit} more accurately  captures the features of GW spectrum from the sound shell model  and  can be used by the community (see Appendix \ref{app-b} for the fit function code).

\section{Physical Behavior of Fit and FOPT Parameters}\label{sec:fitfopt}
Having demonstrated the appropriateness of our fit function, we proceed towards a deeper physical understanding of the different parameters that we have introduced.
We perform a scan over the following input parameters:  the wall velocity $v_{w}$ and the strength of transition $\alpha$, keeping   the nucleation temperature $T_{n} = 100$~ GeV and the  bubble nucleation rate $\beta/H = 1$ fixed. Using these parameters, we calculate: $(i)$ the maximum  velocity $v_{\rm max}$, $(ii)$ the width of velocity profile $\delta_{\xi}$ (defined as the width of the velocity profile in a window $95~\%$ of $v_{\rm max}$), $(iii)$ the corresponding value of $\xi$ as $\xi_{\rm max}$, and $(iv)$ the difference between the values of the enthalpy before the wall and after it i.e. $\delta_{w}$.  Our results are displayed  in Figs.~\ref{fig:scan_vw_alpha_at} - \ref{fig:vwprof-plots-vw} (also see Appendix \ref{app-a} for more scanning plots). 

We begin with a discussion of Fig.~\ref{fig:scan_vw_alpha_at}. Here, we depict values of  ${\tilde r}_b$ (top panel),   $\tilde{b}$ (middle panel), and $\tilde{a}$ (bottom panel) on the plane of $\{v_w, \alpha\}$ for three regimes:  deflagration (red), detonation (green) and  hybrid (blue). The value of $\alpha$ is chosen to vary from $10^{-3} - 1$, since we are interested in studying the global behavior of the parameters. For the plot of ${\tilde r}_b$, it is clear that there is some structure: for example, in the deflagration regime, the values of   ${\tilde r}_b$ for a given $\alpha$ are small at small values of $v_w$, increase around $v_w \sim 0.2-0.4$, before decreasing again. Similarly, in the detonation regime, the values of ${\tilde r}_b$ for a given value of $v_w$ decrease with increasing $\alpha$. A somewhat opposite behavior can be discerned in the hybrid regime, where increasing $\alpha$ corresponds to lower ${\tilde r}_b$. In contrast to the results for  ${\tilde r}_b$, it is difficult to discern any structure in the scans of $\tilde{b}$ and $\tilde{a}$. It is therefore fruitful to study the dependencies in a series of two dimensional plots, which we turn to next.

\begin{figure}[h]
  \centering
  \includegraphics[width=0.8\linewidth]{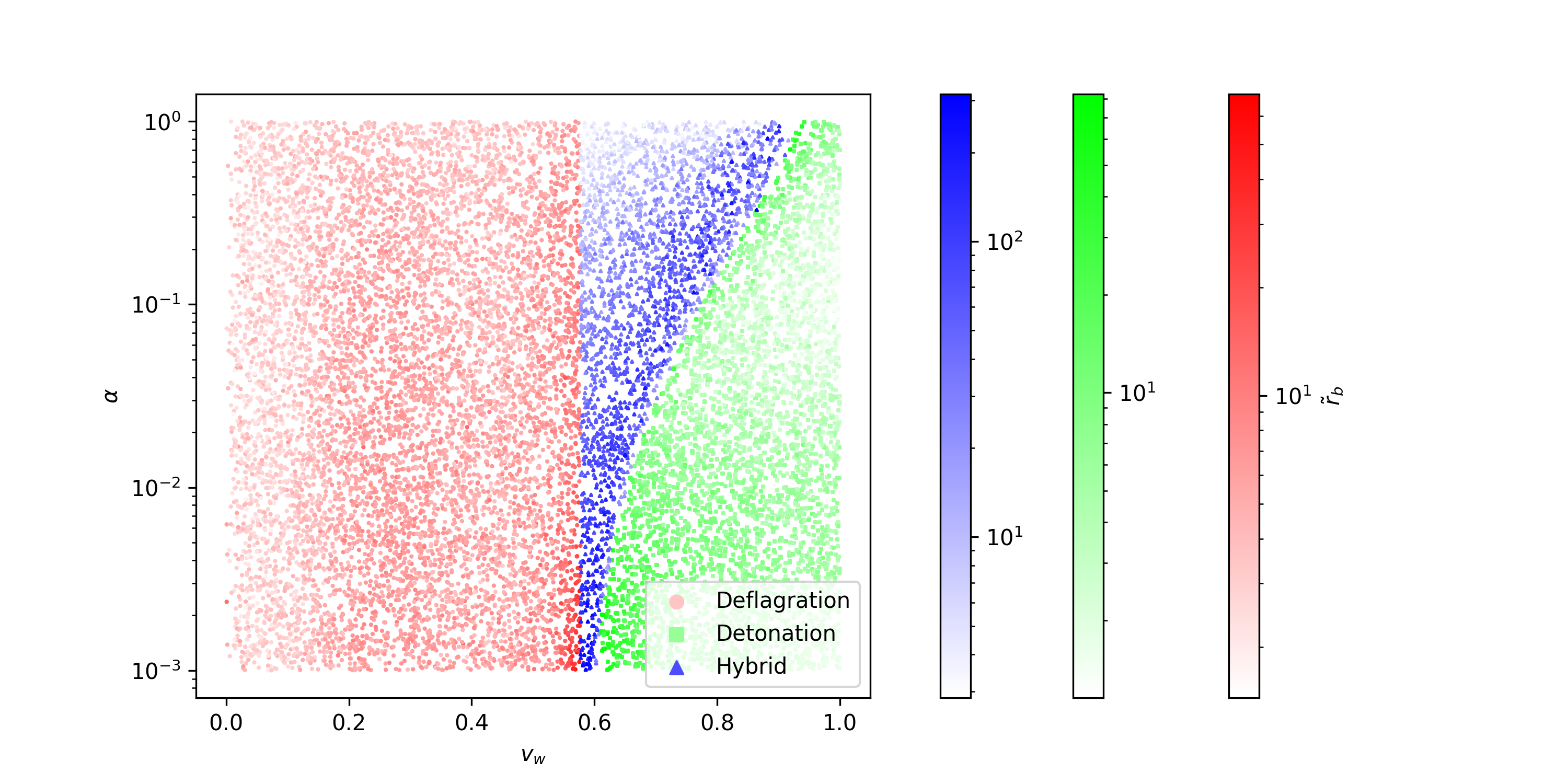}
  \includegraphics[width=0.8\linewidth]{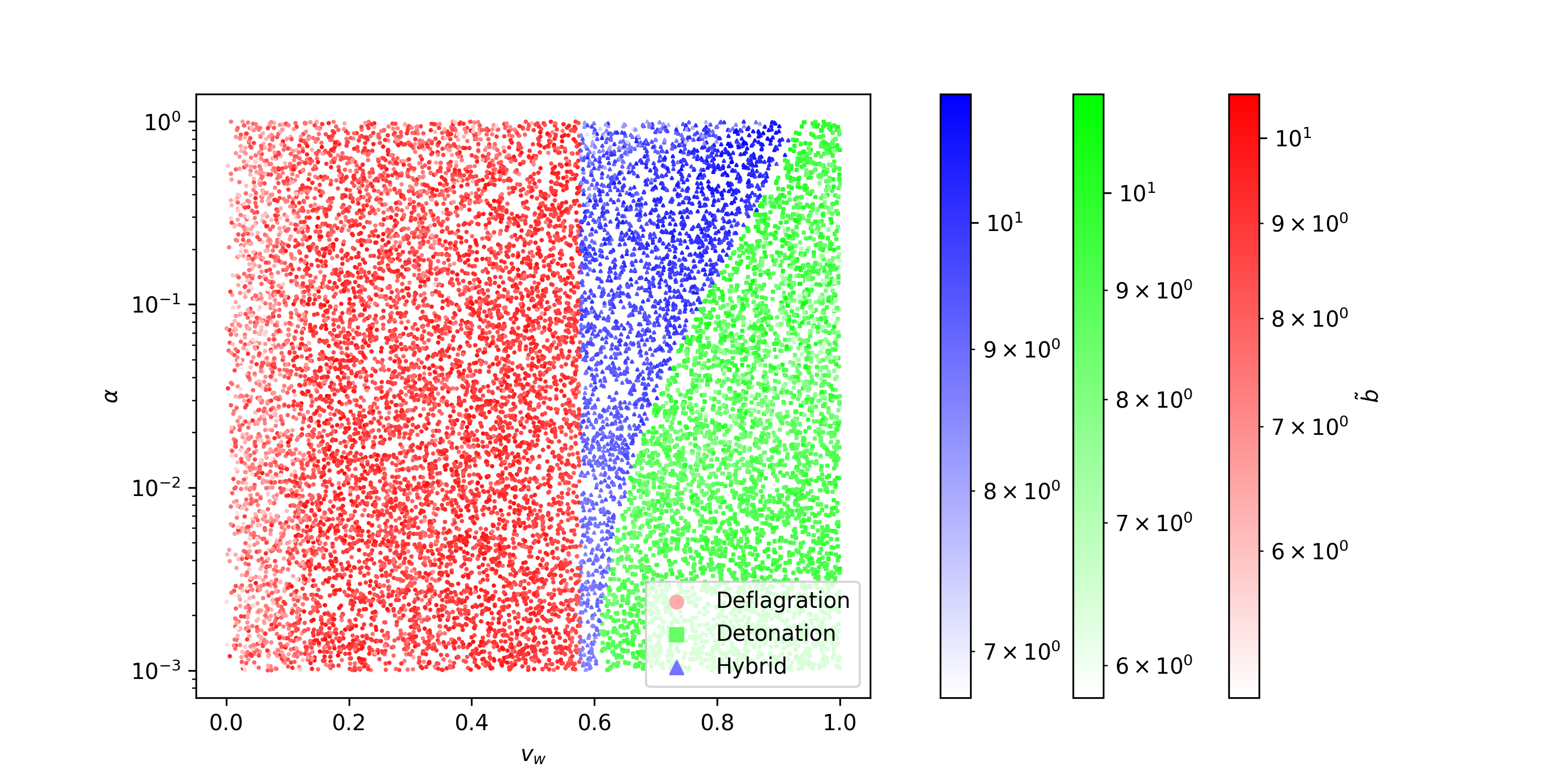}
  \includegraphics[width=0.8\linewidth]{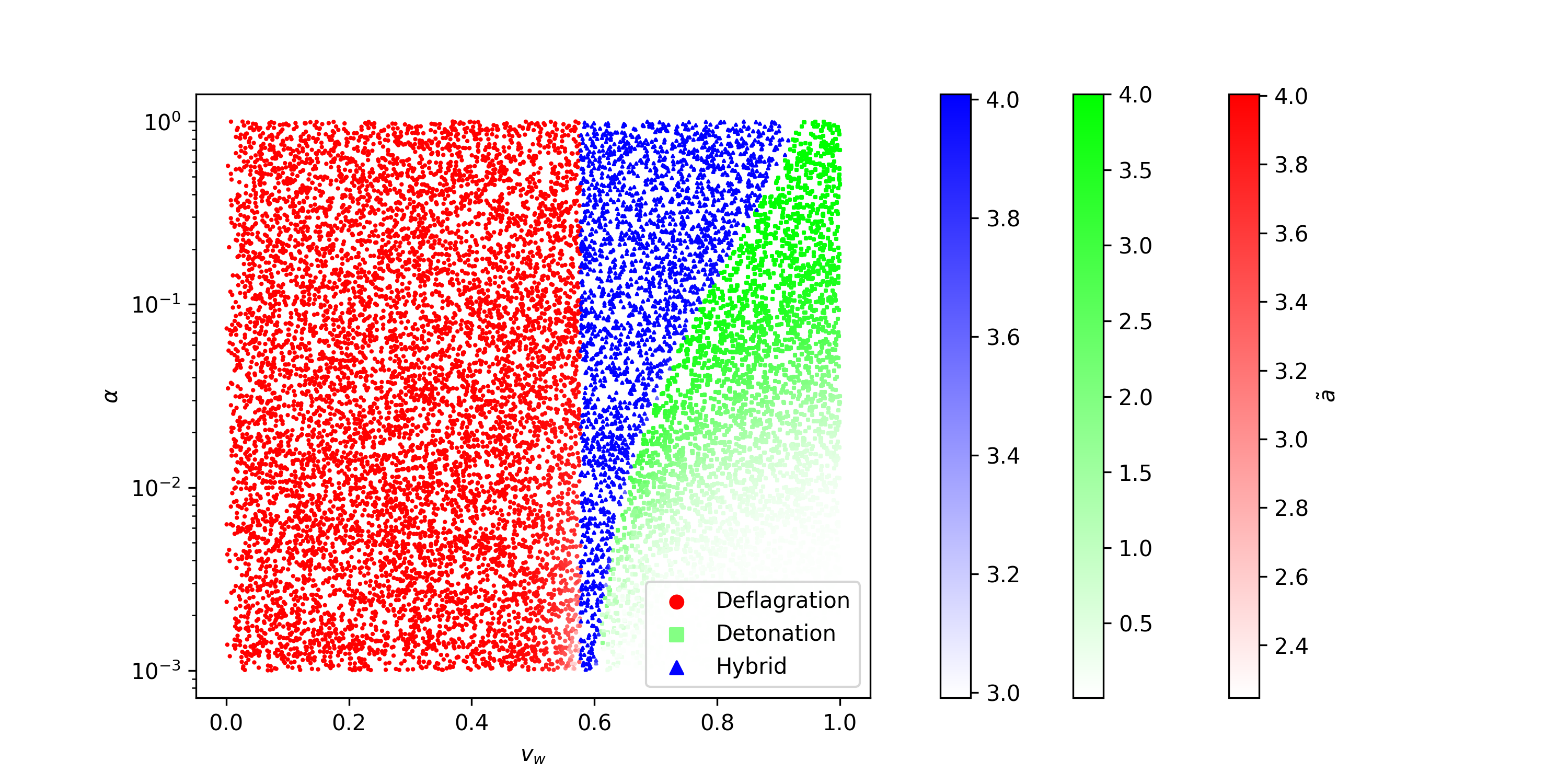}
\caption{From top to bottom: scatter plots for $v_w$, $\alpha$ and $\tilde{r}_b$;  for $v_w$, $\alpha$ and $\tilde{b}$; and for $v_w$, $\alpha$ and $\tilde{a}$, respectively. The type of FOPT is  shown in the legend. Here we assume $T_{n} = 100$~ GeV and $\beta/H = 1$.}
  \label{fig:scan_vw_alpha_at}
\end{figure}

In Fig.~\ref{fig:vwprof-plots-alpha}, we show the  velocity and enthalpy profiles with respect to $\xi$ for different choices of $\alpha$. The scan has been performed over different values of wall velocity $0<v_w<1$. The velocity and enthalpy profiles exhibit the behavior expected for the deflagration, hybrid, and detonation regimes, respectively \cite{Alves:2018jsw}. In the deflagration mode, the plasma in front of the bubble wall flows outward while remaining static inside the bubble. The velocity profiles to the left of the maximum in all panels of Fig.~\ref{fig:vwprof-plots-alpha} exhibit this behavior. As $v(\xi)$ increases, a discontinuity appears and $v(\xi) \rightarrow 0$; this is the shock front, beyond which a supersonic deflagration (hybrid) mode develops when the velocity exceeds the sound speed. The value of $\alpha$ determines the maximum value of $v_{\rm max}$ and maximum of $w_{\rm max}$ where the velocity profile maximises over different values of $\xi$. 
Each value of $v_{\rm max}$ occurs when $\xi = v_w$.

Fig.~\ref{fig:alpha-fixed} depicts the behavior of the various physical quantities, as well as the fit parameters, as a function of $v_w$ for several benchmark values of $\alpha = 0.001, 0.01, 0.1, 0.3,$ and $0.5$.
The behaviour of $v_{max}$ w.r.t. $v_w$ is shown in the bottom left panel of Fig.~\ref{fig:alpha-fixed}. It reaches a maximum that depends on the value $\alpha$; a similar pattern is exhibited by the thermal efficiency factor $\kappa$ defined in Eq.~\ref{eq:kappa} and depicted in the top right panel. 
The results of this panel, as well as the top right panel depicting $\kappa$, are consistent with the left panels of Fig.~\ref{fig:vwprof-plots-alpha}. The peak frequency $f_p$ is depicted in the bottom right panel of Fig.~\ref{fig:alpha-fixed} and shows a local maximum where $v_{\rm max}$ is maximized. The dependence of $\Omega_p$ w.r.t. $v_w$  is shown in the second row of the left panel of Fig.~\ref{fig:alpha-fixed}.  $\Omega_p$ is a decreasing function of $v_w$ in the deflagration regime; on reaching $v_{\rm max}$, it rises somewhat and then continues falling in the hybrid and detonation regimes. 
We have defined the width of velocity  profile $\delta_{\xi}$ as the difference between the values of $\xi$ when $\xi_{\rm max} = v_w$ and $v(\xi_{\rm width})=~0.95~v_{\rm max}$. This quantity is shown in the  left panel of the first row. Its behavior matches with expectations from the velocity profiles shown in Fig.~\ref{fig:vwprof-plots-alpha}.  The value of $\xi_{\rm max}$  increases monotonously with $v_w$, independently of $\alpha$, as shown in the right of panel of the fourth row  of Fig.~\ref{fig:alpha-fixed}.

We now turn to a discussion of the fit parameters $\tilde{b}$ and $\tilde{a}$. Firstly, it should be noted that smaller values of $\tilde{b}$ correspond to a situation where more energy resides in the contribution from the thickness of the sound shell rather than from the bubble separation. Conversely, larger values of $\tilde{b}$ correspond to more energy residing in the contribution from the bubble separation. While it is difficult to discern structure, we do find a linearly increasing trend in $\tilde{b}$ as $v_w$ increases, shown by trend lines, albeit crude. Physically, it is reasonable that as $v_w$ increases, more energy should reside in the bubble separation contribution. The behavior of the fit parameter $\tilde{a}$ in the left panel of  the third row is interesting. It should be noted that smaller values of $\tilde{a}$ correspond to sharper IR spectra, while larger values of $\tilde{a}$ correspond to shallower IR spectra. It is clear that the value of $\tilde{a}$ is insensitive to variations in $v_w$ in the deflagration mode, but swiftly decreases after $v_{\rm max}$, for all values of $\alpha$. 

We now turn to a discussion of Fig.~\ref{fig:vwprof-plots-vw}, where a scan over $\alpha$ between $0$ and $1$ for fixed  values of $v_w = 0.1, 0.4, 0.7, 0.95$ was performed. The plot of the ${\tilde a}$  parameter versus $\alpha$ is shown in the  fourth row, left panel.  ${\tilde a}$ appears to be constant with a value of around $4$, which reduces when it reaches the maximum of $v_{\rm max}$. This is consistent with the behavior in Fig.~\ref{fig:alpha-fixed}. The plot of ${\tilde b}$ versus $\alpha$ is shown in the second row of the right panel of Fig.~\ref{fig:vw-fixed}. It is difficult to discern a specific pattern here; nevertheless, we have made a crude linear fit. For $v_w = 0.1, 0.4,$ and $0.7$, we see that ${\tilde b}$ decreases very mildly with increasing $\alpha$.

\section{Conclusions}\label{conclu}
The sound shell model  currently provides the most sophisticated semi-analytical framework for calculating the GW spectrum in models of BSM physics. In conjunction with numerical simulations, a precision GW spectrum frontier is then already somewhat  within reach for particle physicists; one can then ask why fit functions of the type studied in this paper are even necessary, or why the physical interpretation of fit parameters is an interesting question. The response is that full calculations of the sound shell model, not to speak of actual simulations, are difficult and expensive; to perform large-scale scans over the parameter space of   particle physics models and obtain the predicted GW spectrum, portable and simple fits  to the spectrum, and the interpretation of the fit parameters, become necessary. Providing such a fit -- currently the one that most faithfully captures the full results coming from the sound shell model -- has been the goal of this paper. We have shared the results of our work   in a set of {\tt CSV}, {\tt Python} and {\tt Mathematica} files\footnote{\href{https://github.com/SFH2024/precise-fit-fopt-gw}{{\tt GitHub} link}.}. These files can be used by the BSM community to  efficiently depict the GW spectrum at various points in the   parameter space of their models.

Studies in GW physics are rapidly entering a precision frontier. Indeed, as pointed out by a subset of the present authors in \cite{Guo:2021qcq}, uncertainties of several orders of magnitude can be introduced in the calculation of the GW spectrum if one neglects careful treatments of the source lifetime; mean bubble separation; employing a beyond the bag model approximation when solving the hydrodynamics equations and explicitly calculating the fraction of energy in the fluid from these equations rather than using a fit. The incorporation of these effects constitutes, currently, the ``highest level of diligence" (in the parlance of \cite{Guo:2021qcq}) in obtaining the GW spectrum. Calculating the full GW spectrum at the highest level of diligence, and providing the relevant fit functions, constitute concrete future directions.

\begin{figure}[H]
  \centering
  \includegraphics[width=0.35\linewidth]{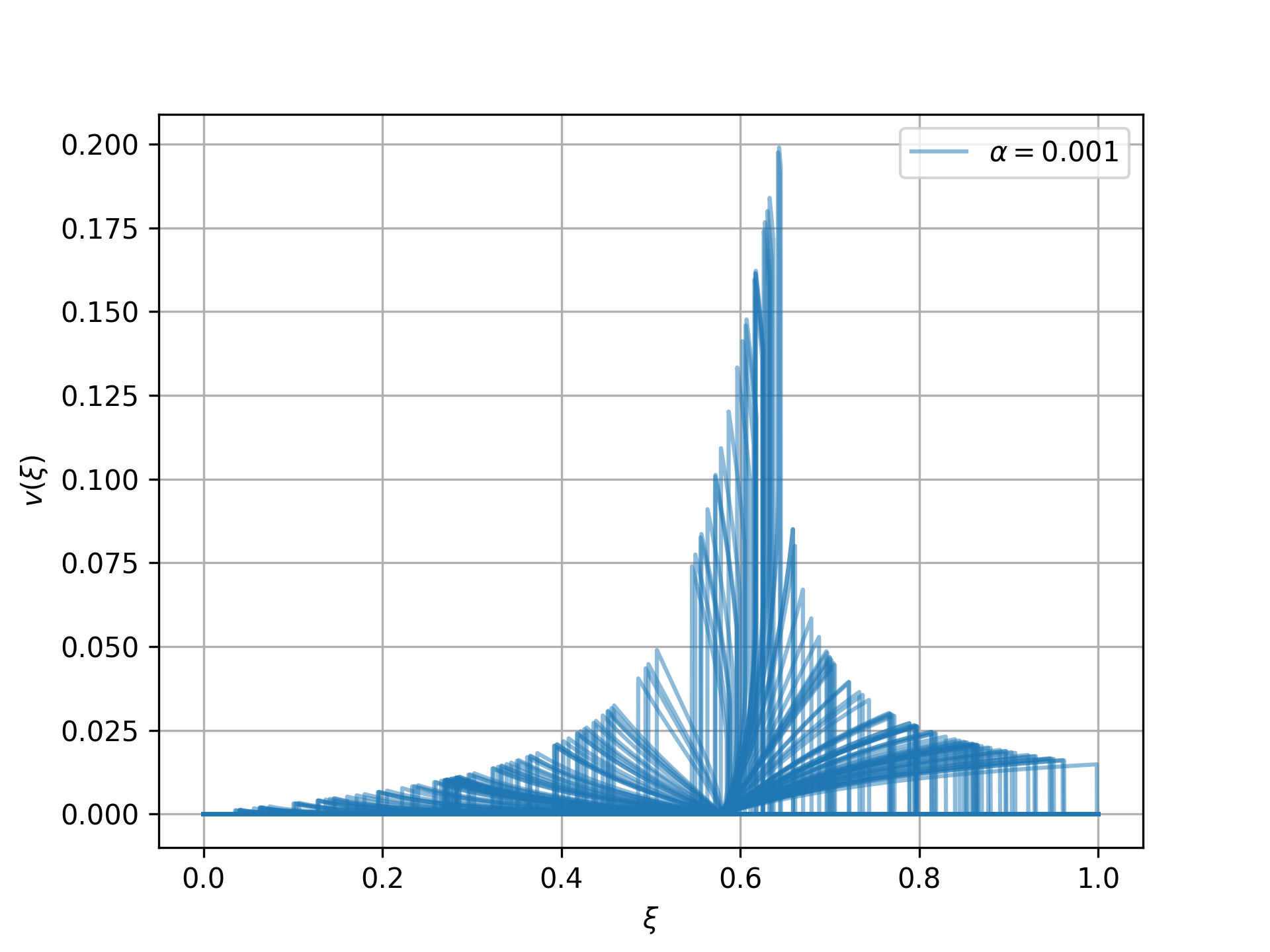}
  \includegraphics[width=0.35\linewidth]{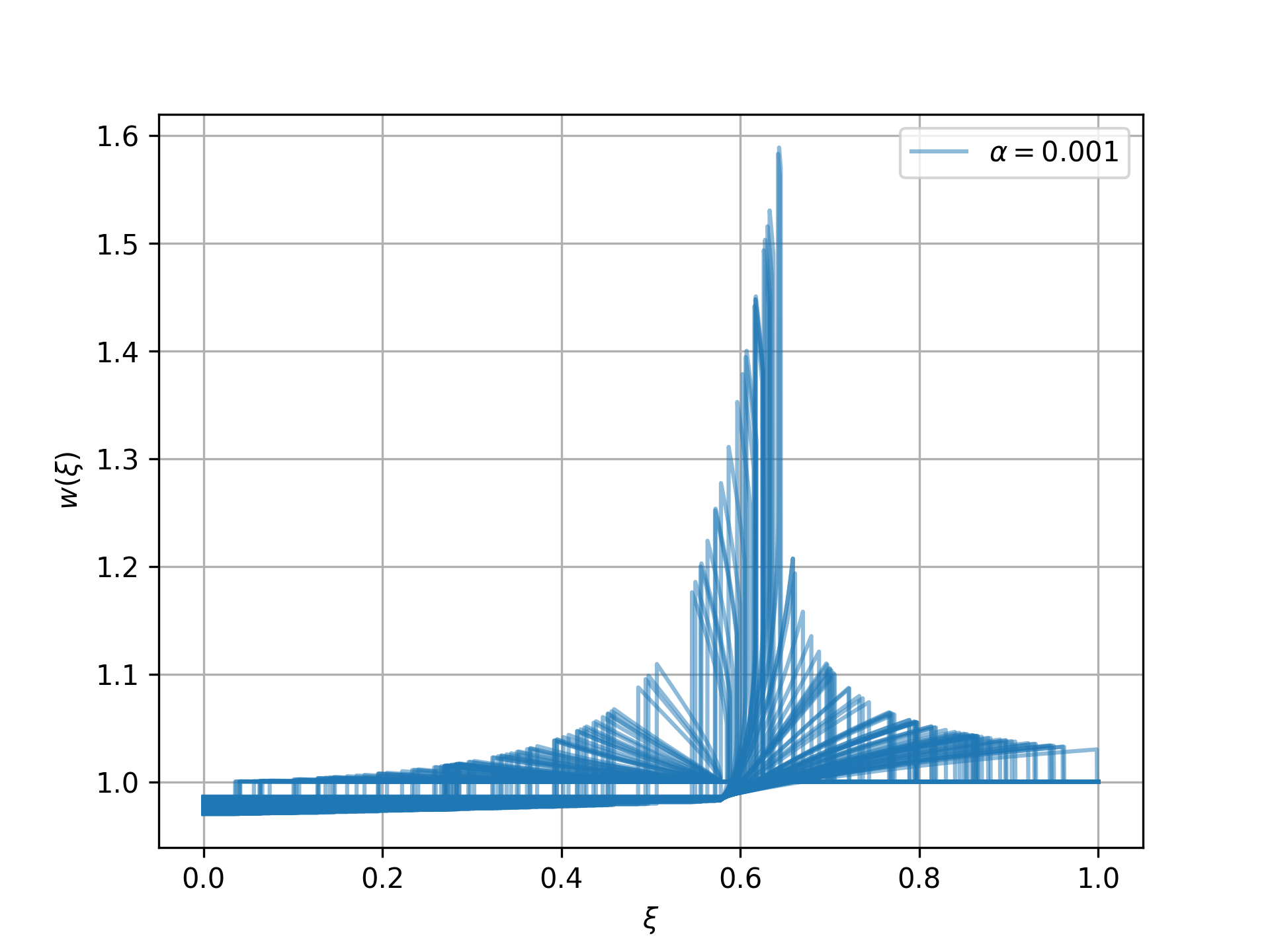}
  \includegraphics[width=0.35\linewidth]{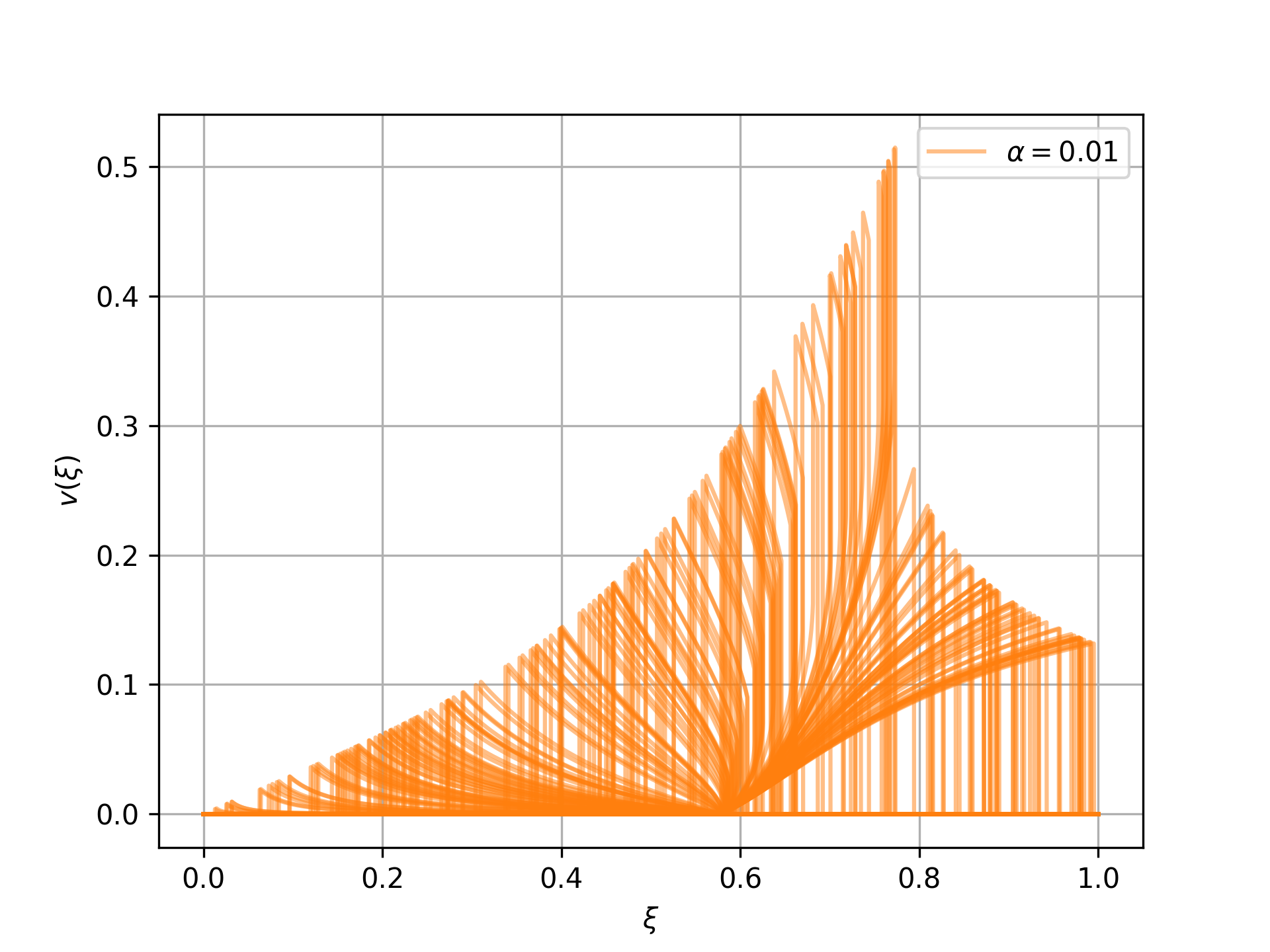}
  \includegraphics[width=0.35\linewidth]{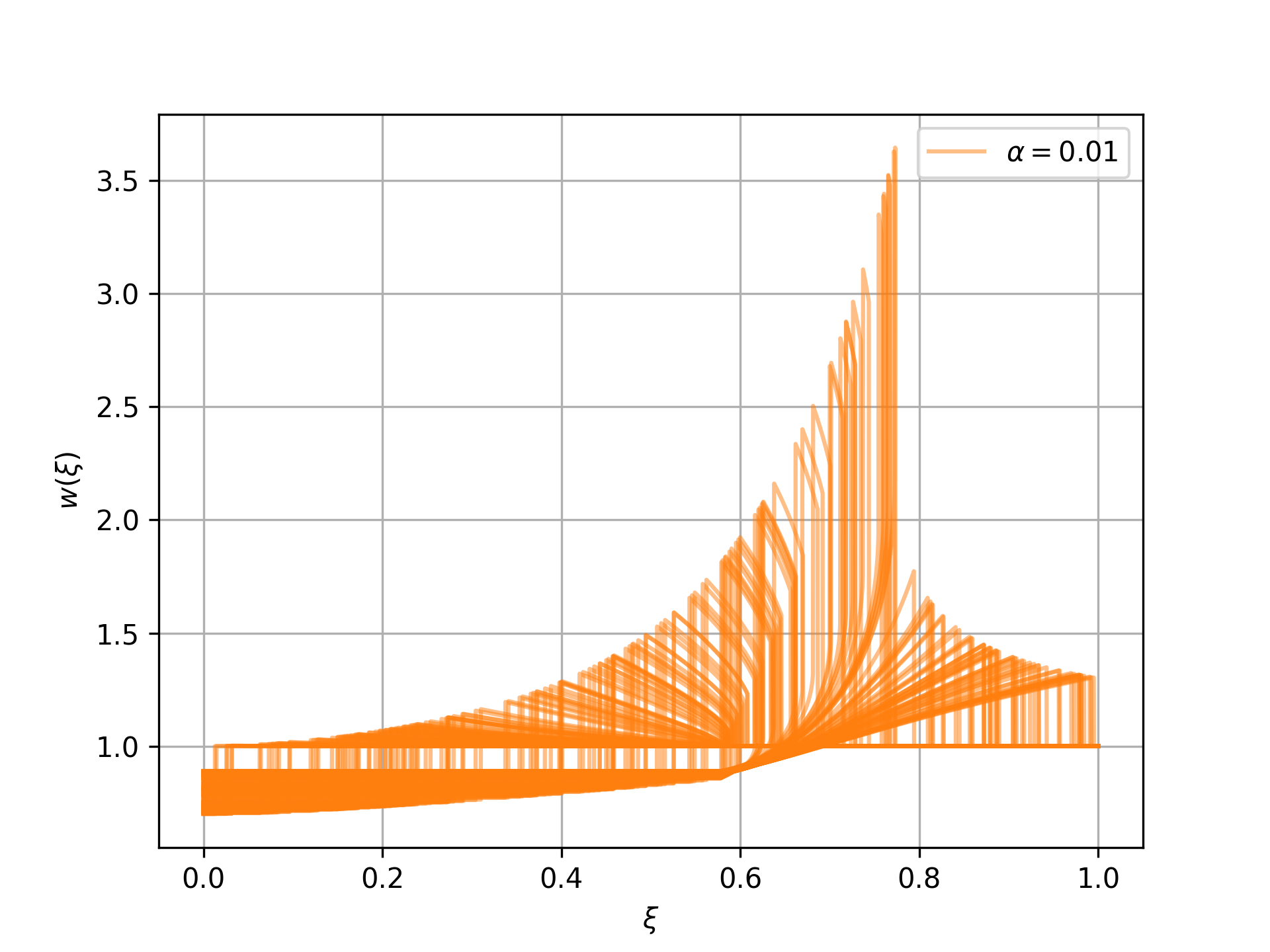}  \includegraphics[width=0.35\linewidth]{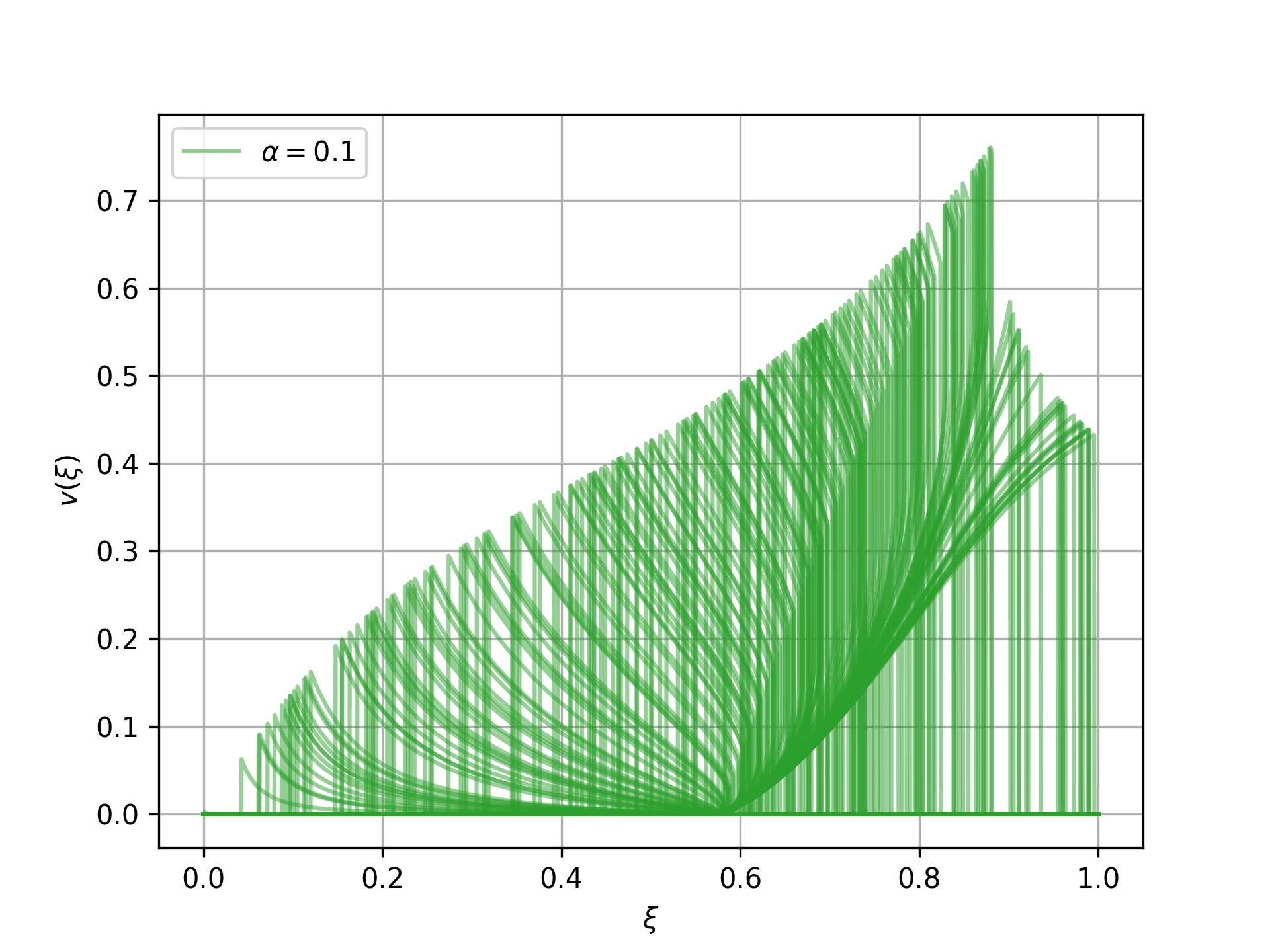}  \includegraphics[width=0.35\linewidth]{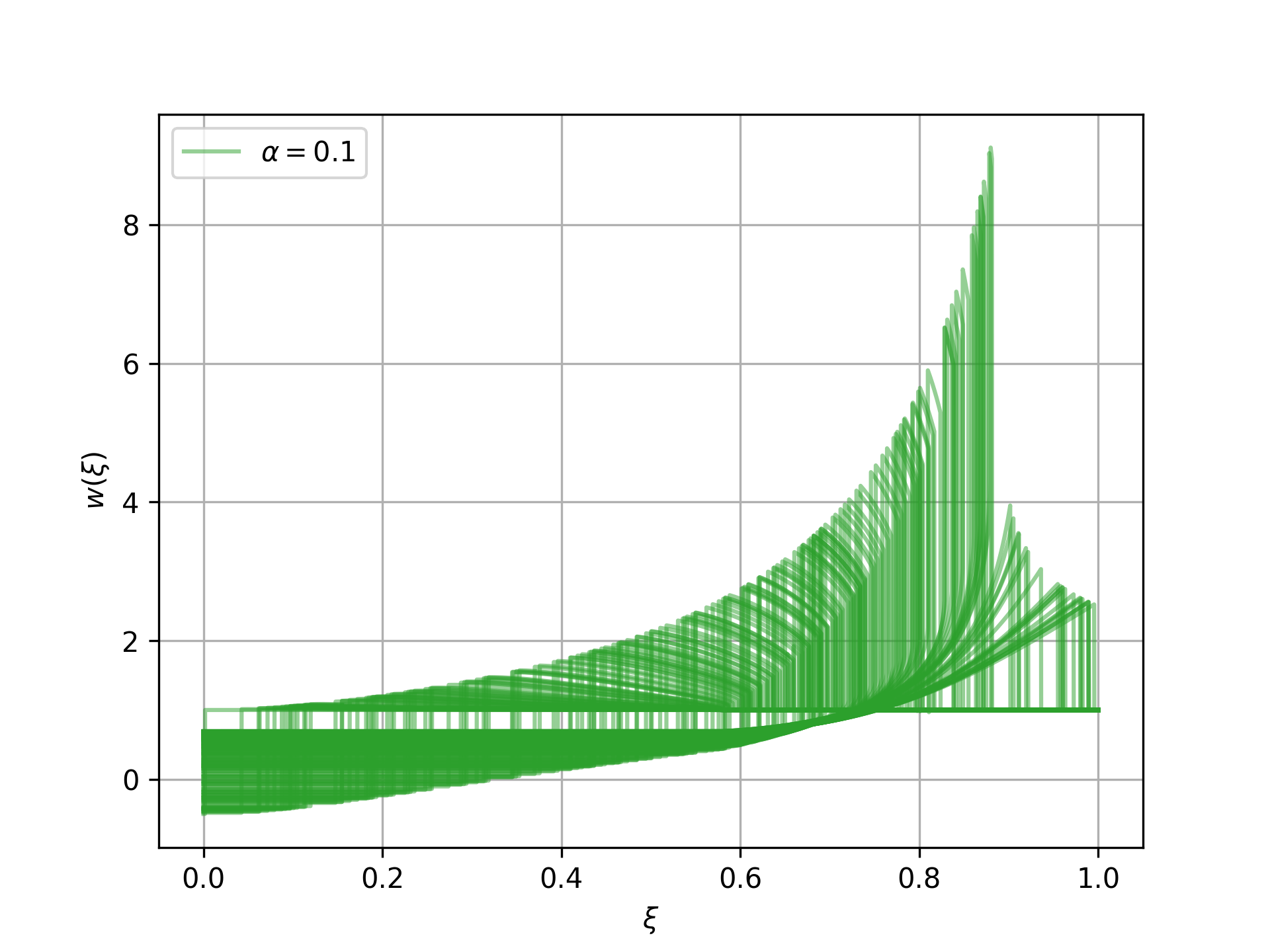}  \includegraphics[width=0.35\linewidth]{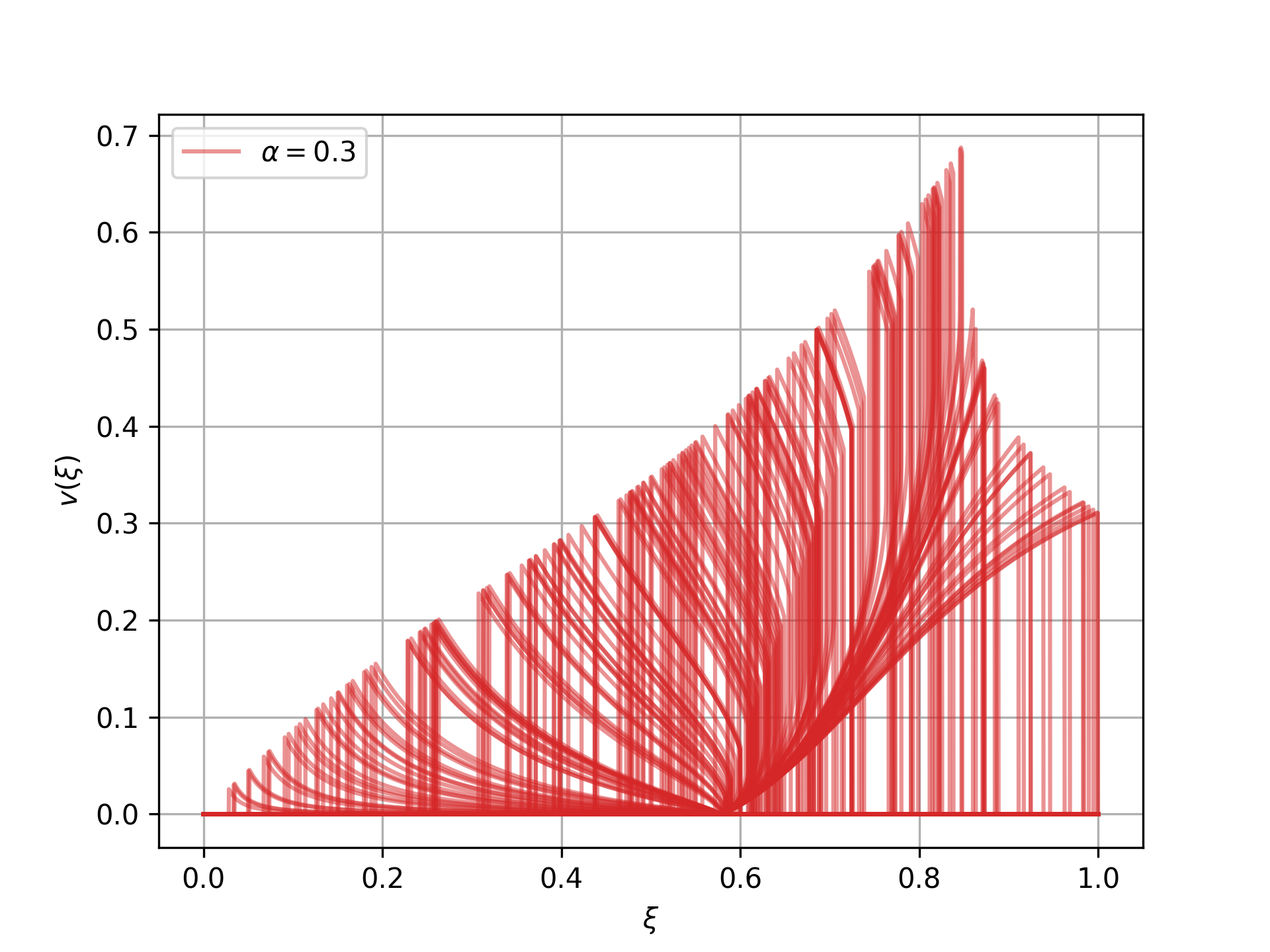}  \includegraphics[width=0.35\linewidth]{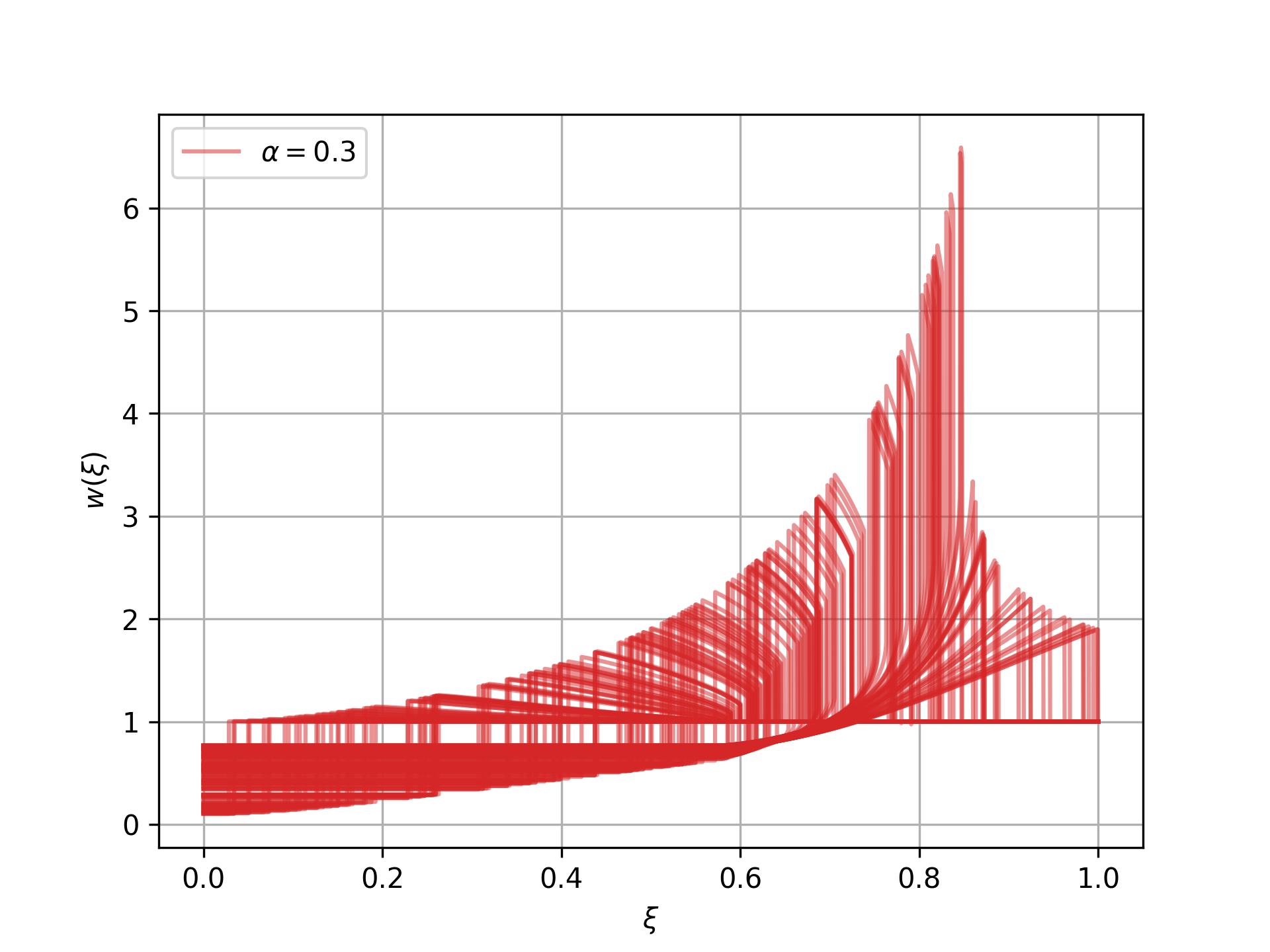}  \includegraphics[width=0.35\linewidth]{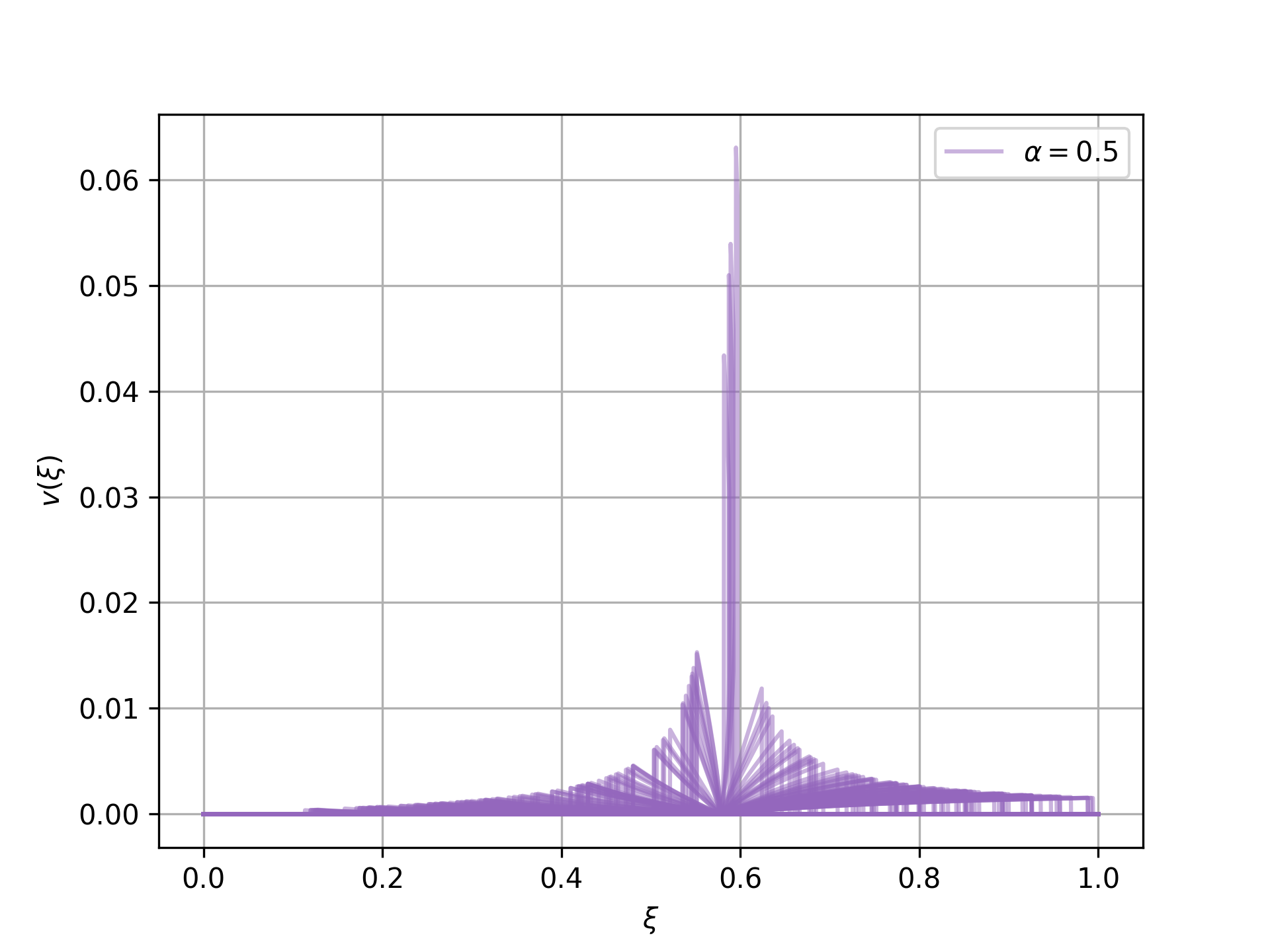}  \includegraphics[width=0.35\linewidth]{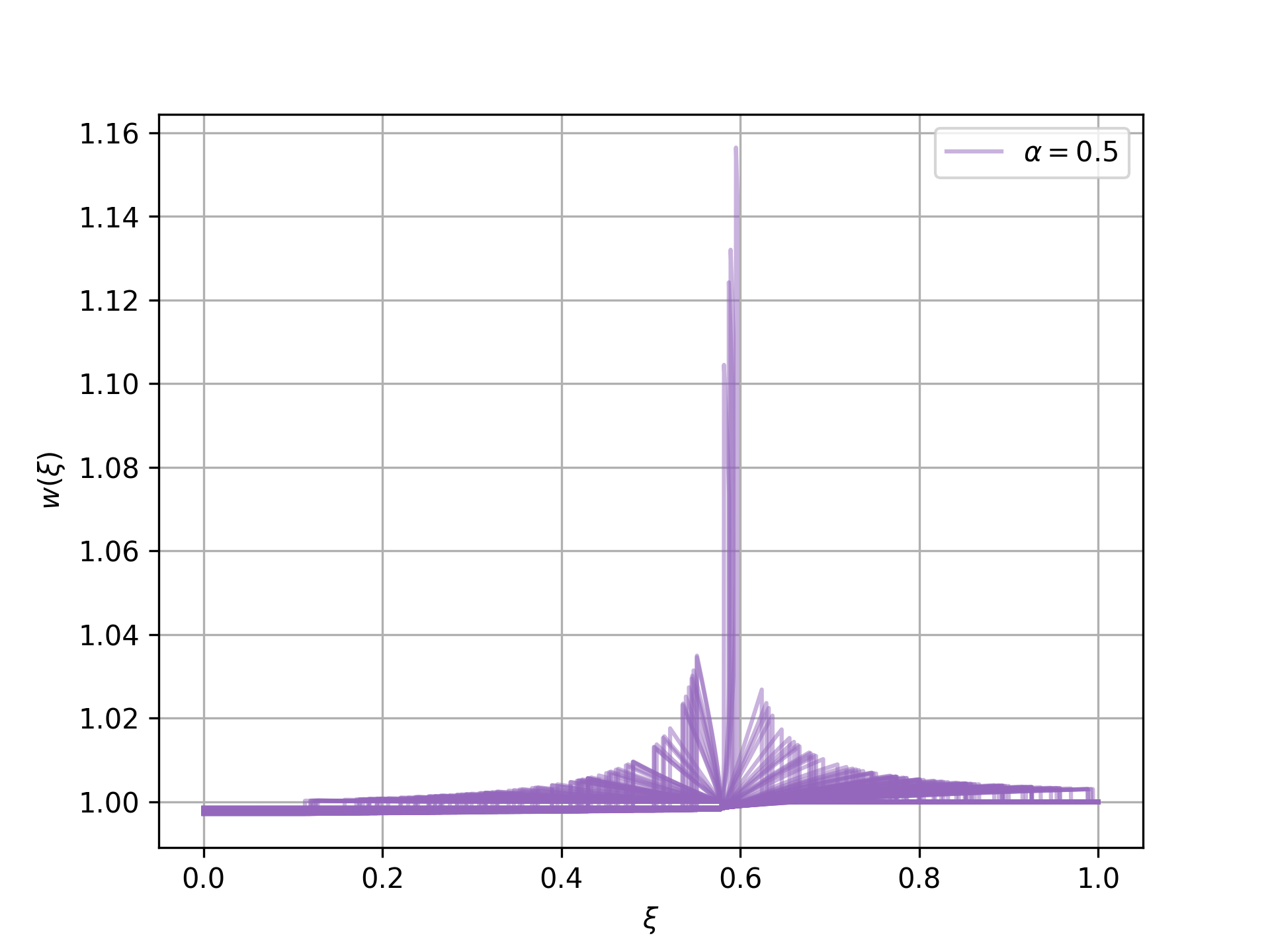}
    \caption{Velocity and ethalpy profiles for fixed values of $\alpha$. Each row of these panels corresponds to  a color that are blue, orange, green, red and violet equivalent to a specific value of $\alpha = 0.001,\, 0.01,\,0.1,\,0.3,\,0.5$ shown in the legends.}
  \label{fig:vwprof-plots-alpha}
\end{figure}

\begin{figure}[H]
  \centering
  \includegraphics[width=0.35\linewidth]{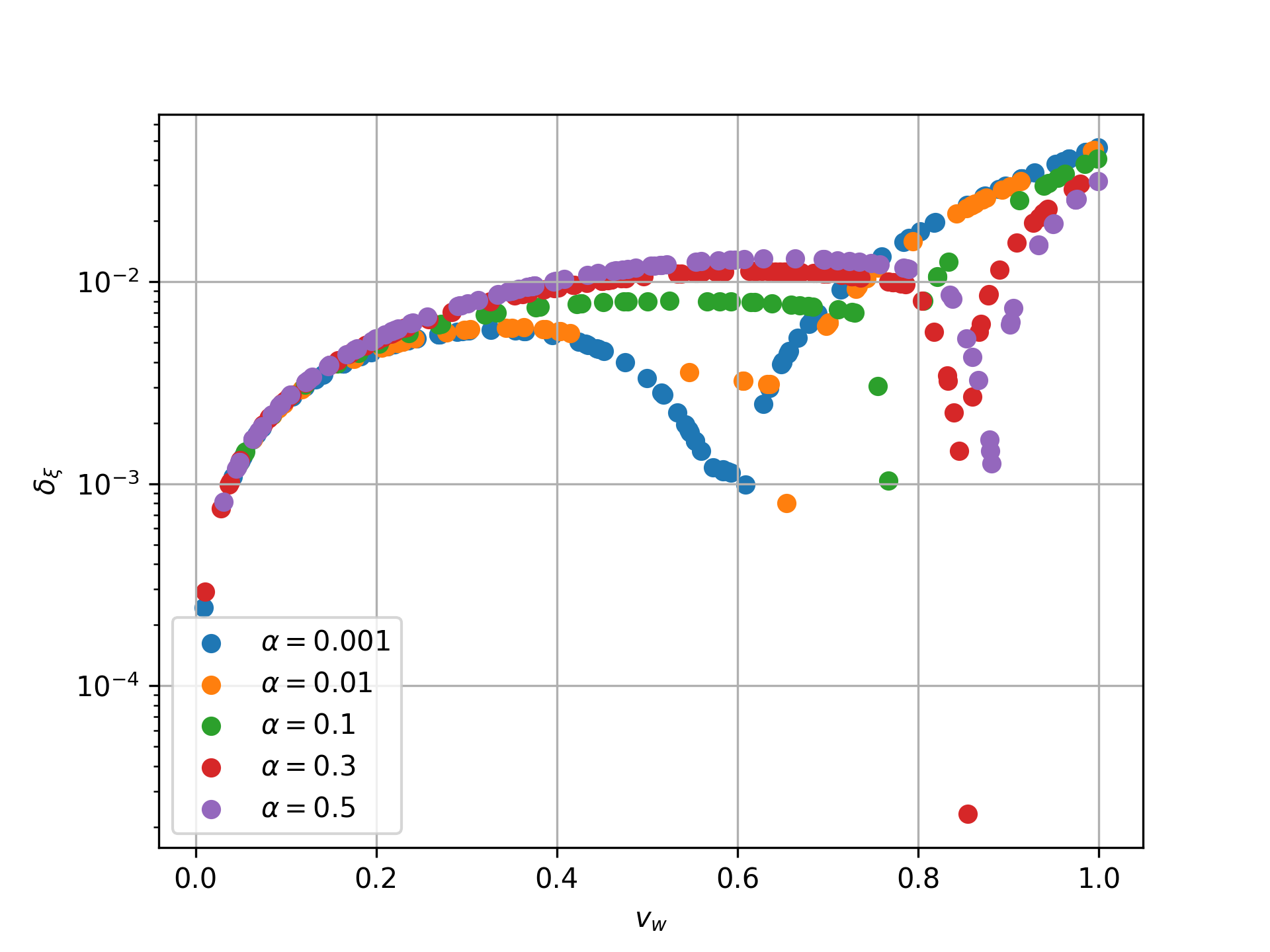}
  \includegraphics[width=0.35\linewidth]{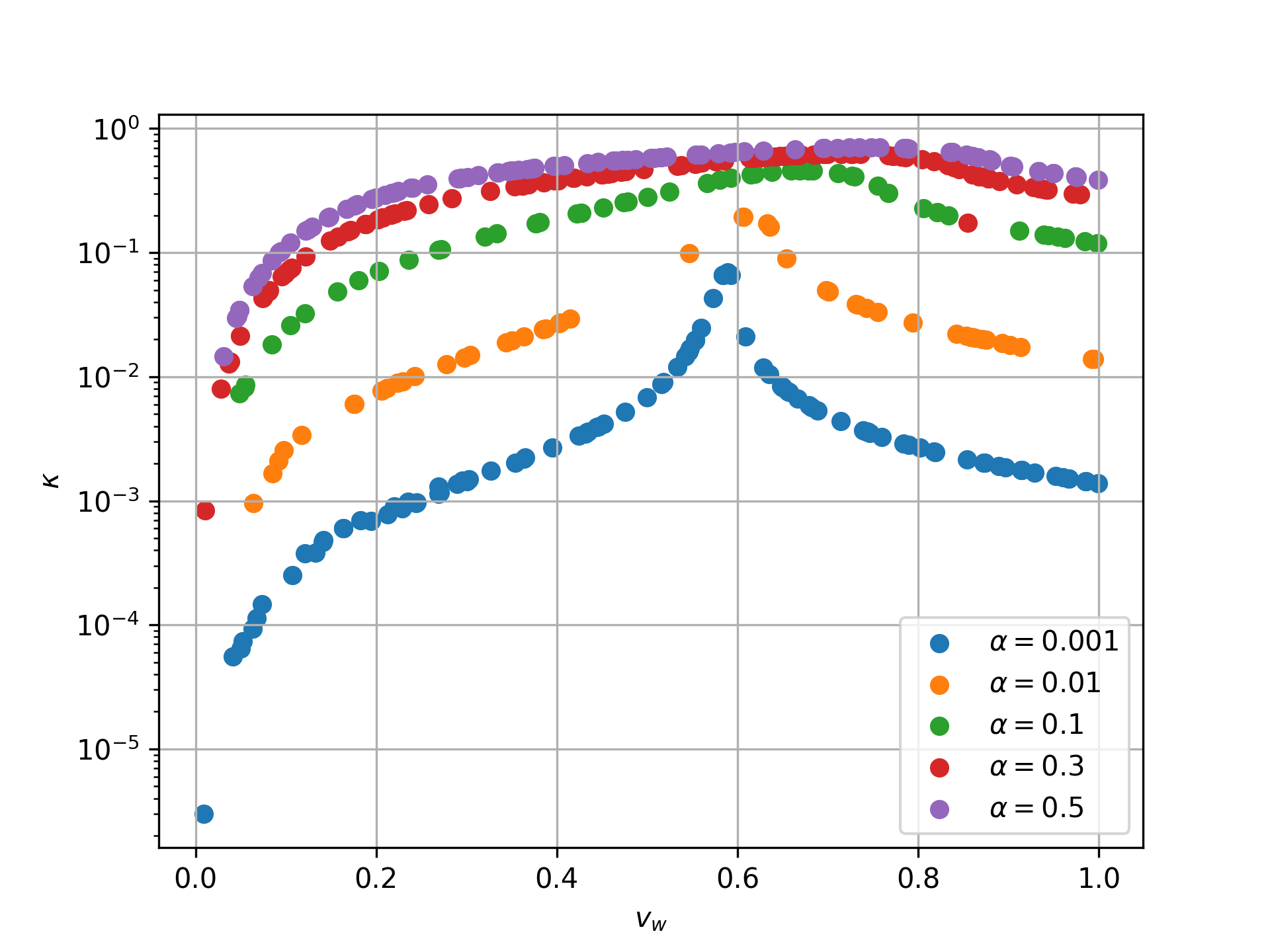}
  \includegraphics[width=0.35\linewidth]{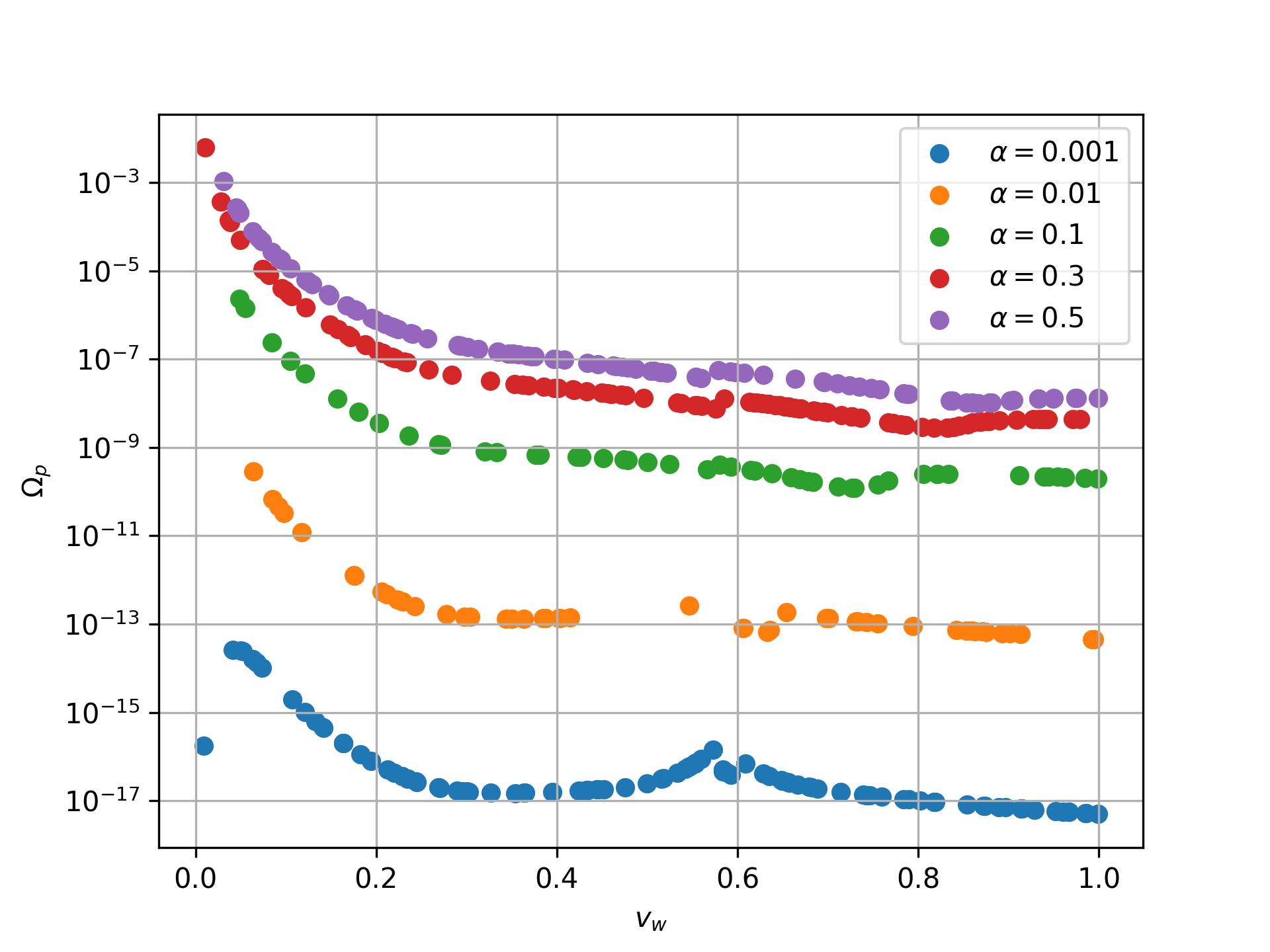}
  \includegraphics[width=0.35\linewidth]{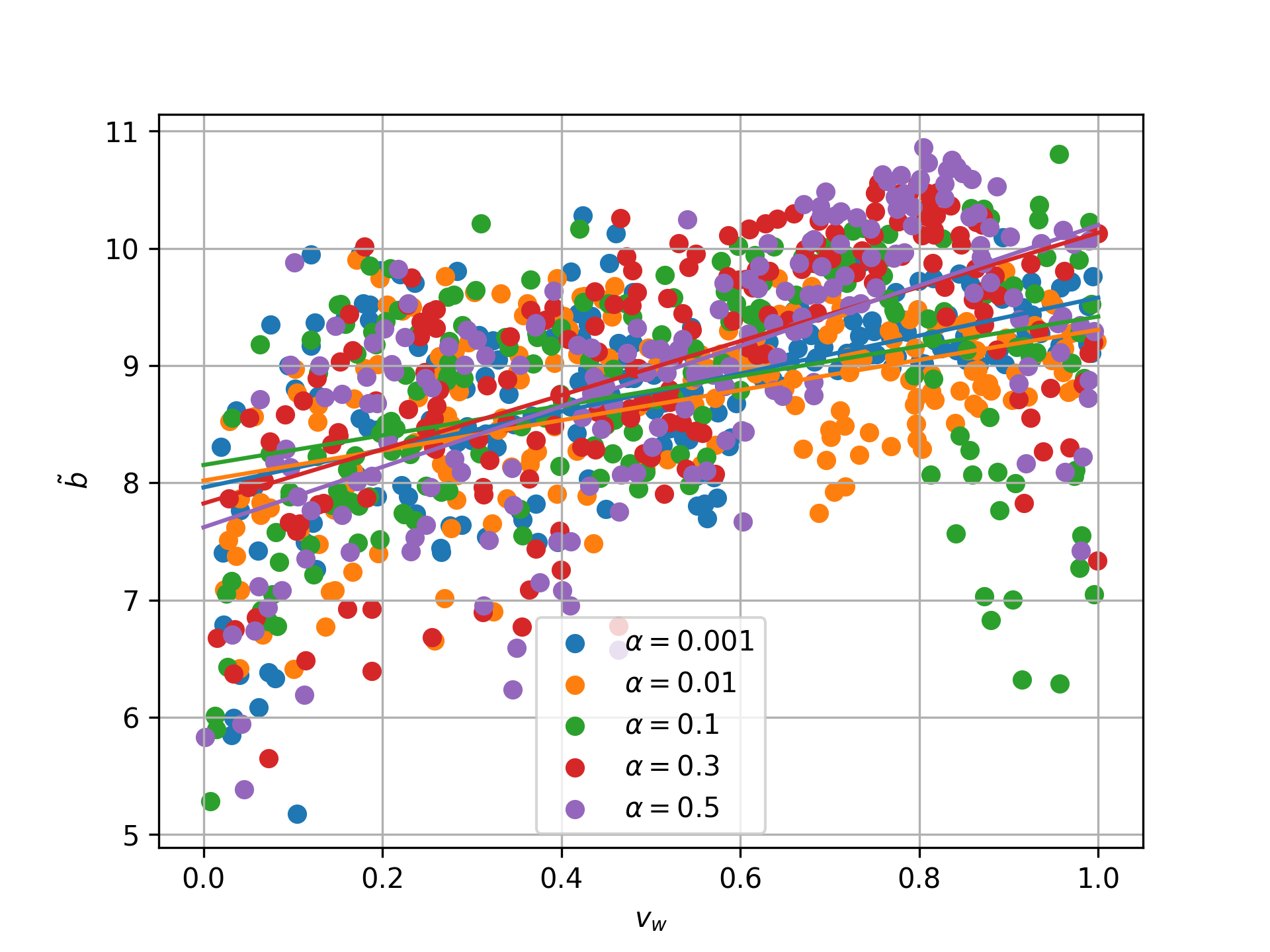}
  \includegraphics[width=0.35\linewidth]{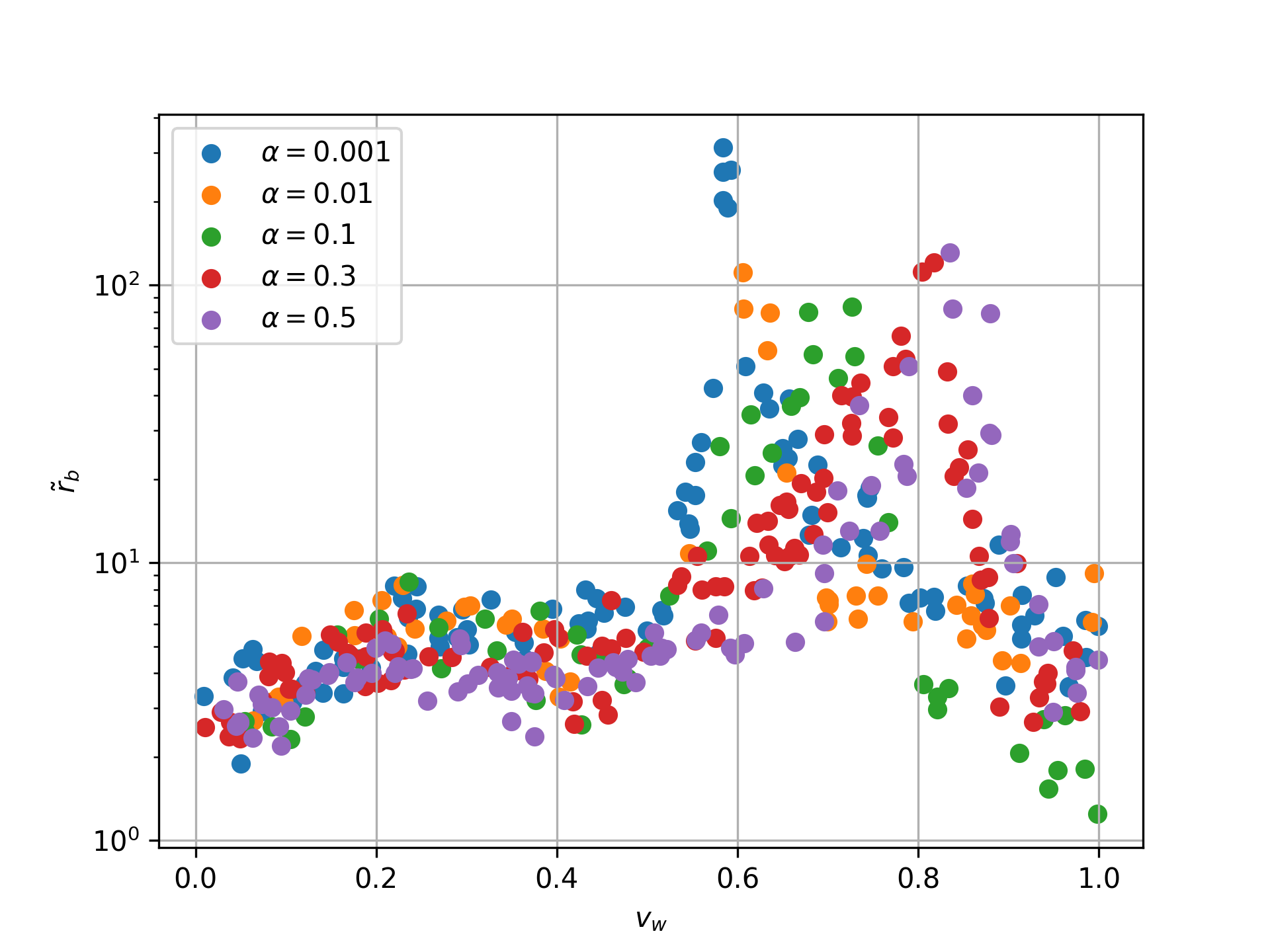}
  \includegraphics[width=0.35\linewidth]{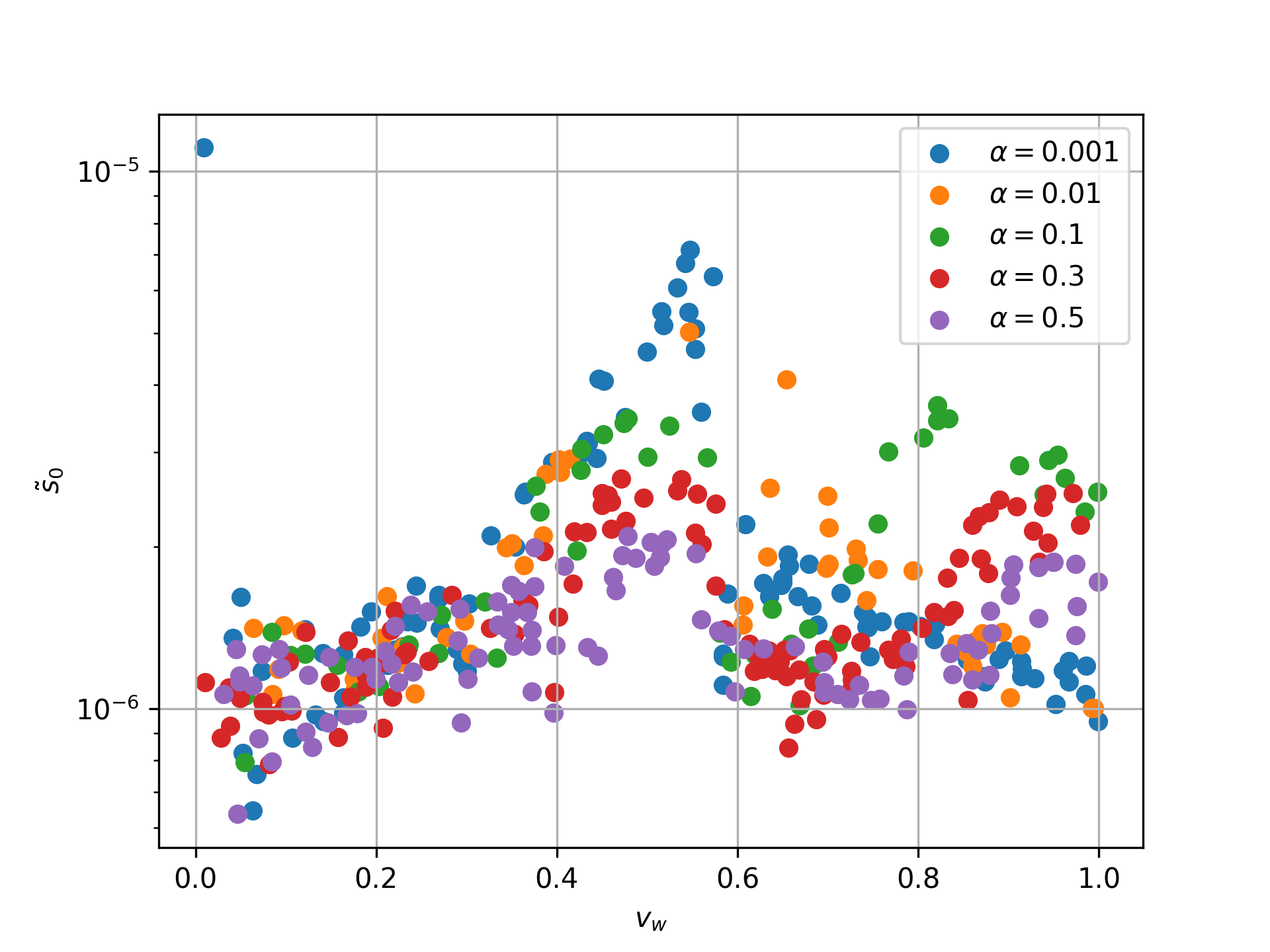}
  \includegraphics[width=0.35\linewidth]{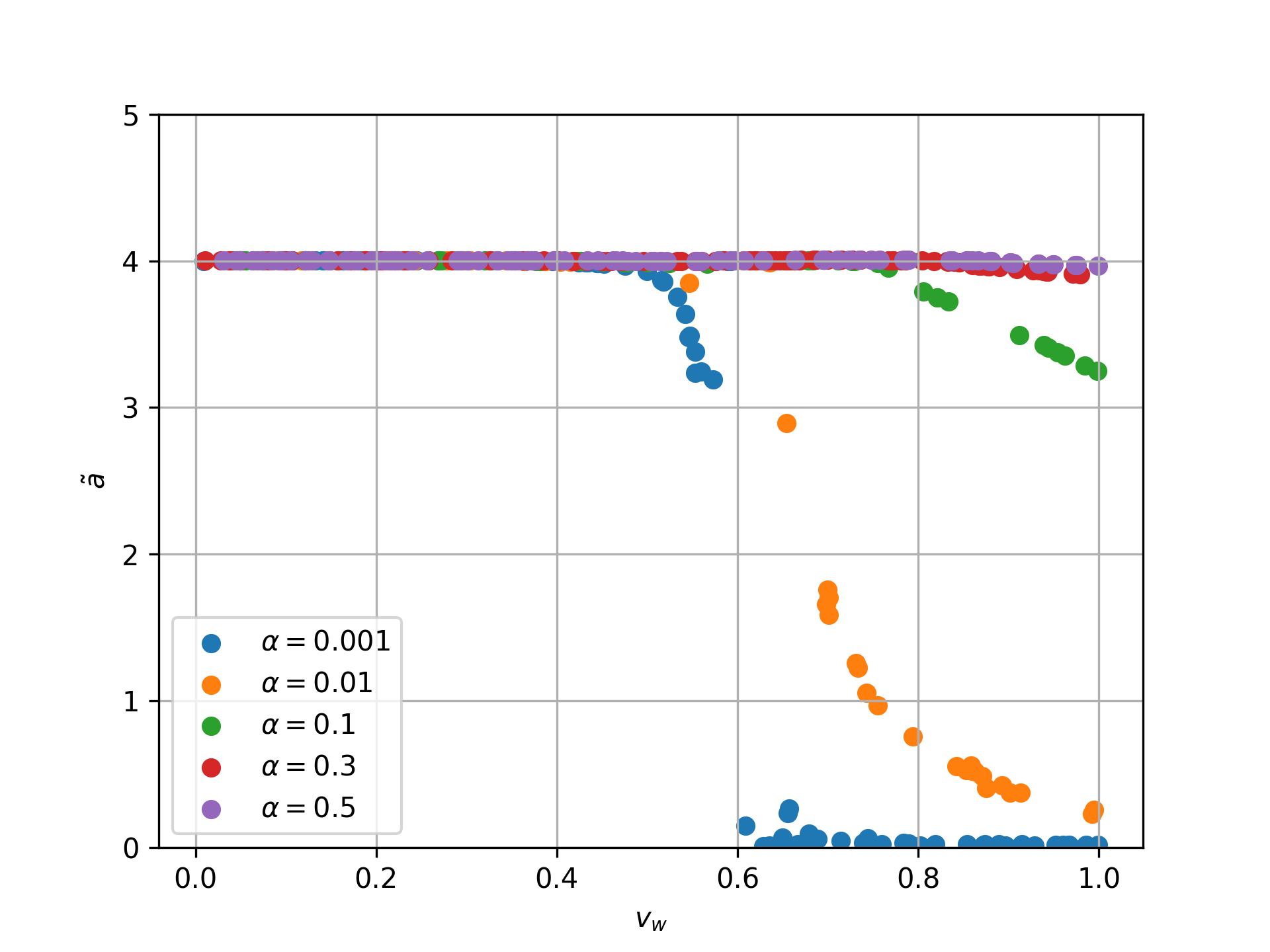}
  \includegraphics[width=0.35\linewidth]{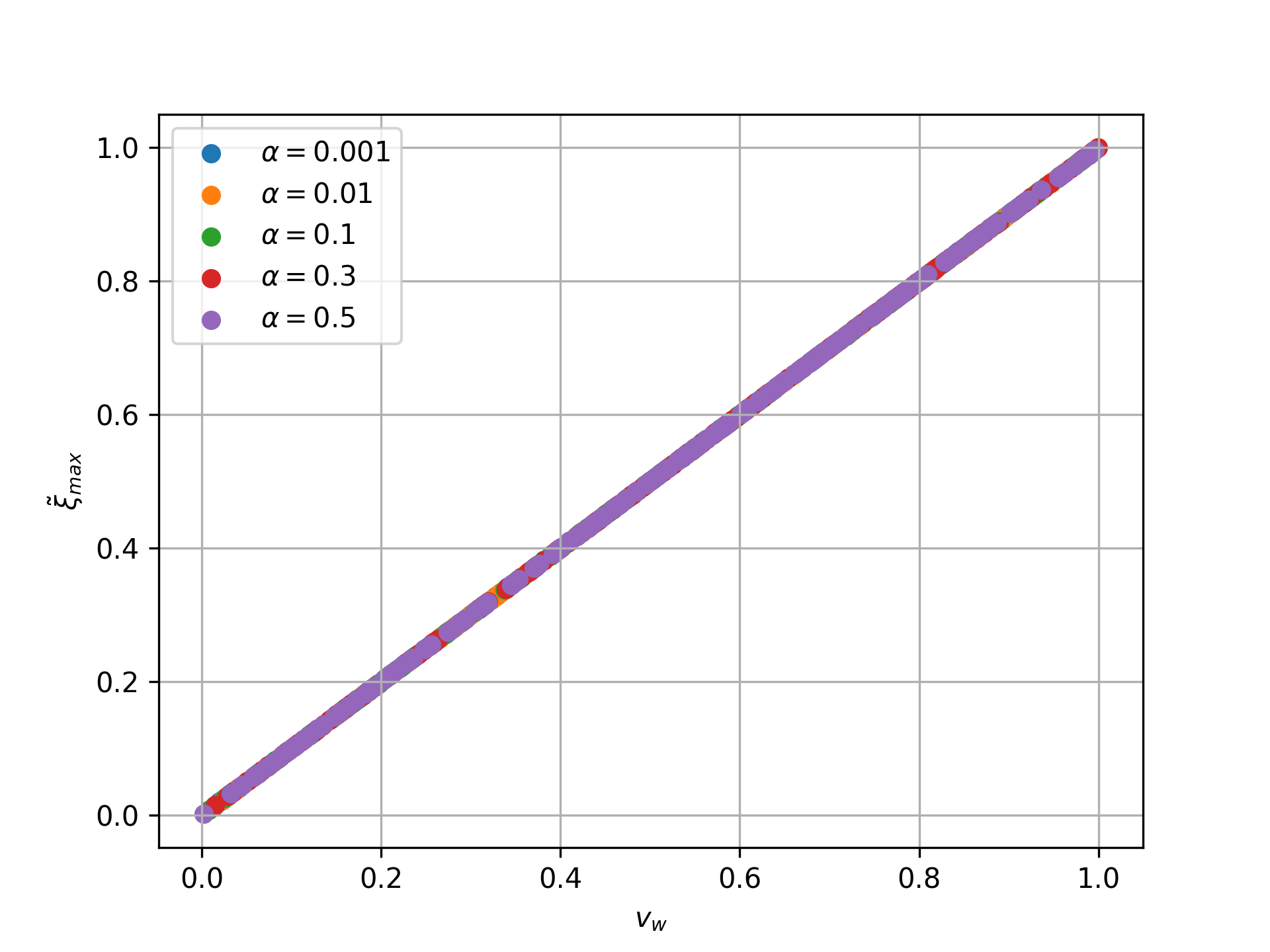}
  \includegraphics[width=0.35\linewidth]{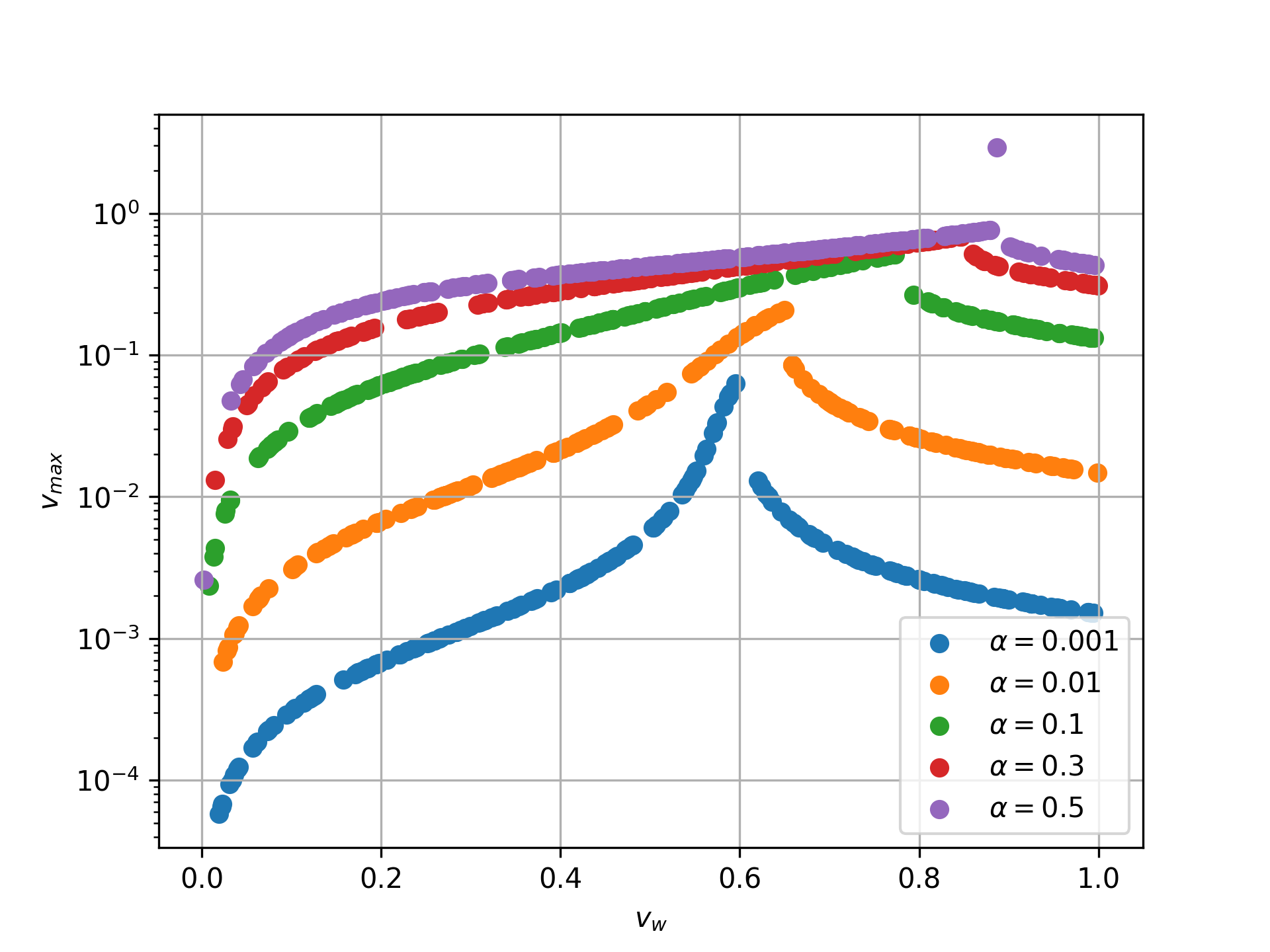}
  \includegraphics[width=0.35\linewidth]{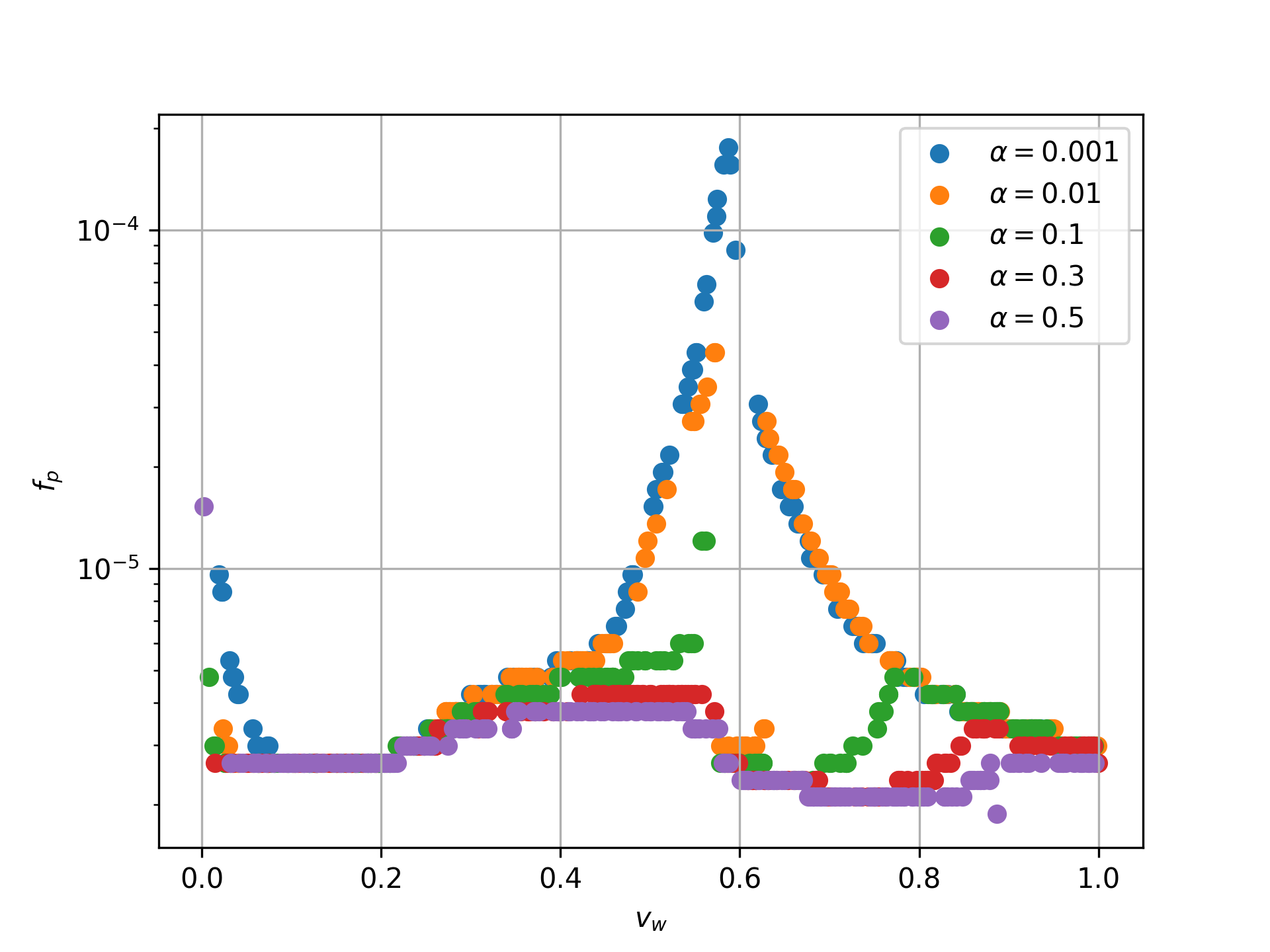}
    \caption{Plots for different variables with respect to $\alpha$ for fixed $\alpha$'s are shown. Each of colors blue, orange, green, red and violet corresponds to a specific value of $\alpha = 0.001,\, 0.01,\,0.1,\,0.3,\,0.5$ shown in the legends. We describe them in the main text. Here we assume $T_{n} = 100$~ GeV and $\beta/H = 1$.}
  \label{fig:alpha-fixed}
\end{figure}

\begin{figure}[H]
  \centering
  \includegraphics[width=0.35\linewidth]{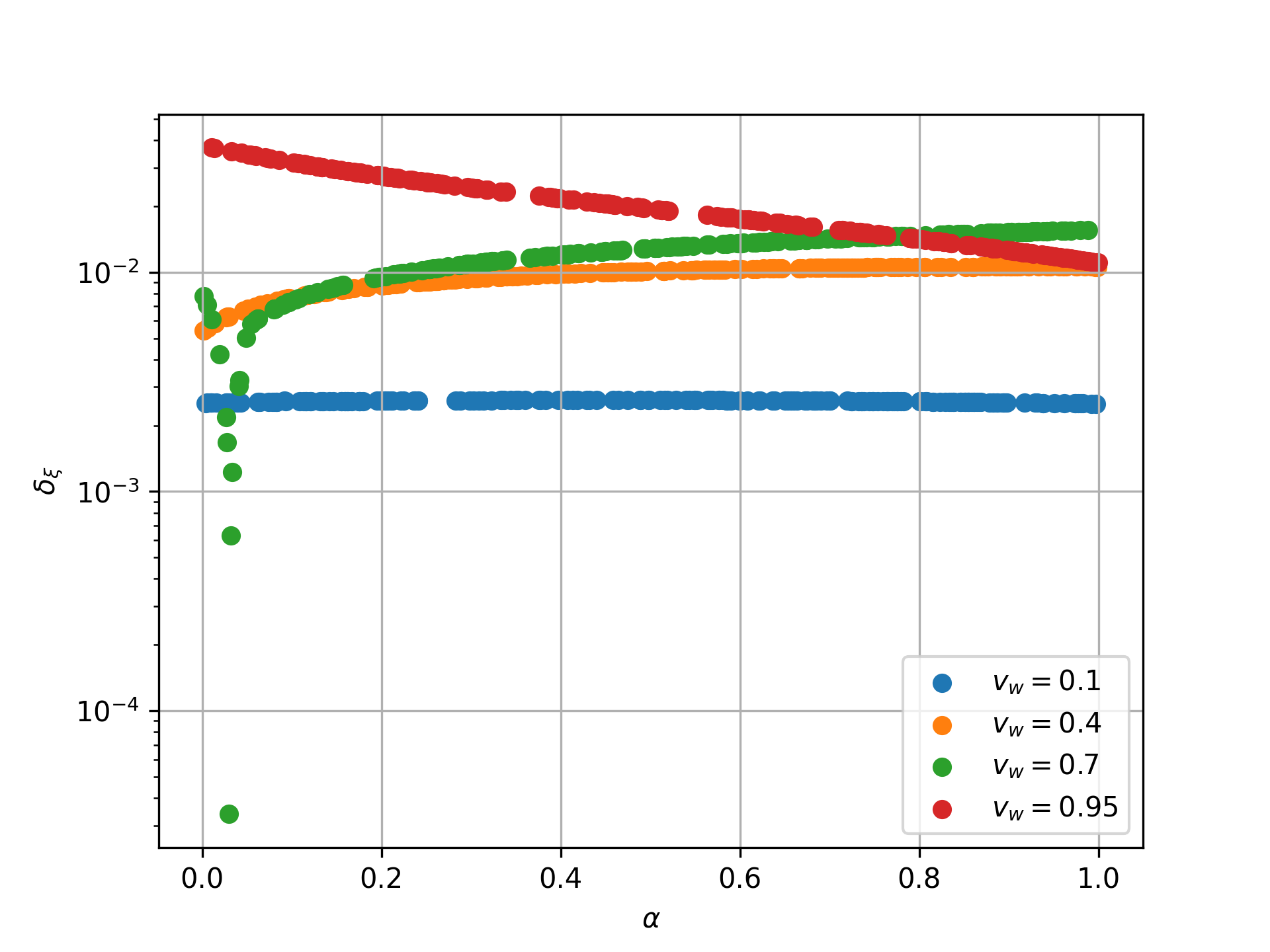}
  \includegraphics[width=0.35\linewidth]{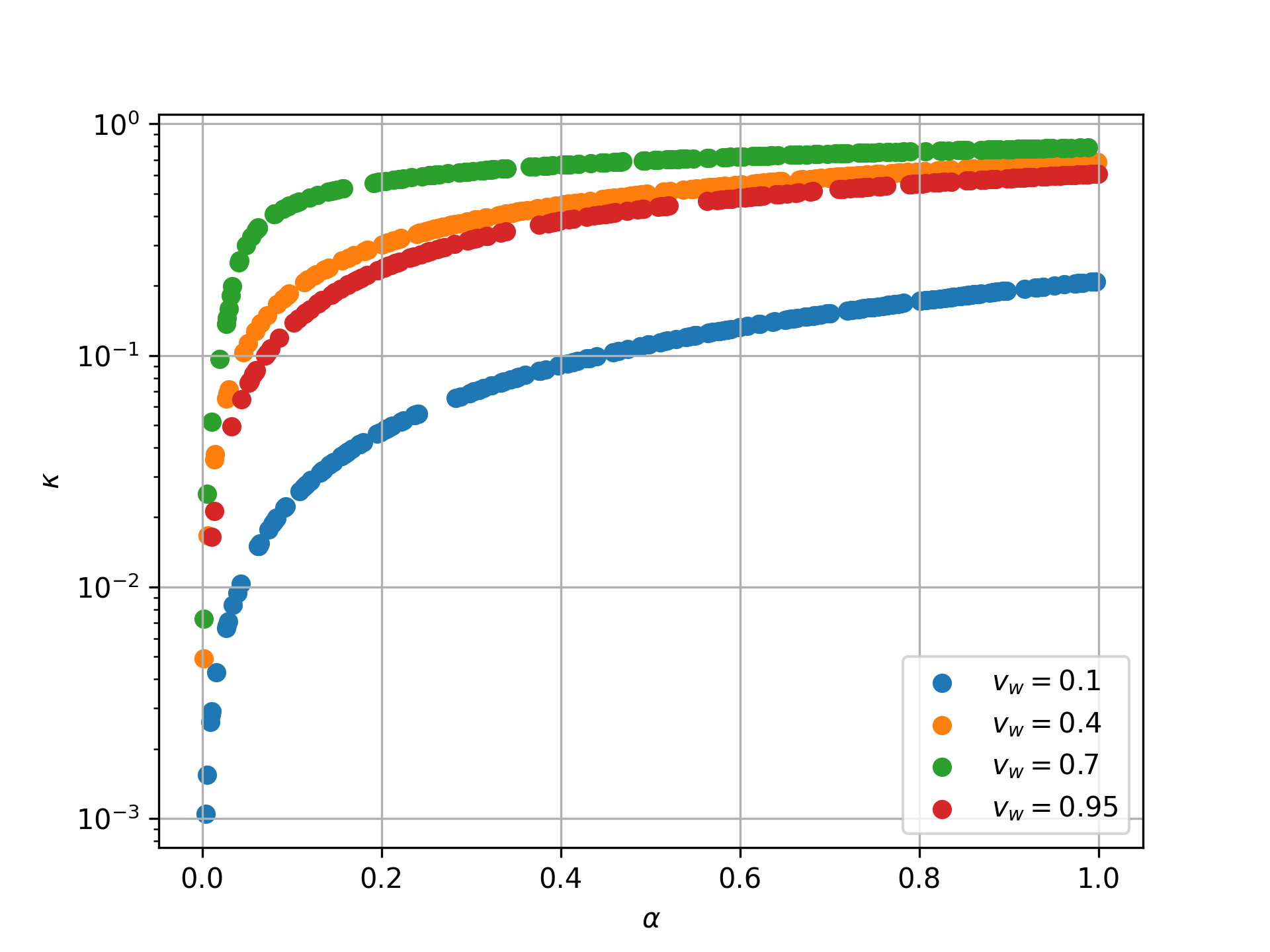}
  \includegraphics[width=0.35\linewidth]{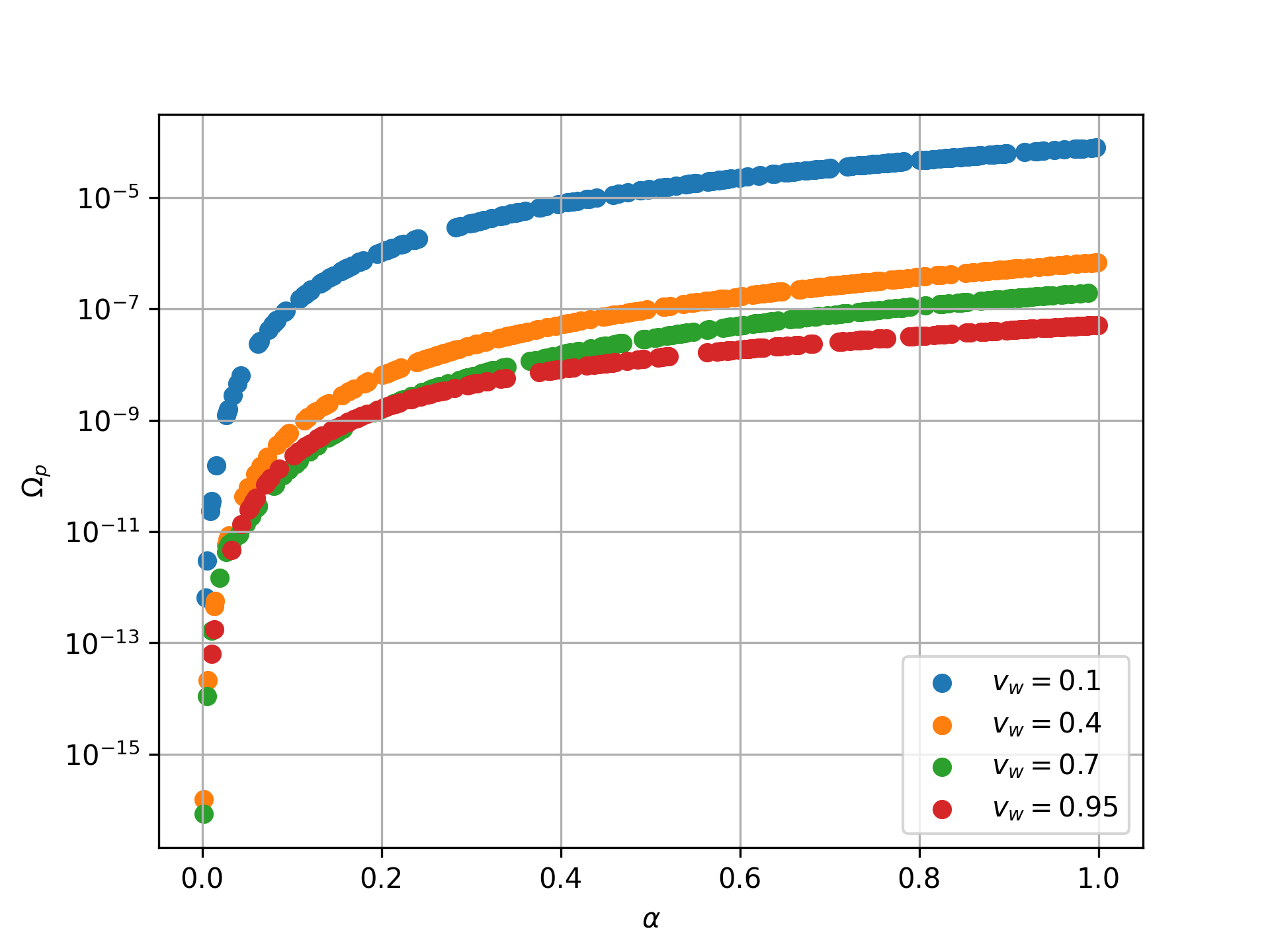}
  \includegraphics[width=0.35\linewidth]{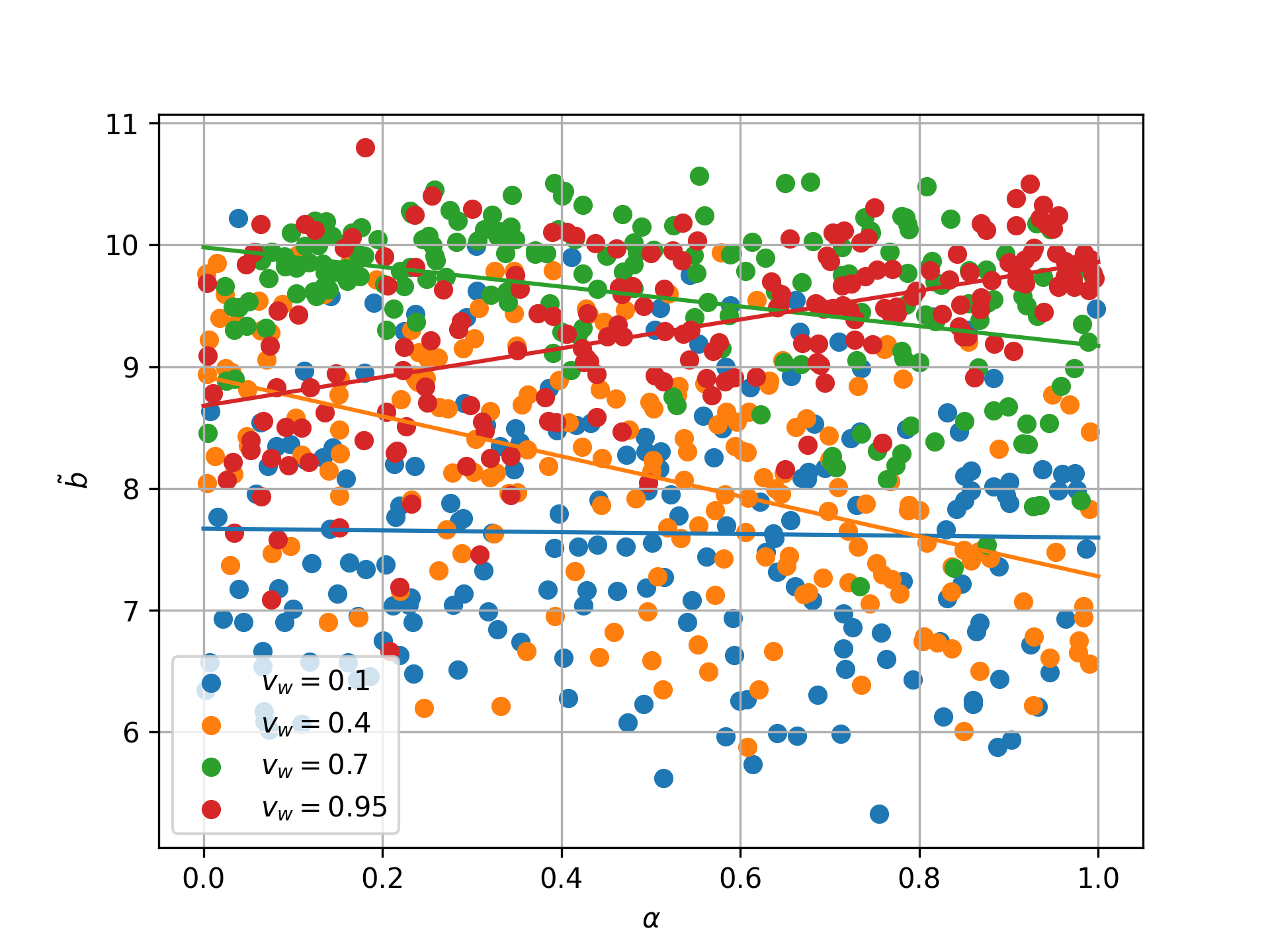}
  \includegraphics[width=0.35\linewidth]{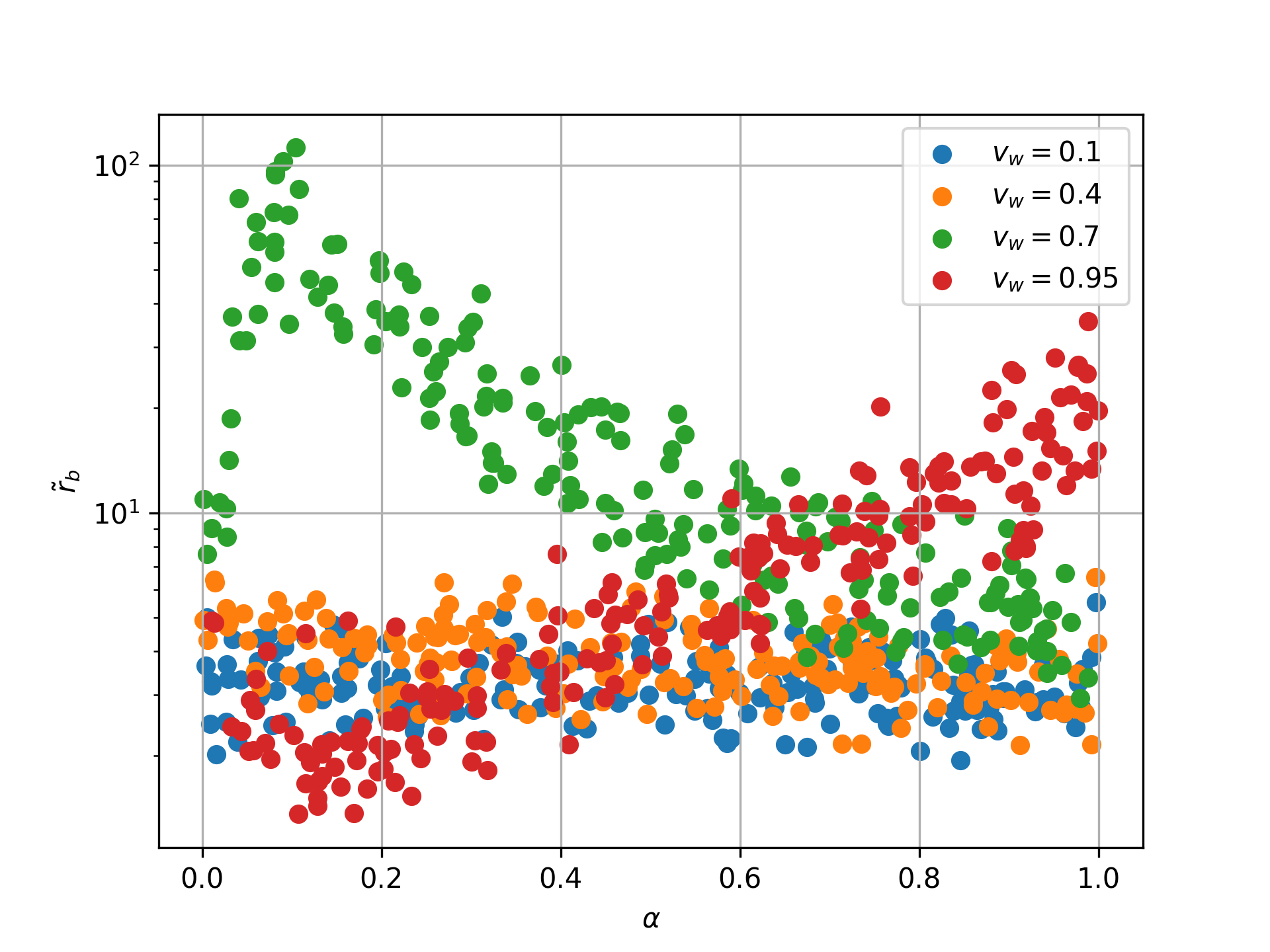}
  \includegraphics[width=0.35\linewidth]{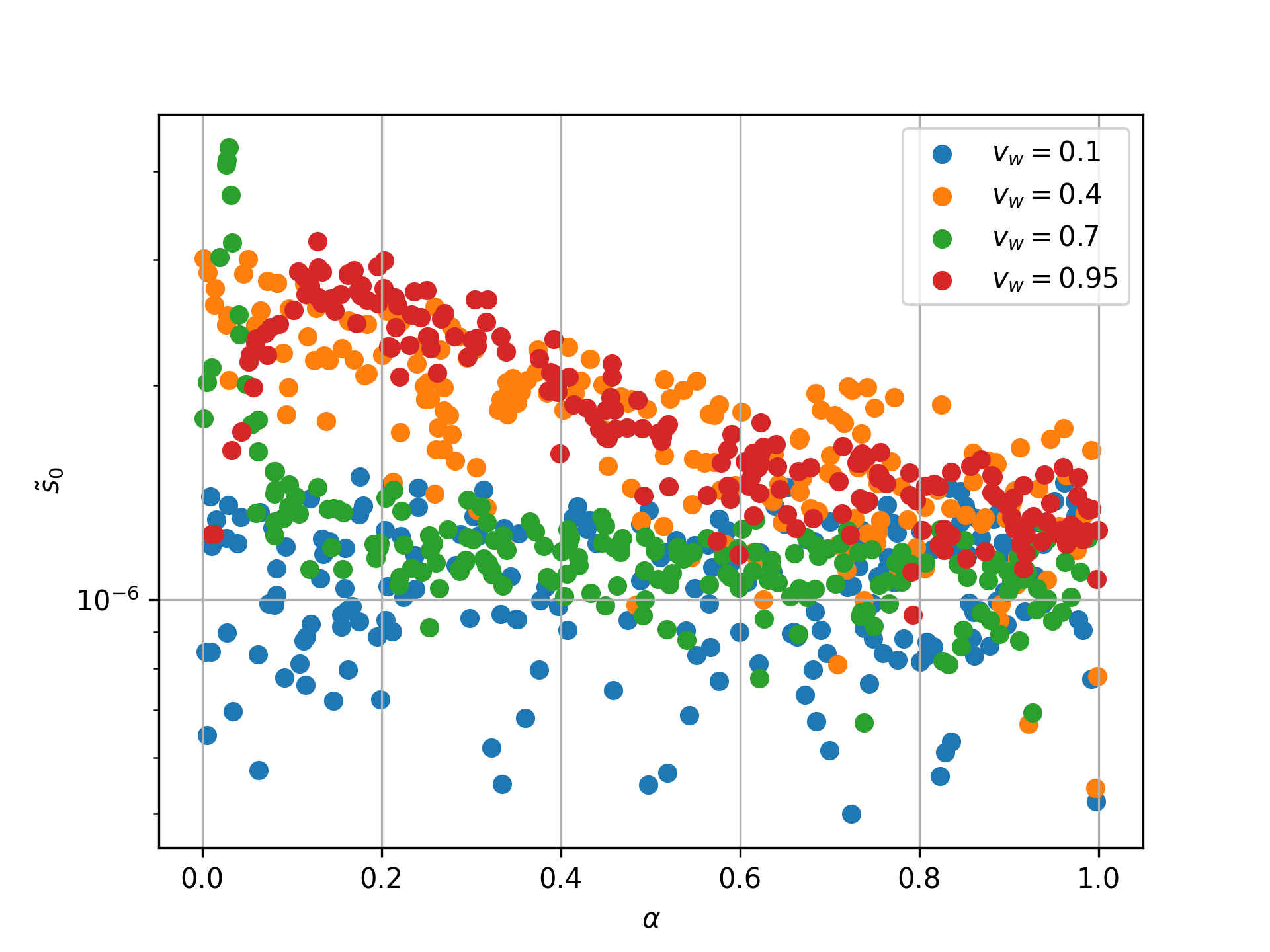}
  \includegraphics[width=0.35\linewidth]{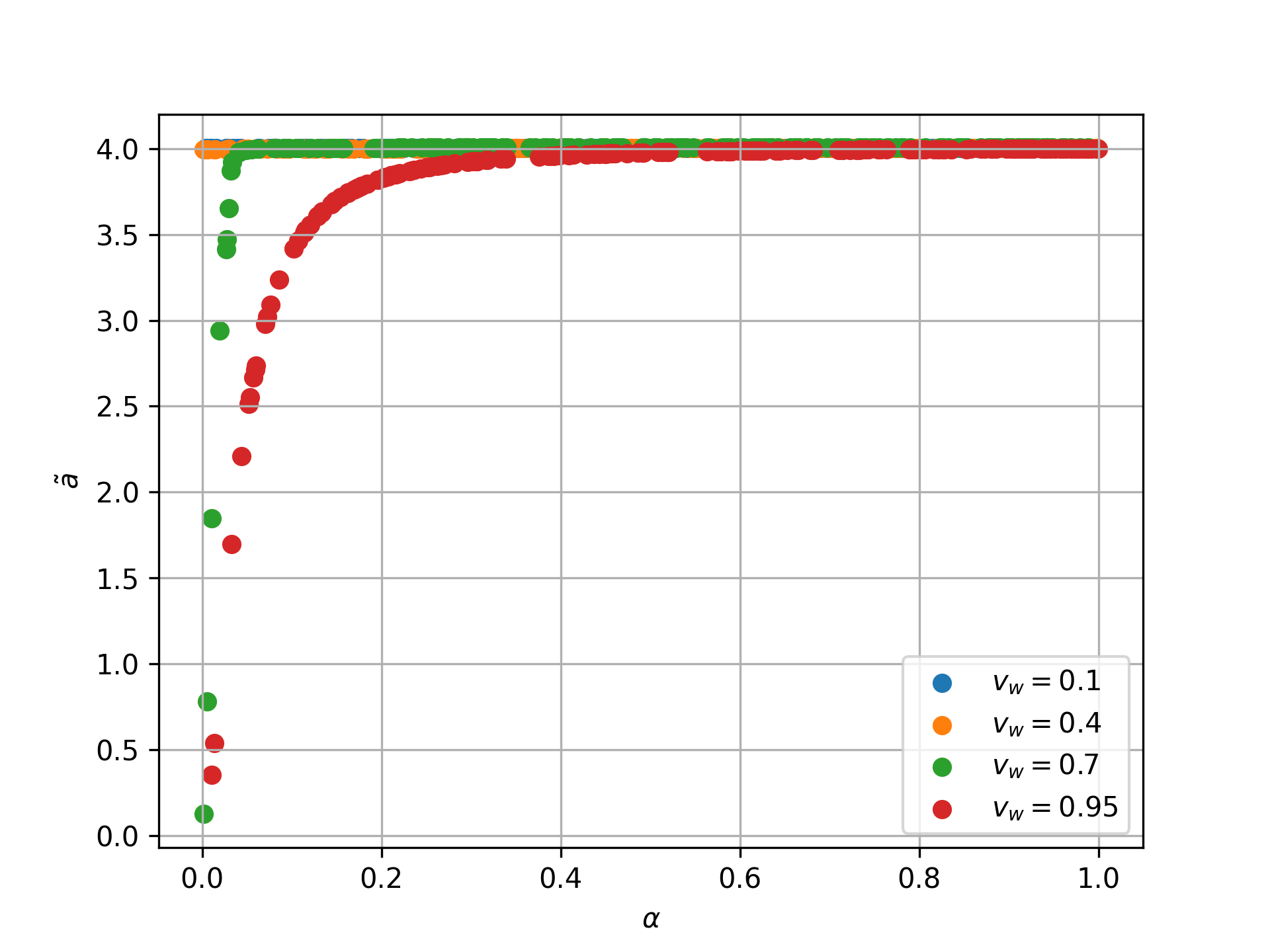}
  \includegraphics[width=0.35\linewidth]{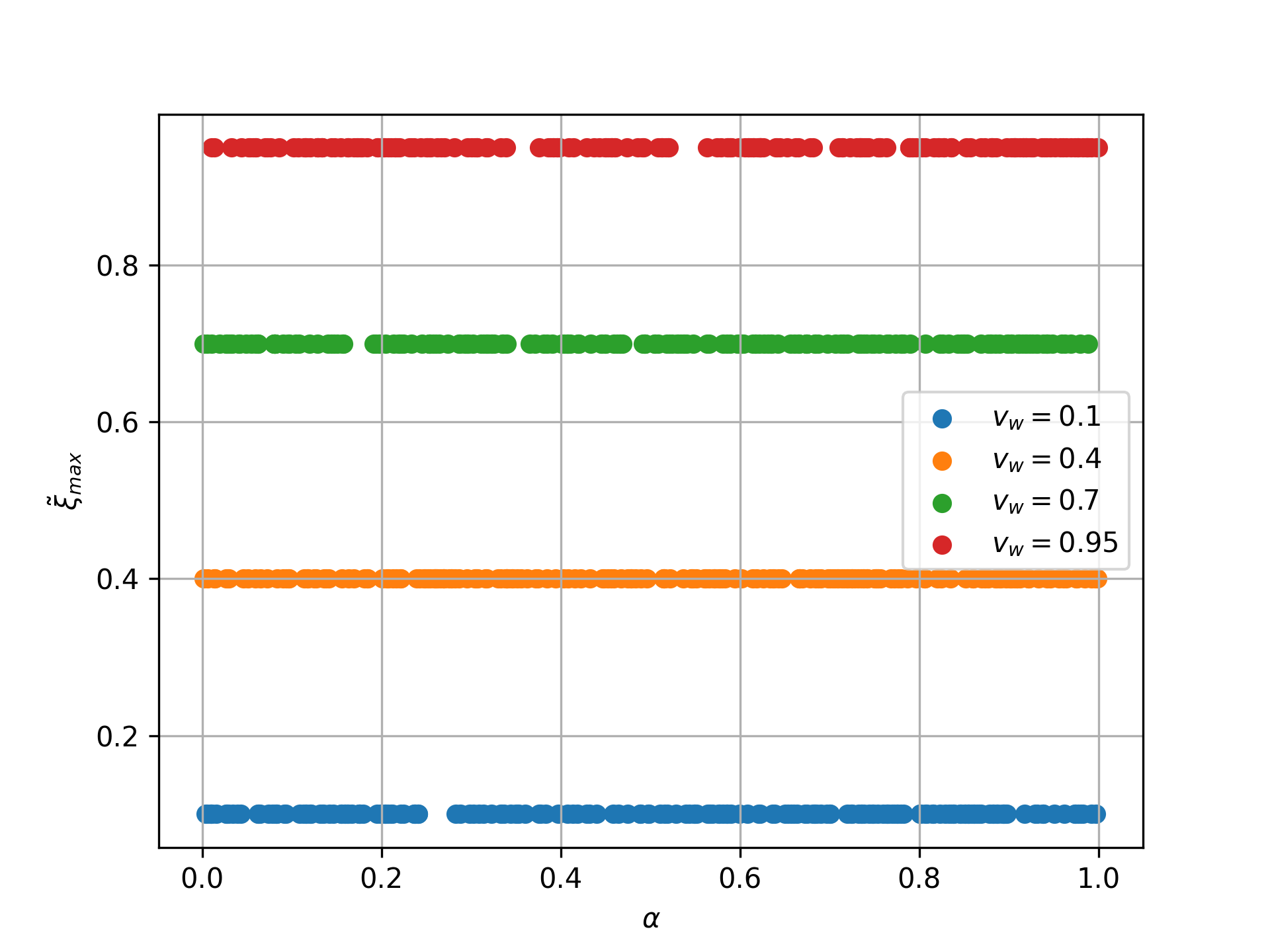}
  \includegraphics[width=0.35\linewidth]{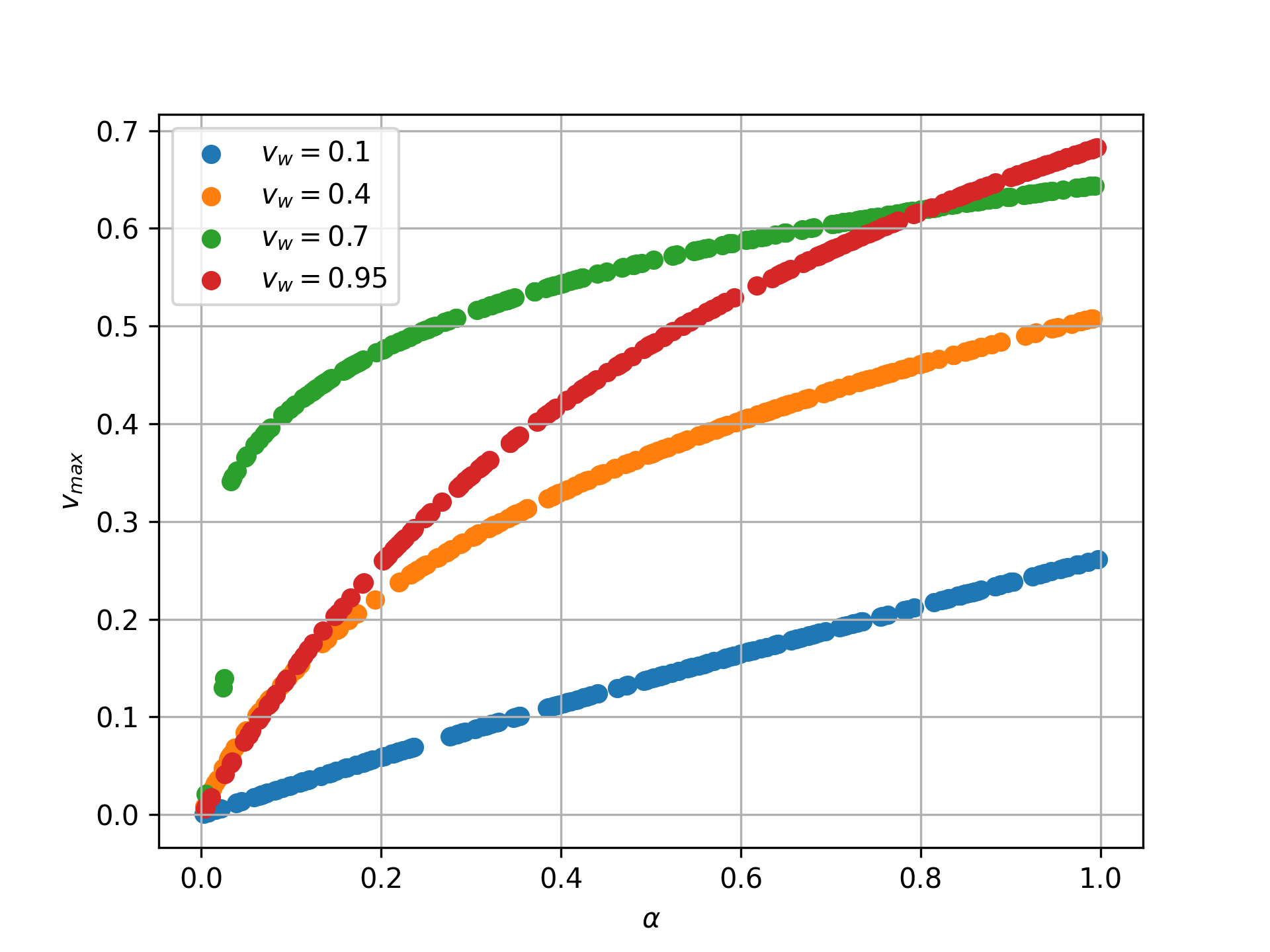}
  \includegraphics[width=0.35\linewidth]{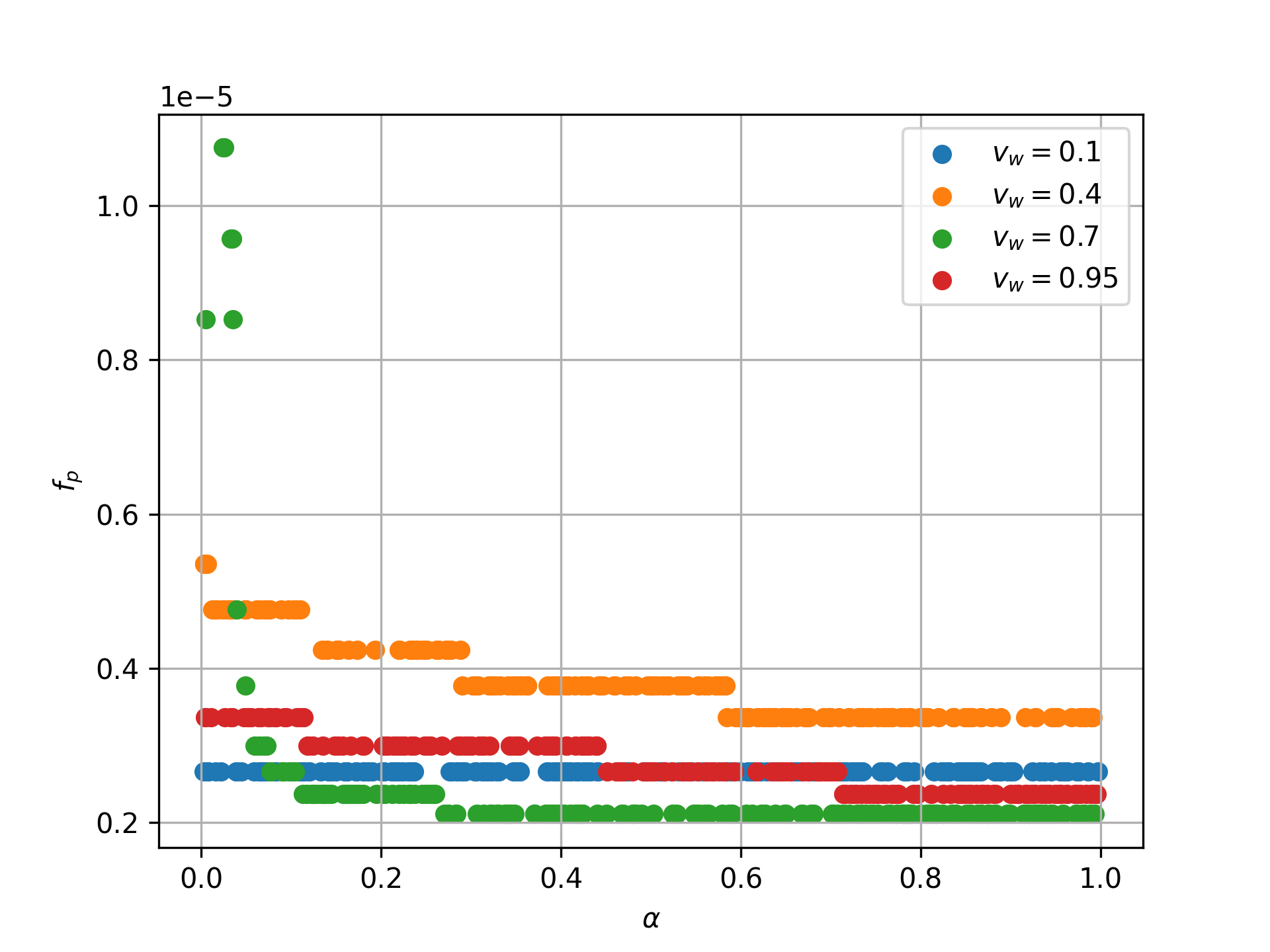}
    \caption{Plots for different variables with respect to $\alpha$ for fixed $v_w$'s are shown.  Each of colors blue, orange, green and red corresponds to a specific value of $v_w = 0.1,\,0.4,\,0.7,\,0.95$. We describe them in the main text. Here we assume $T_{n} = 100$~ GeV and $\beta/H = 1$.}
  \label{fig:vw-fixed}
\end{figure}

\begin{figure}[H]
  \centering
  \includegraphics[width=0.35\linewidth]{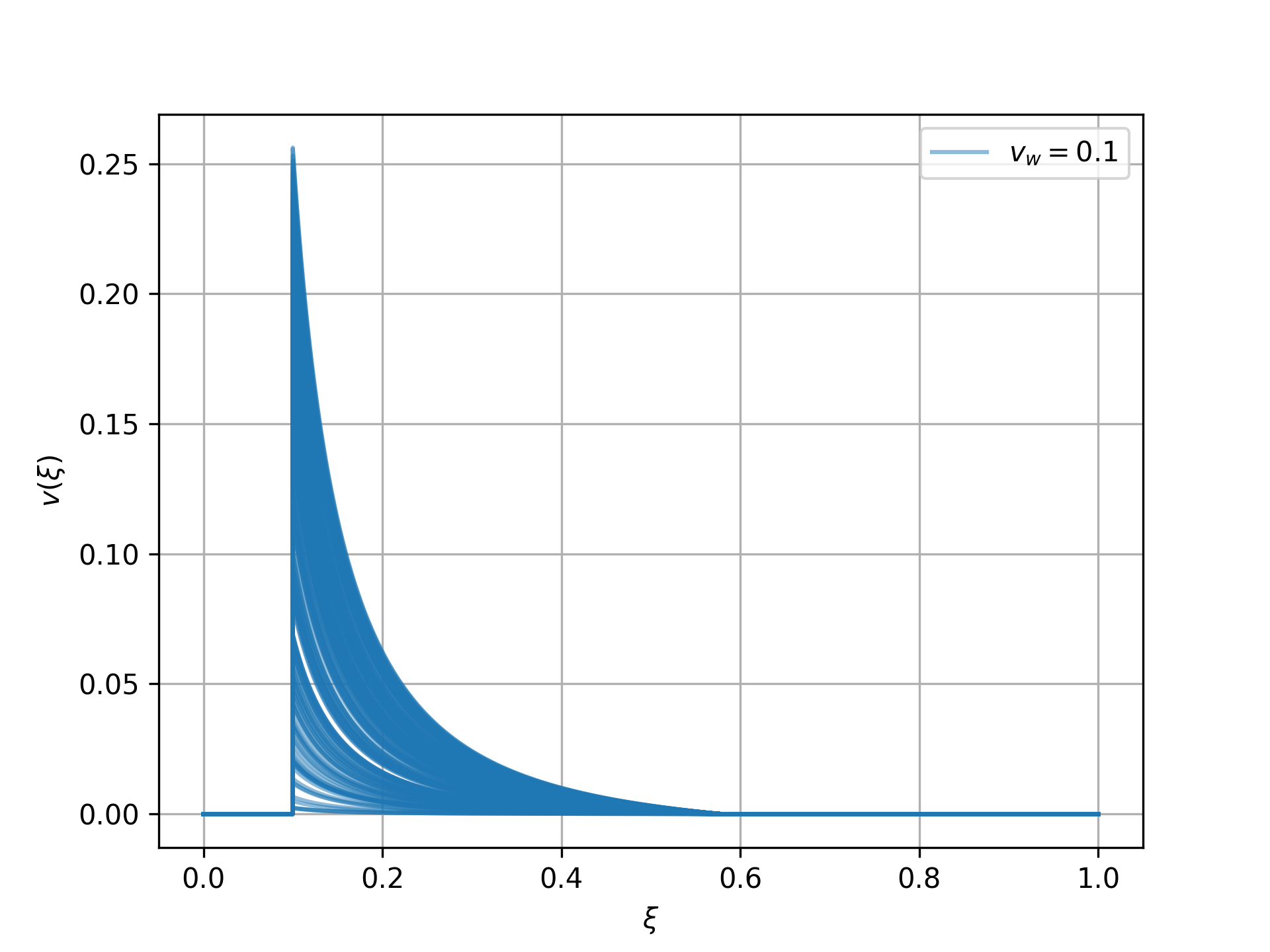}
  \includegraphics[width=0.35\linewidth]{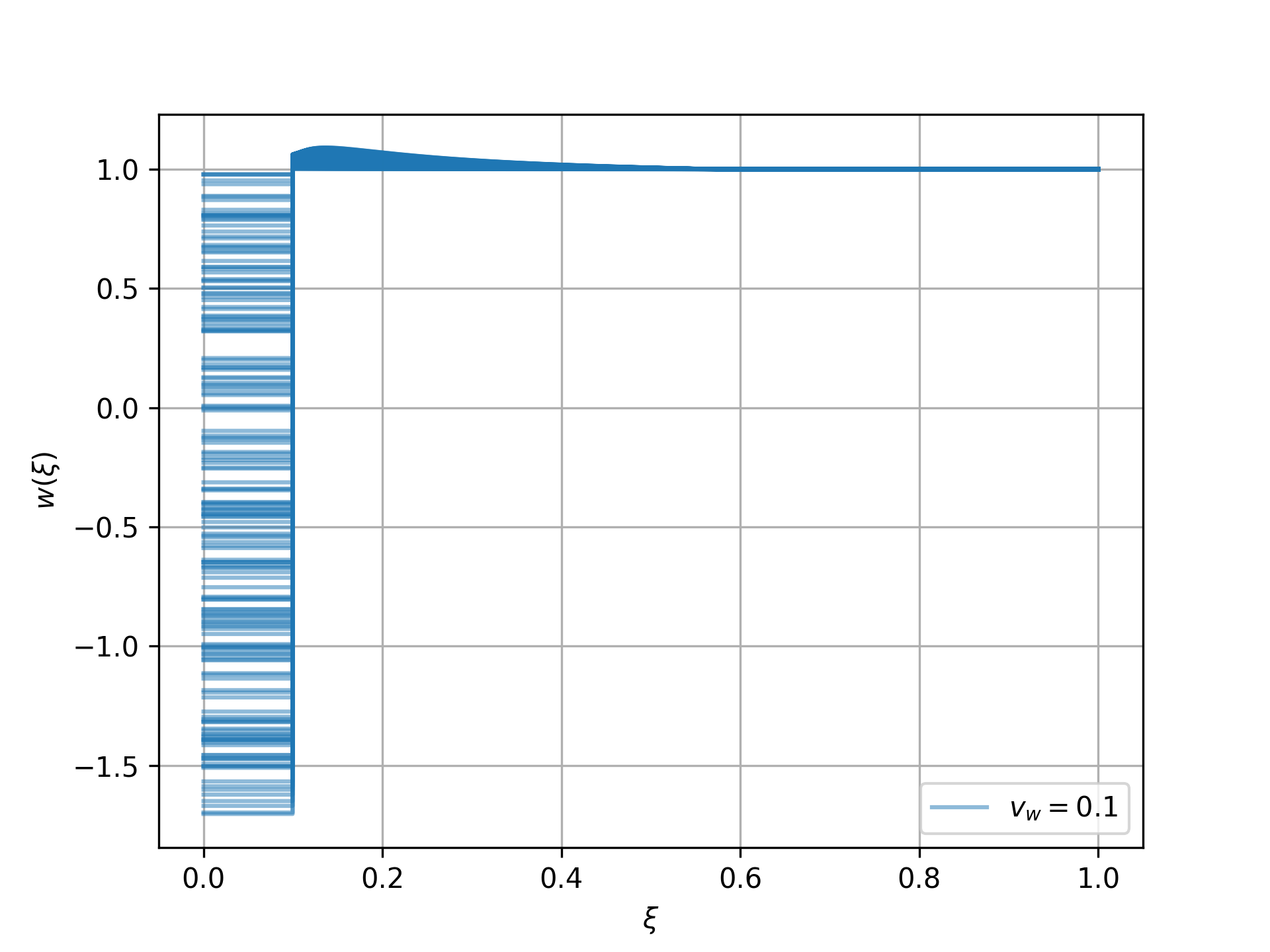}
  \includegraphics[width=0.35\linewidth]{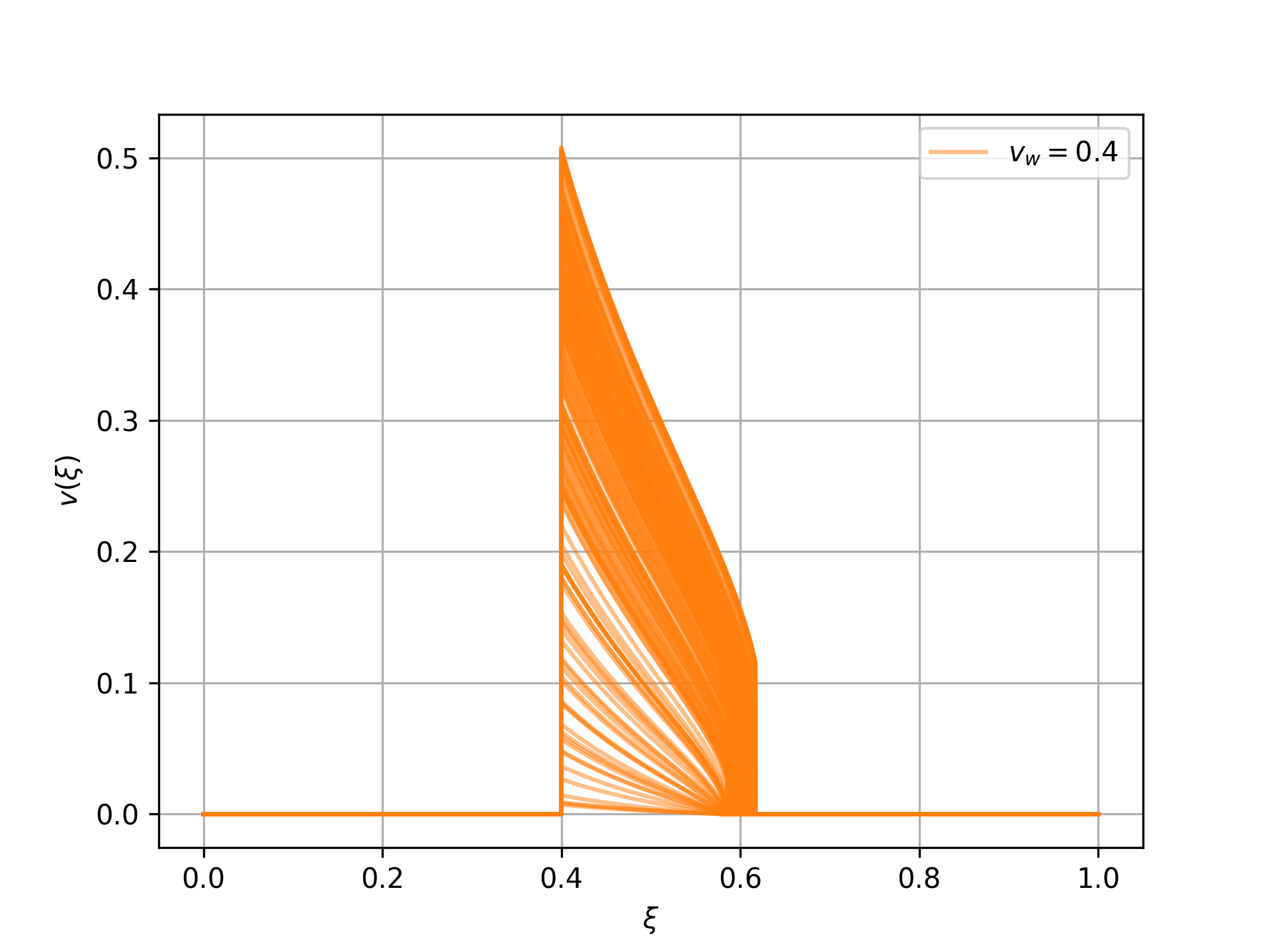}
  \includegraphics[width=0.35\linewidth]{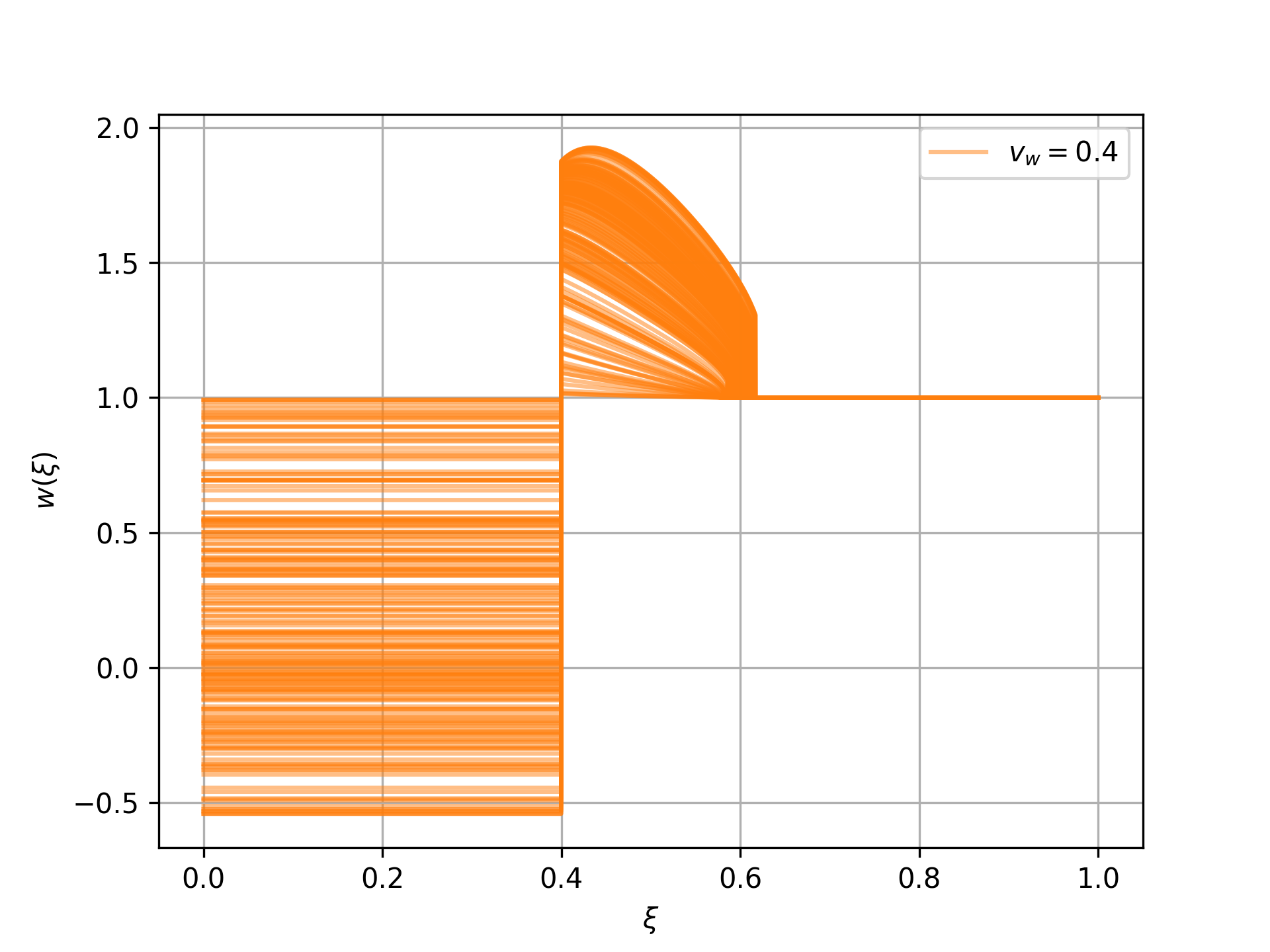}  \includegraphics[width=0.35\linewidth]{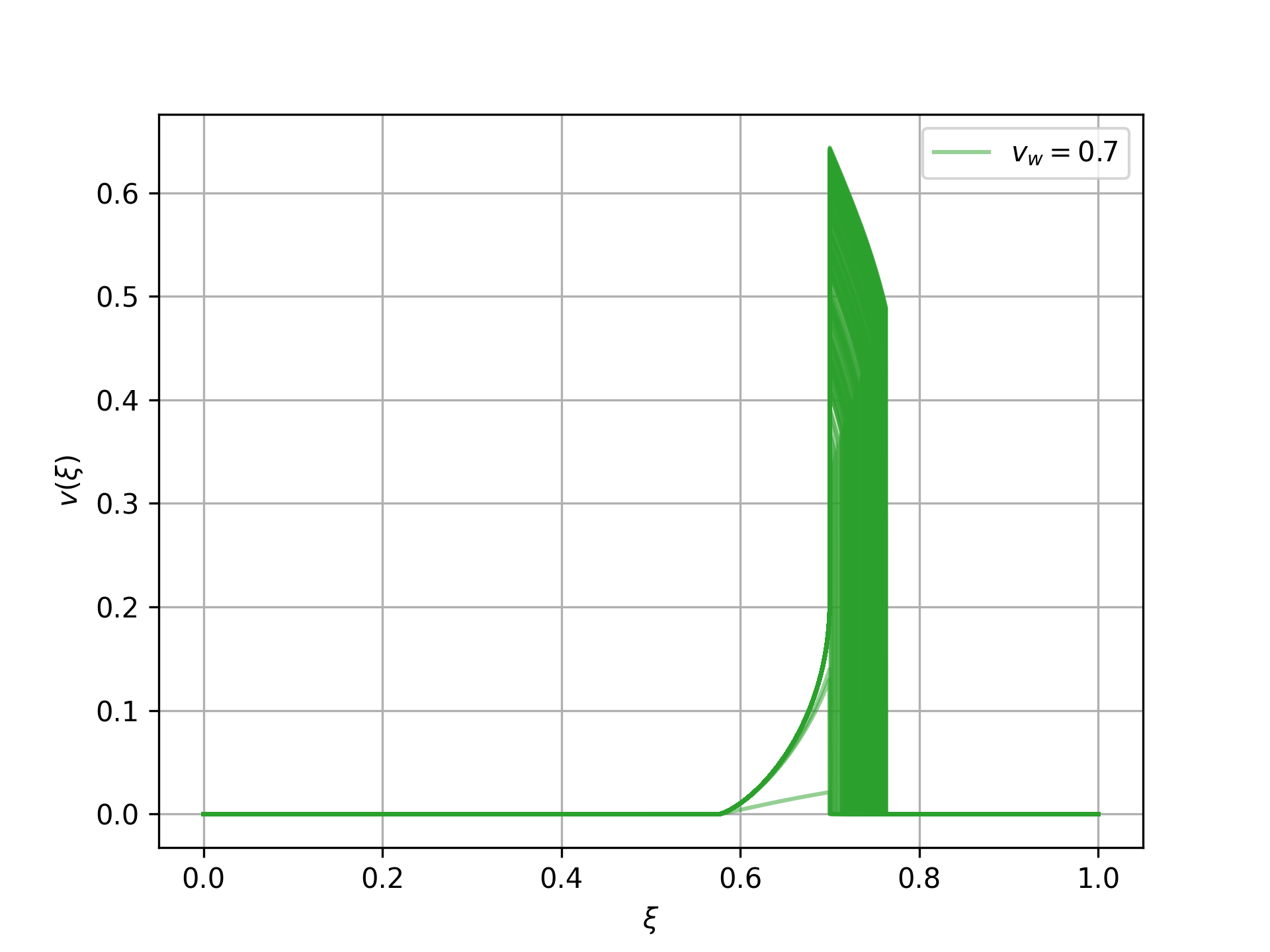}  \includegraphics[width=0.35\linewidth]{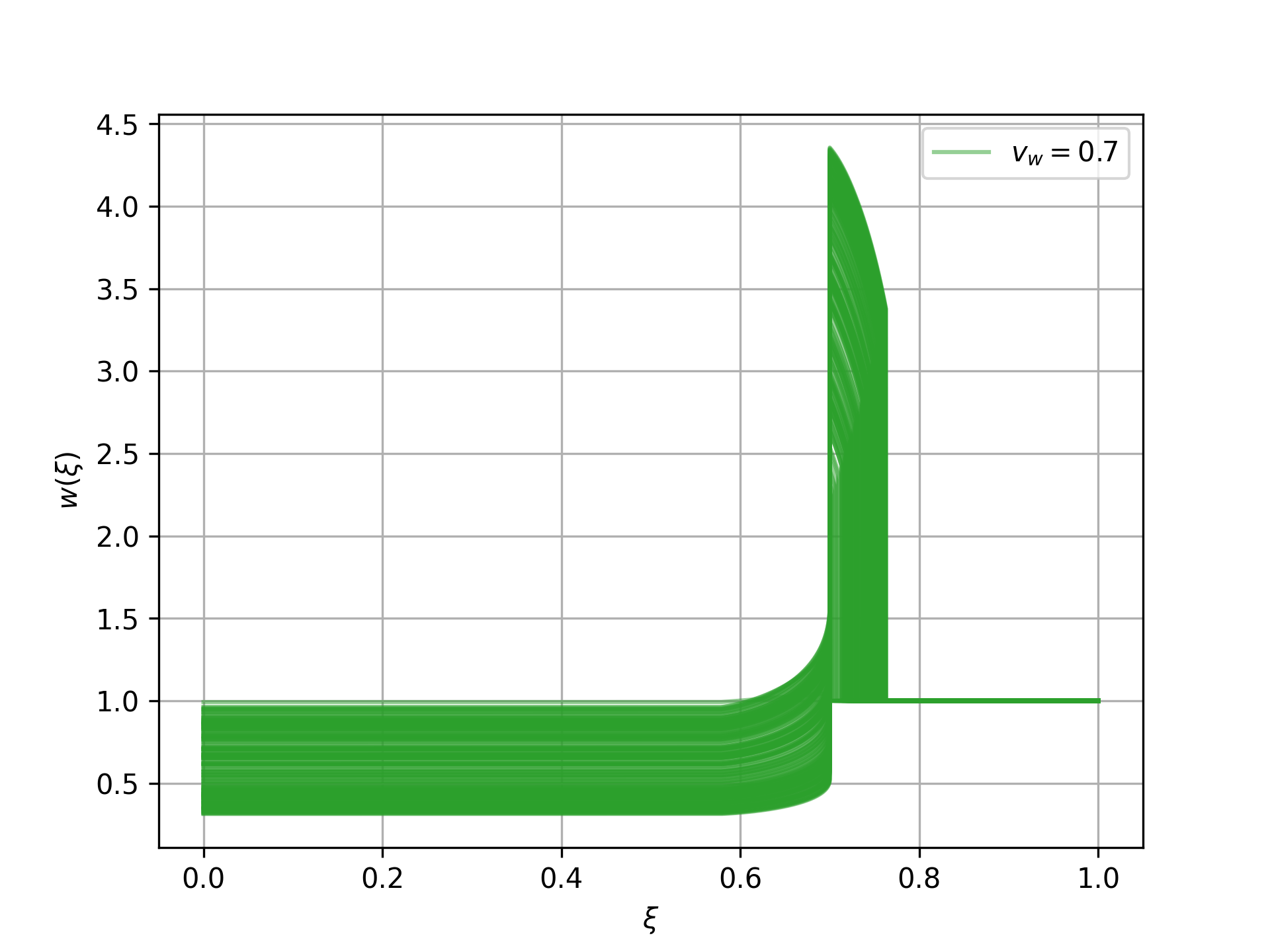}   \includegraphics[width=0.35\linewidth]{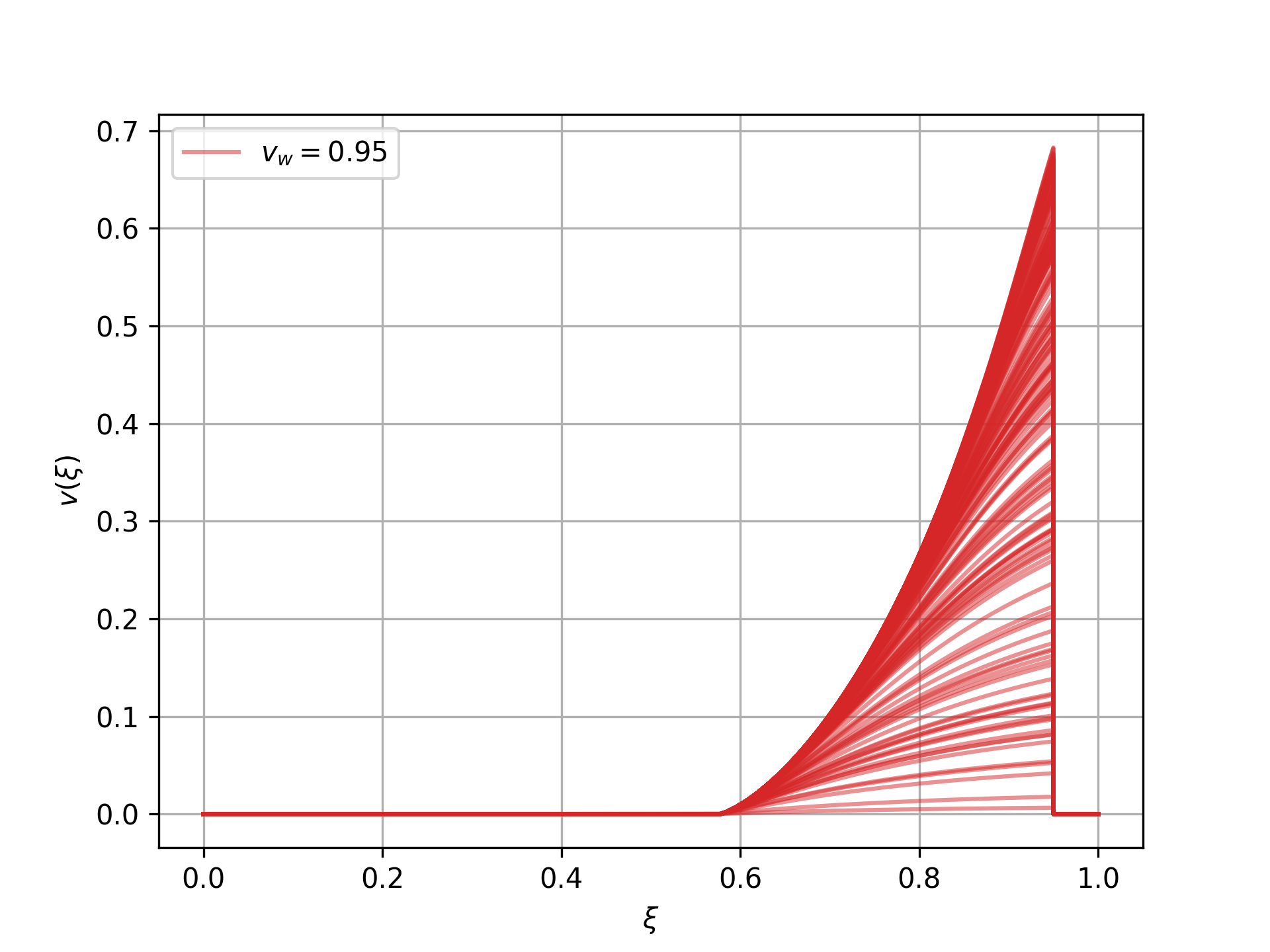}  \includegraphics[width=0.35\linewidth]{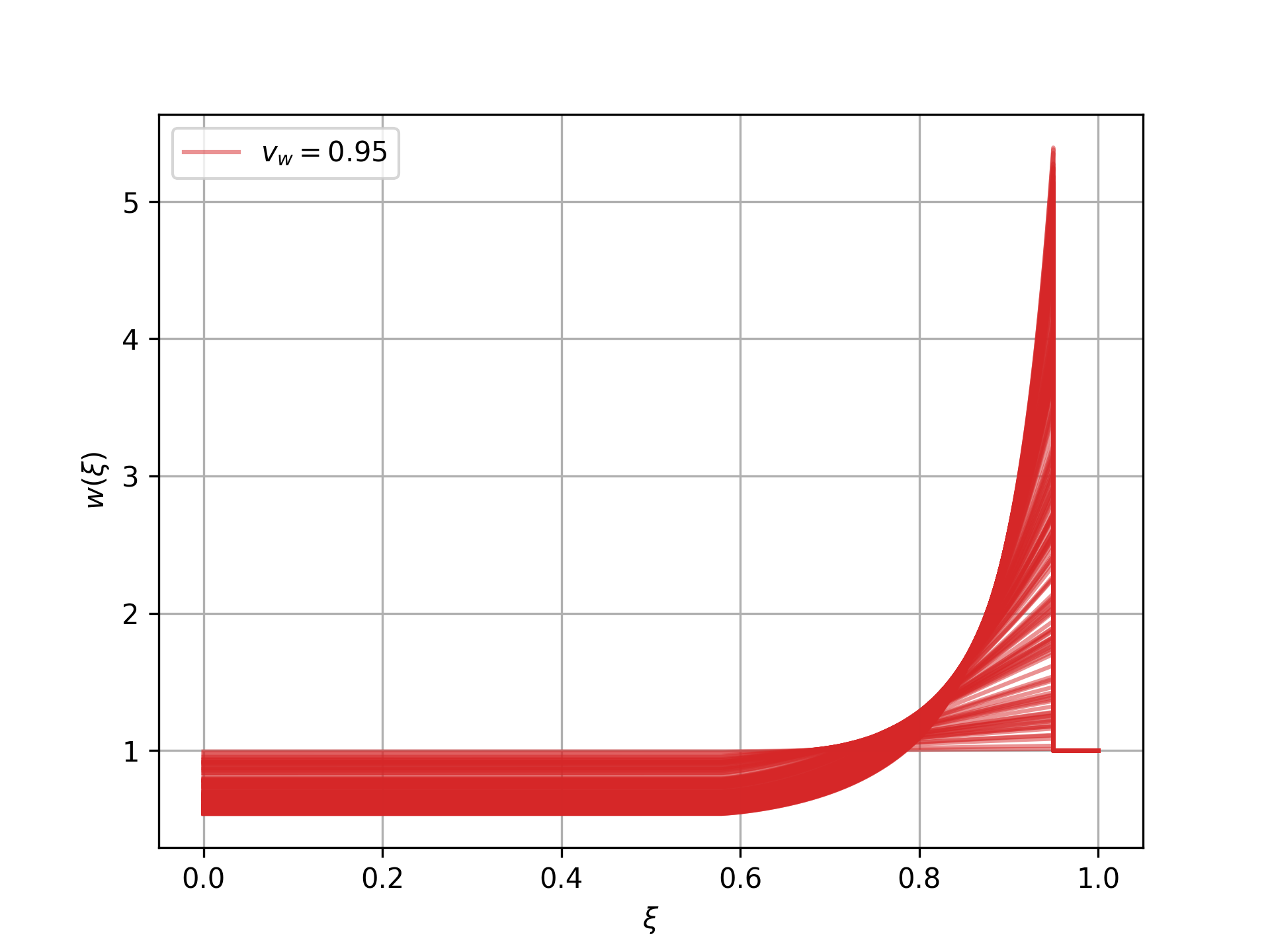}
    \caption{Velocity and enthalpy profiles for fixed values of  wall velocity $v_w$. Each row of these panels corresponds to  a color that are blue, orange, green and red  equivalent to a specific value of $v_w = 0.1,\,0.4,\,0.7,\,0.95$ shown in the legends.}
  \label{fig:vwprof-plots-vw}
\end{figure}

The precision GW frontier has several other challenges that lie outside the scope of this work. The transitions between the different power law regimes in the GW spectrum, the existence of new regimes, the shape of the GW spectrum near the peak, and the peak frequency  depend on several factors: fluctuations in the local temperature affect the distribution of nucleated bubbles \cite{Jinno:2021ury};  energy lost to vorticity can suppress the spectrum \cite{Cutting:2019zws}; and dissipative effects encoded by the shear viscosity, bulk viscosity, and thermal conduction can also suppress the spectrum \cite{Guo:2023koq}. Calculating these effects would make further material progress on the precision frontier.

\acknowledgments 
We are grateful to Jose Miguel No for collaboration during the initial stages of the project, and to 
Daniel Vagie for useful discussions.  F.H. and K.S. thank the organizers of the Mitchell Conference in  May 2024 at  Texas A \& M University for their hospitality and support during the final stages of this project. 
They are also grateful to the organizers of workshop of Center for Theoretical Underground Physics and Related Areas (CETUP* - 2024), The Institute for Underground Science at Sanford
Underground Research Facility (SURF), Lead, South Dakota 
for their hospitality and financial support.
K.S. would like to thank the Aspen Center for Theoretical Physics, supported by the National Science Foundation grant PHY-2210452, for hospitality during the course of this work.

\appendix

\section{Other Scanning Plots}
\label{app-a}
\begin{figure}[h]
  \centering
  \includegraphics[width=0.8\linewidth]{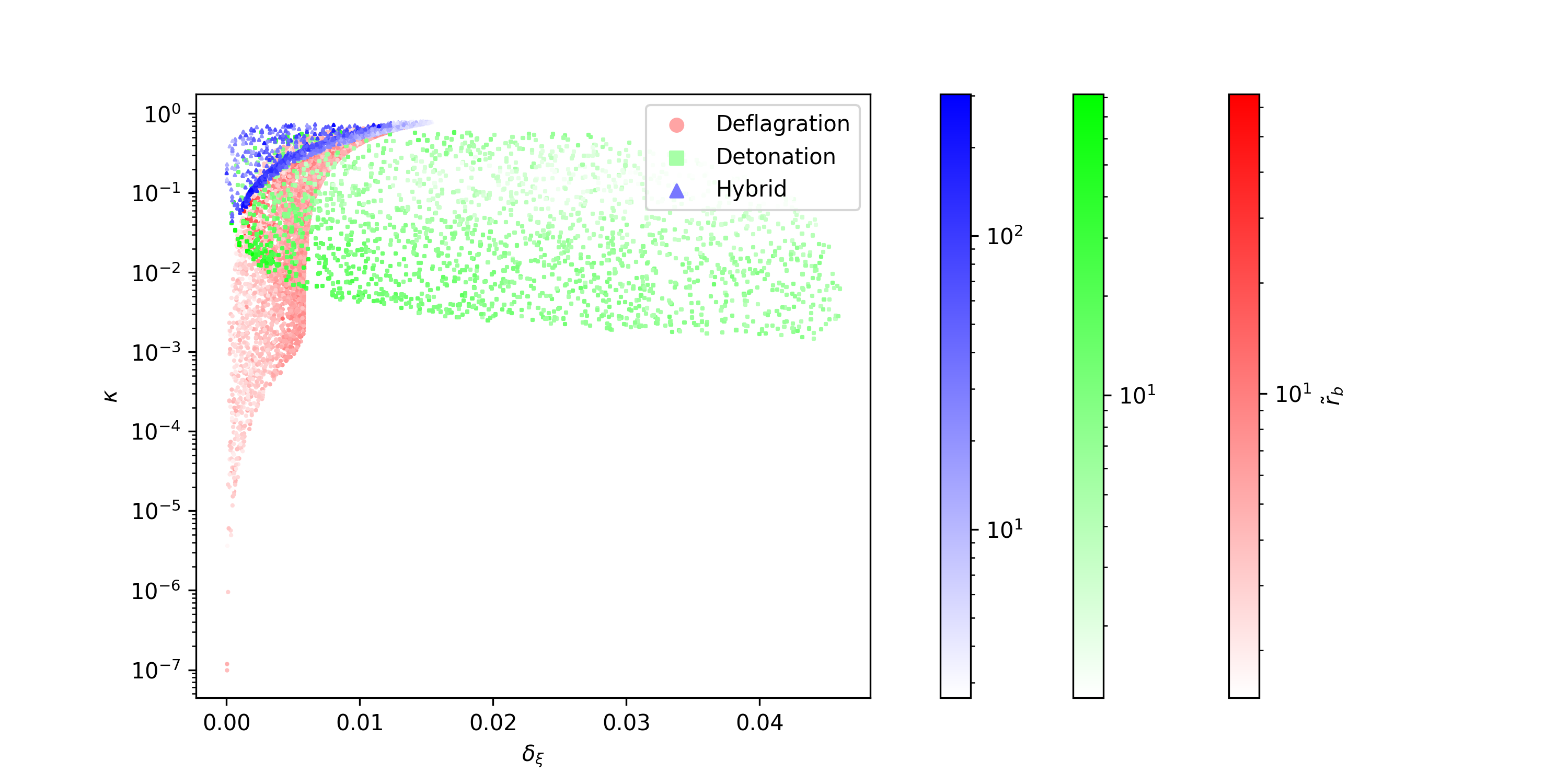}
  \includegraphics[width=0.8\linewidth]{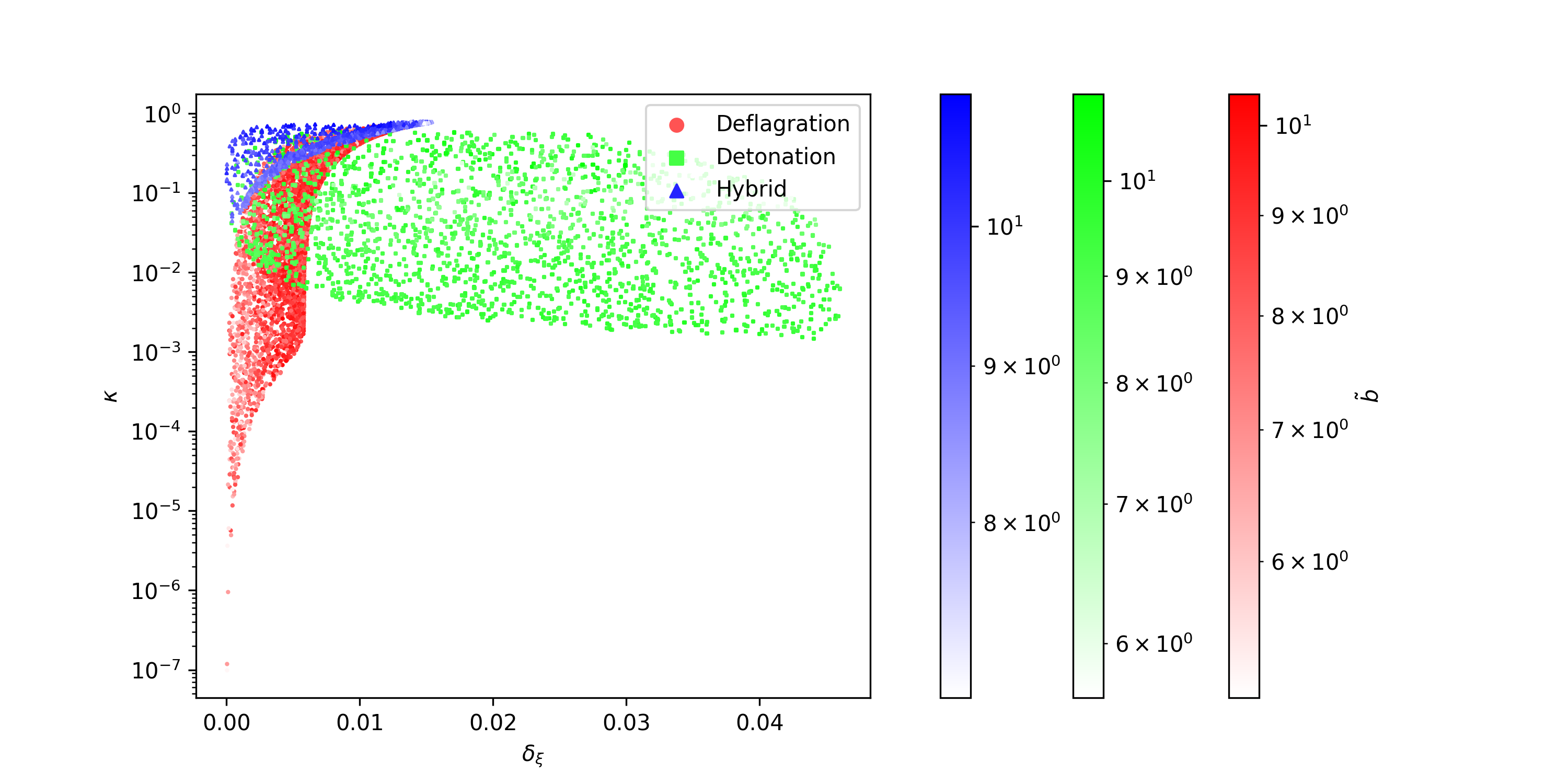}
  \includegraphics[width=0.8\linewidth]{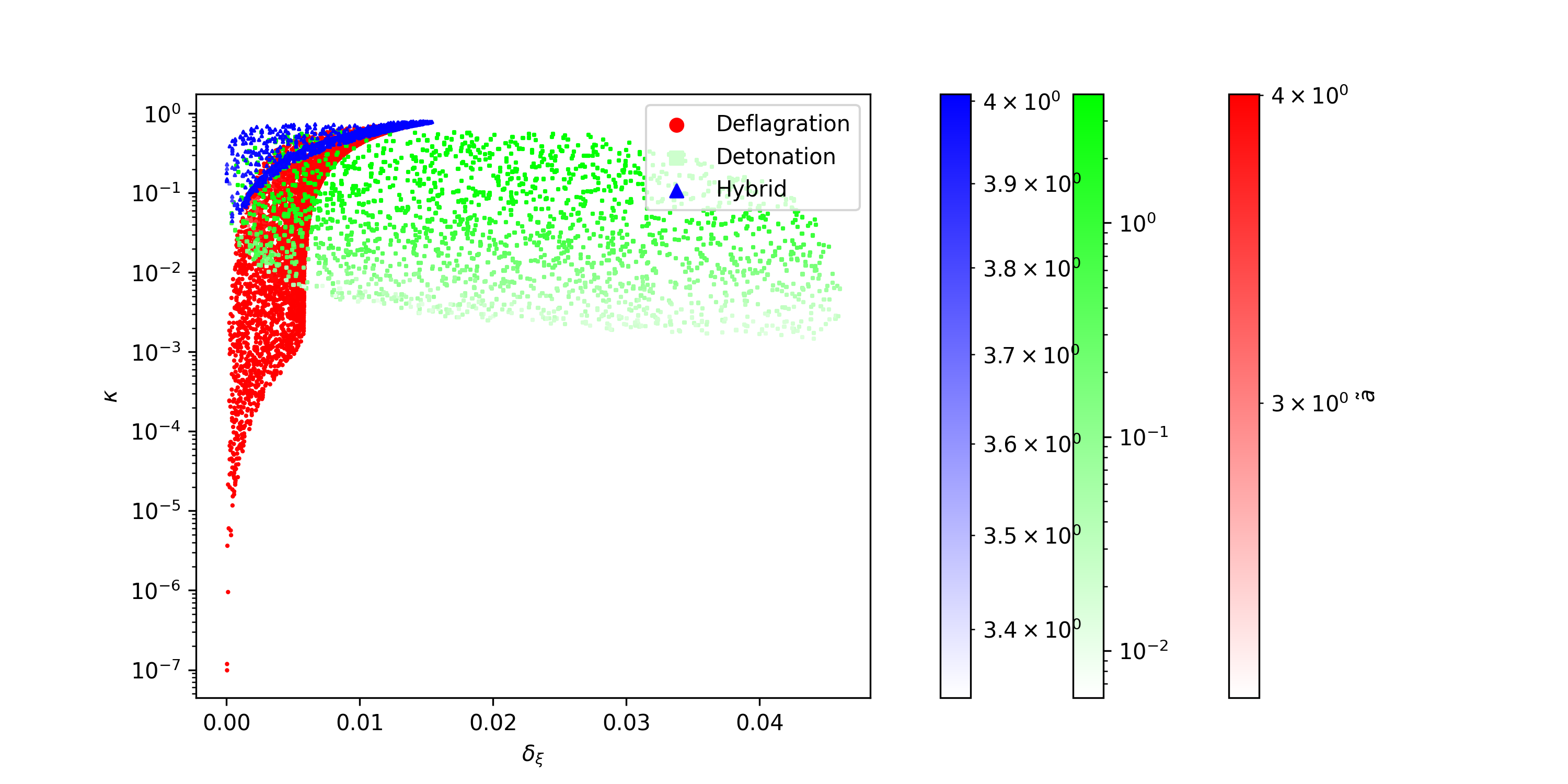}
    \caption{From top to bottom: scatter plots for $\delta_\xi$, $\kappa$ and $\tilde{r}_b$; for $\delta_\xi$, $\kappa$ and $\tilde{b}$; and for $\delta_\xi$, $\kappa$ and $\tilde{a}$. The type of FOPT is  shown in the legend. Here we assume $T_{n} = 100$~ GeV and $\beta/H = 1$.}
  \label{fig:scan_kappa_delta_xi}
\end{figure}

We have shown other scanning plots over $\delta_\xi$ and $\kappa$ parameter space in this section in Fig.~\ref{fig:scan_kappa_delta_xi}. We also represented the variation of some of our new fit parameters: $\tilde{r}_b$, $\tilde{b}$ and $\tilde{a}$ as introduced in Eq~(\ref{eq:newfit}).
In first panel we have shown the plot for $\tilde{r}_b$ i.e. the ratio of frequencies between two peaks.
As Fig.~\ref{fig:scan_kappa_delta_xi} shows large values of $\tilde{r}_b$ are accumulating between $0\lesssim \delta_{\xi} \lesssim 0.01$  and $10^{-2}\lesssim \kappa \lesssim 0.5$ for all regimes and $\tilde{r}_b$ decreases as $\delta_\xi$ and $\kappa$ are changing from these regions. 
Also, the second panel shows the changes parameter $\tilde{b}$ w.r.t. $\delta_\xi$ and $\kappa$. 
For deflagration and hybrid regimes increasing $\delta_\xi$ and $\kappa$ makes $\tilde{b}$ larger. However, for the detonation regime $\tilde {b}$ slightly reduces by decreasing of $\kappa$.  
In the last panel we examine the variation of parameter $\tilde{a}$ that deviates from $4$ depending on the values of $\delta_\xi$ and $\kappa$. 
As it can be seen the points in deflagration and hybrid regimes are almost independent of $\delta_\xi$ and $\kappa$. However, as $\kappa$ increases $\tilde{a}$ also increases for detonation regime.

\section{GW Spectrum for a Specific Choice of FOPT Paramters}
\label{app-b}
Here we briefly explain how one can use our script to calculate the spectrum of GW from sound waves of FOPT. We provide two scripts written in {\tt Python} and {\tt Mathematica}.

\subsection{{\tt Mathematica} script}
\label{sec:math}
The {\tt Mathematica} script is displayed below:

\begin{lstlisting}[language=Mathematica, caption=Mathematica code to calculate the spectrum of GW from sound waves of FOPT.]
(*Load the CSV file*)
data = Import[NotebookDirectory[] <> "data-fopt-fit.csv"];

(*Extract and clean column names from the first row*)
columnNames = StringTrim /@ First[data];

(*Remove the header row to isolate the data*)
data = Rest[data];

(*Print column names to check*)
Print[columnNames];

(*Assign each column to a separate variable using safe access to \
Position*)
vw = If[Length[Position[columnNames, "vw"]] > 0, 
   data[[All, Position[columnNames, "vw"][[1, 1]]]], {}];
alpha = If[Length[Position[columnNames, "alpha"]] > 0, 
   data[[All, Position[columnNames, "alpha"][[1, 1]]]], {}];
betaH = If[Length[Position[columnNames, "betaH"]] > 0, 
   data[[All, Position[columnNames, "betaH"][[1, 1]]]], {}];
Tn = If[Length[Position[columnNames, "Tn"]] > 0, 
   data[[All, Position[columnNames, "Tn"][[1, 1]]]], {}];
fp = If[Length[Position[columnNames, "f_p"]] > 0, 
   data[[All, Position[columnNames, "f_p"][[1, 1]]]], {}];
Omp = If[Length[Position[columnNames, "Om_p"]] > 0, 
   data[[All, Position[columnNames, "Om_p"][[1, 1]]]], {}];
at = If[Length[Position[columnNames, "at"]] > 0, 
   data[[All, Position[columnNames, "at"][[1, 1]]]], {}];
bt = If[Length[Position[columnNames, "bt"]] > 0, 
   data[[All, Position[columnNames, "bt"][[1, 1]]]], {}];
rbt = If[Length[Position[columnNames, "rbt"]] > 0, 
   data[[All, Position[columnNames, "rbt"][[1, 1]]]], {}];
s0t = If[Length[Position[columnNames, "s0t"]] > 0, 
   data[[All, Position[columnNames, "s0t"][[1, 1]]]], {}];
omega0t = 
  If[Length[Position[columnNames, "omega_0t"]] > 0, 
   data[[All, Position[columnNames, "omega_0t"][[1, 1]]]], {}];
K = If[Length[Position[columnNames, "K"]] > 0, 
   data[[All, Position[columnNames, "K"][[1, 1]]]], {}];

(*Define the data points*)
dataVWAlphaFP = Transpose[{vw, alpha, fp}];
dataVWAlphaOmP = Transpose[{vw, alpha, Omp}];
dataVWAlphaAt = Transpose[{vw, alpha, at}];
dataVWAlphaBt = Transpose[{vw, alpha, bt}];
dataVWAlphaRbt = Transpose[{vw, alpha, rbt}];
dataVWAlphaS0t = Transpose[{vw, alpha, s0t}];
dataVWAlphaOmega0t = Transpose[{vw, alpha, omega0t}];
dataVWAlphaK = Transpose[{vw, alpha, K}];


(*Create interpolating functions*)
intFP = Interpolation[dataVWAlphaFP, InterpolationOrder -> 1];
intOmP = Interpolation[dataVWAlphaOmP, InterpolationOrder -> 1];
intAt = Interpolation[dataVWAlphaAt, InterpolationOrder -> 1];
intBt = Interpolation[dataVWAlphaBt, InterpolationOrder -> 1];
intRbt = Interpolation[dataVWAlphaRbt, InterpolationOrder -> 1];
intS0t = Interpolation[dataVWAlphaS0t, InterpolationOrder -> 1];
intOmega0t = 
  Interpolation[dataVWAlphaOmega0t, InterpolationOrder -> 1];
intK = Interpolation[dataVWAlphaK, InterpolationOrder -> 1];

(*Evaluate at a point*)
vw0 = 0.5;
alpha0 = 0.5;
fp0 = intFP[vw0, alpha0];
at0 = intAt[vw0, alpha0];
bt0 = intBt[vw0, alpha0];
rbt0 = intRbt[vw0, alpha0];
s0t0 = intS0t[vw0, alpha0];
Omegap0 = intOmega0t[vw0, alpha0];
K0 = intK[vw0, alpha0];

(*Degrees of freedom*)
gstarData = Import[NotebookDirectory[] <> "gstar.txt", "Table"];
gHighTemp = {{1.*10^14*gstarData[[1, 1]], gstarData[[1, 2]]}};
dataG1 = Join[gHighTemp, gstarData];
TData = dataG1[[All, 1]];
gstarData = dataG1[[All, 2]];
gstarFun = Interpolation[Transpose[{TData, gstarData}]];

(*Frequency domain*)
fdomainDefault = 10^Range[-24, 24, 48/999] // N;

(*Generating fit parameters from the interpolated function*)
fb0 = rbt0*fp0;
(*This part can be modified based on different nucleation \
tamperatures and bubble nucleation rates.*)
betaHDefault = 1;
betaHNew = 10;
TnDefault = 100;
TnNew = 1000;

rstarDefault = (8 \[Pi])^(1/3)*vw0/betaHDefault;
rstarNew = (8 \[Pi])^(1/3)*vw0/betaHNew;
gstarDefault = gstarFun[TnDefault];
gstarNew = gstarFun[TnNew];

xDefault = rstarDefault/(K0^(1/2));
JDefault = rstarDefault*(1 - 1/(1 + 2*xDefault)^(1/2));
FgwDefault = 3.57*10^-5*(100/gstarDefault)^(1/3);
xNew = rstarNew/(K0^(1/2));
JNew = rstarNew*(1 - 1/(1 + 2*xNew)^(1/2));
FgwNew = 3.57*10^-5*(100/gstarNew)^(1/3);

f0Default = 2.6*10^-6*(TnDefault/100)*(gstarDefault/100)^(1/6);
kR = fdomainDefault*(rstarDefault/f0Default);


(*GW spectrum-default*)
omegaFitDefault = 
   Omegap0*(fdomainDefault/
       s0t0)^9*((2 + 
        rbt0^(-12 + bt0))/((fdomainDefault/s0t0)^
         at0 + (fdomainDefault/s0t0)^bt0 + 
        rbt0^(-12 + bt0)*(fdomainDefault/s0t0)^12)); // Quiet



f0New = 2.6*10^-6*(TnNew/100)*(gstarNew/100)^(1/6);
fdomainNew = kR/(rstarNew/f0New);
omegaFitNew = (omegaFitDefault/(FgwDefault*JDefault))*(FgwNew*JNew);


(*Plotting*)
ListLogLogPlot[{Transpose[{fdomainDefault, omegaFitDefault}], 
  Transpose[{fdomainNew, omegaFitNew}]}, 
 PlotRange -> {{1.*10^-12, 1.*10^4}, {1.*10^-22, 1.*10^-2}}, 
 Frame -> True, 
 FrameLabel -> {"f (Hz)", 
   "\!\(\*SubscriptBox[\(\[CapitalOmega]\), \
\(GW\)]\)\!\(\*SuperscriptBox[\(h\), \(2\)]\)"}, 
 PlotLabel -> "FOPT - GW", GridLines -> Automatic, Joined -> True, 
 PlotLegends -> {"Tn = 100 GeV, beta/H = 1", 
   "Tn = 1000 GeV, beta/H = 10"}, FrameStyle -> 18, ImageSize -> 800]



\end{lstlisting}
\subsection{{\tt Python} script}
\label{sec:pyth}
We have written the following code in {\tt Python} to calculate GW from the phase transitions: 

\begin{lstlisting}[language=Python, caption=Python code to calculate the spectrum of GW from sound waves of FOPT.]
# Python script for this manuscript
import numpy as np
from scipy.interpolate import LinearNDInterpolator
from scipy.interpolate import interp1d
import matplotlib.pyplot as plt
import pandas as pd

# Load the CSV file into a DataFrame
df = pd.read_csv('data-fopt-fit.csv')

# Assign each column to a separate variable
vw = df['vw'].to_numpy()
alpha = df['alpha'].to_numpy()
betaH = df['betaH'].to_numpy()
Tn = df['Tn'].to_numpy()
f_p = df['f_p'].to_numpy()
Om_p = df['Om_p'].to_numpy()
at = df['at'].to_numpy()
bt = df['bt'].to_numpy()
rbt = df['rbt'].to_numpy()
s0t = df['s0t'].to_numpy()
omega_0t = df['omega_0t'].to_numpy()
K = df['K'].to_numpy()

# Interpolatoin over data (t : tilde)
int_f_p = LinearNDInterpolator(list(zip(vw, alpha)), f_p)
int_Om_p = LinearNDInterpolator(list(zip(vw, alpha)), Om_p)
int_at = LinearNDInterpolator(list(zip(vw, alpha)), at)
int_bt = LinearNDInterpolator(list(zip(vw, alpha)), bt)
int_rbt = LinearNDInterpolator(list(zip(vw, alpha)), rbt)
int_s0t = LinearNDInterpolator(list(zip(vw, alpha)), s0t)
int_omega_0t = LinearNDInterpolator(list(zip(vw, alpha)), omega_0t)
int_K = LinearNDInterpolator(list(zip(vw, alpha)), K)

vw0 = 0.5
alpha0 = 0.5 

# Generatng fit parameters from the interpolated function
f_p0 = int_f_p(vw0, alpha0)
at0 = int_at(vw0, alpha0)
bt0 = int_bt(vw0, alpha0)
rbt0 = int_rbt(vw0, alpha0)
s0t0 = int_s0t(vw0, alpha0)
omega_p0 = int_omega_0t(vw0, alpha0)
K0 = int_K(vw0, alpha0)

# Degrees of freedom
# arXiv: https://arxiv.org/pdf/1503.03513
filename = "gstar.txt"
data_g = np.loadtxt(filename, skiprows=0)
g_high_temp = np.array([[1.e14 * data_g[0, 0], data_g[0, 1]]])
data_g1 = np.concatenate((g_high_temp, data_g))
T_data = data_g1[:, 0]  # [GeV]
gstar_data = data_g1[:, 1]  # [GeV]
gstar_fun = interp1d(T_data, gstar_data)
#
f_b0 = rbt0 * f_p0
# This part can be modified based on different nucleation tamperatures and bubble nucleation rates.
# beta over H
betaH_default = 1
betaH_new = 10
# Nucleation temperature
Tn_default = 100 # GeV
Tn_new = 1000 # GeV
# 
rstar_default = (8*np.pi)**(1/3)*vw0/betaH_default
rstar_new = (8*np.pi)**(1/3)*vw0/betaH_new
# Degrees of Freedom
gstar_default = gstar_fun(Tn_default)
gstar_new = gstar_fun(Tn_new)
#
x_default = rstar_default/(K0**(1/2))
J_default = rstar_default*(1-1/(1+2*x_default)**(1/2))
Fgw_default = 3.57*1e-5*(100/gstar_default)**(1/3)
x_new = rstar_new/(K0**(1/2))
J_new = rstar_new*(1-1/(1+2*x_new)**(1/2))
Fgw_new = 3.57*1e-5*(100/gstar_new)**(1/3)

# Frequency
fdomain_default = 10**np.linspace(-24,24,1000) 

f0_default = 2.6*1.e-6*(Tn_default/100)*(gstar_default/100)**(1/6)
omega_p0_default = omega_p0
kR = fdomain_default * (rstar_default/f0_default) 

# GW spectrum - default
omega_fit_default = omega_p0_default * (fdomain_default/s0t0)**9 * (
    (2 + rbt0**(-12 + bt0)) / 
    ((fdomain_default/s0t0)**at0 + (fdomain_default/s0t0)**bt0 + 
     rbt0**(-12 + bt0) * (fdomain_default/s0t0)**12))

f0_new = 2.6*1.e-6*(Tn_new/100)*(gstar_new/100)**(1/6)
fdomain_new = kR / (rstar_new/f0_new) 
omega_fit_new = (omega_fit_default  / (Fgw_default*J_default) ) * (Fgw_new*J_new)


# Plotting
plt.figure(figsize=(10, 6))
plt.loglog(fdomain_default, omega_fit_default, label='FOPT - Sound Waves - GW - $T_n = 100$ GeV and $\\beta /H = 1$')
plt.loglog(fdomain_new, omega_fit_new, label='FOPT - Sound Waves - GW')

# Set limits for frequency and GW relic axes
plt.xlim(1.e-12, 1.e4)  
plt.ylim(1.e-22, 1.e-2)  
plt.legend()
plt.xlabel('f (Hz)')
plt.ylabel('$\Omega_{GW}h^2$(f)')
plt.grid(True)
plt.show()

\end{lstlisting}

Using the above script and the files provided along with this paper one can generate the predicted GW spectrum by inserting the input parameters from a beyond Standard Model scenario and the cosmological scale that the phase transition occurs at. These input parameters are $\alpha$, $v_w$, $\beta/H$ and $T_n$: with these, one can produce the GW spectrum from the above {\tt Python} and {\tt Mathematica} scripts \footnote{\href{https://github.com/SFH2024/precise-fit-fopt-gw}{{\tt GitHub} link}.}.

\bibliography{biblio.bib}
\end{document}